\documentclass{article}
\usepackage{graphicx,emumine}
\def\huntsf{in C. Meegan, R. Preece \& T. Koshut, Eds.,
{\it Gamma-Ray Bursts 4th Huntsville Symposium} AIP Conf. Proc.   428
(New York: AIP)}
\def\huntso{in  W. S. Paciesas, \&  G. J. Fishman Eds.,
{\it  Gamma-Ray Bursts} AIP Conf. Proc.  265,
(New York: AIP) .} 
\def\huntstw{in Fishman, G. J., \& Brainerd, J. J.,  Eds.,
{\it  Gamma-Ray Bursts 2nd Huntsville Symposium} AIP Conf. Proc. 307
(New York: AIP) .} 
\def\huntst{ in Kouveliotou, C.Briggs M.S. \& G.J. Fishman Eds.  
{\it Gamma-Ray Bursts 3rd Huntsville Symposium}, AIP Conf. Proc. 384,
(New York: AIP).}
\def\etall{{\it et al. }}
\def\etal,{{\it et al., }}
\def\O{\Omega}
\def\zm{z_{max}}
\def\ga{\gamma}
\def\g100{\gamma_{100}}
\def\E52{E_{52}}
\def\d10{\delta_{10}}
\def\D12{\Delta_{12}}

\def\tge{{c_\gamma}}

\def\z1{{(1+z)/ 1.835}}

\def\t2day{({t/ 2{\rm days}})}

\def\sles{\lower2pt\hbox{$\buildrel {\scriptstyle <}
   \over {\scriptstyle\sim}$}}
\def\sgreat{\lower2pt\hbox{$\buildrel {\scriptstyle >}
   \over {\scriptstyle\sim}$}}
\def\E051{(E/10^{51}{\rm ergs})}
\def\R07{(R_i/10^7{\rm cm})}
\def\hmpc{\ {\rm h^{-1}Mpc}}
\def\pmb#1{\setbox0=\hbox{#1}
 \kern-.025em\copy0\kern-\wd0
 \kern.s05em\copy0\kern-\wd0}

\begin{document}

\title{Gamma-Ray Bursts and the Fireball Model}
\author{Tsvi Piran} 
\affil{Racah Institute for Physics, The Hebrew University, 
Jerusalem,  91904, Israel\footnote{permanent address} \\
and \\
Physics Department, Columbia University, New York, NY 10027, USA}

\begin{abstract}
Gamma-ray bursts (GRBs) have puzzled astronomers since their
accidental discovery in the late sixties. The BATSE detector on the
COMPTON-GRO satellite has been detecting one burst per day for the
last six years. Its findings have revolutionized our ideas about the
nature of these objects. They have shown that GRBs are at cosmological
distances.  This idea was accepted with difficulties at first. The
recent discovery of an X-ray afterglow by the Italian/Dutch satellite
BeppoSAX has led to a detection of high red-shift absorption lines in
the optical afterglow of GRB970508 and in several other bursts and to
the identification of host galaxies to others. This has confirmed the
cosmological origin. Cosmological GRBs release $\sim
10^{51}-10^{53}$ergs in a few seconds making them the most
(electromagnetically) luminous objects in the Universe. The simplest,
most conventional, and practically inevitable, interpretation of these
observations is that GRBs result from the conversion of the kinetic
energy of ultra-relativistic particles or possibly the electromagnetic
energy of a Poynting flux to radiation in an optically thin
region. This generic ``fireball" model has also been confirmed by the
afterglow observations. The ``inner engine" that accelerates the
relativistic flow is hidden from direct observations. Consequently it
is difficult to infer its structure directly from current
observations. Recent studies show, however, that this ``inner engine''
is responsible for the complicated temporal structure observed in
GRBs.  This temporal structure and energy considerations indicates
that the ``inner engine'' is associated with the formation of a
compact object - most likely a black hole.  
\end{abstract}

\section{Introduction}
\label{sec:intro}

Gamma-ray bursts (GRBs), short and intense bursts of $\sim
100$keV-1MeV photons, were discovered accidentally in the late sixties by
the Vela satellites\cite{Kle}. The mission of these satellites was to
monitor the ``Outer Space Treaty" that forbade nuclear explosions in
space.  A wonderful by-product of this effort was the discovery of
GRBs.

The discovery of GRBs was announced in 1973 \cite{Kle}.  It was
confirmed quickly by Russian observations \cite{Russian} and by
observations on the IMP-6 satellite \cite{Cline73}.  Since then,
several dedicated satellites have been launched to observe the bursts
and numerous theories were put forward to explain their origin.  
Claims of observations of cyclotron spectral lines and 
of discovery of optical archival counterparts
led  in
the mid eighties to a consensus  that GRBs originate from Galactic
neutron stars. This model was accepted quite generally and was even
discussed in graduate textbooks \cite{Hillier84,Murthy86,Meszaros92}
and encyclopedia articles \cite{Cline92,Luminet88}.  

The BATSE detector on the COMPTON-GRO (Gamma-Ray Observatory) was
launched in the spring of 1991.  It has revolutionized GRB
observations and consequently our basic ideas on their nature. 
BATSE observations of the isotropy of GRB directions, combined
with the deficiency of faint GRBs, 
ruled out the galactic disk neutron star model\footnote{A few GRBs,
now called soft gamma repeaters, compose a different phenomenon, 
are believed to form on galactic neutron stars.}  and make a convincing case
for their  extra-galactic origin at cosmological distances
\cite{Meegan92}. This conclusion was recently confirmed by the
discovery by BeppoSAX
\cite{Costa97a} of an X-ray transient counterparts to several GRBs. This was
followed by a discovery of optical \cite{vanParadijs97a,Bond97} and
radio transients \cite{Frail97a}. 
Absorption line 
 with a redshift $z = 0.835$
were measured in the optical spectrum of the counterpart to GRB970508
\cite{Metzger97a} providing  the first redshift of the
optical transient and the associated GRB.  
Latter,  redshifted emission lines from galaxies
associated with GRB971214 \cite{Kulkarni98a} (with $z = 3.418$) and GRB980703 
\cite{Djorgovski98a} (with $z=0.966$) were discovered. Galaxies has been
discovered at the positions of other bursts. There is little doubt now
that some, and most likely all GRBs are cosmological.

The cosmological origin of GRBs immediately implies that GRB sources
are much more luminous than previously thought. They release $\sim
10^{51}-10^{53}$ergs or more in a few seconds, the most
(electromagnetically) luminous objects in the Universe. This also
implies that GRBs are rare events. 
BATSE observes on average one burst per day. This corresponds, with
the simplest model (assuming that the rate of GRBs does not change
with cosmological time) to  one burst per million years
per galaxy. The average rate changes, of course, if we allow beaming
or a cosmic evolution of the rate of GRBs.   

In spite of those discoveries, the origin of GRBs is still mysterious.
This makes GRBs a unique phenomenon in modern astronomy. While
pulsars, quasars and X-ray sources were all explained within a few
years, if not months, after their discovery, the origin of GRBs
remains unknown after more than thirty years. The fact that GRBs are a
short transient phenomenon which until recently did not have any known
counterpart, is probably the main reason for this situation. Our
inability to resolve this riddle also reflects the accidental and
unexpected nature of this discovery which was not done by an
astronomical mission.  Theoretical astrophysics was not ripe to cope
with GRBs when they were discovered.

A generic scheme of a cosmological GRB model has emerged in the last
few years and most of this review is devoted to an exposition of this
scheme. The recently observed X-ray, optical and radio counterparts
were predicted by this picture
\cite{PacRho93,Katz94,Katz94a,SaP97a,MR97}. This discovery can, to some
extent, be considered as a confirmation of this model
\cite{Wiejers_MR97,Waxman97a,Vietri97,KP97}.  According to this scheme the
observed $\gamma$-rays are emitted when an ultra-relativistic energy
flow is converted to radiation. Possible forms of the energy flow are
kinetic energy of ultra-relativistic particles or electromagnetic
Poynting flux. This energy is converted to radiation in an optically
thin region, as the observed bursts are not thermal.  It has been
suggested that the energy conversion occurs either due to the
interaction with an external medium, like the ISM
\cite{MR1}  or due to internal process, such as internal shocks and
collisions within the flow \cite{NPP,MR4,PacXu}.  Recent work
\cite{SaP97a,Piran97} shows that the external shock scenario is quite
unlikely, unless the energy flow is confined to an extremely narrow
beam, or else the process is highly inefficient. The only alternative is
that the burst is produced by internal shocks. 

The ``inner engine"
that produces the relativistic energy flow is hidden from direct
observations. However, the observed temporal structure reflects
directly this ``engine's'' activity.  This model requires a compact
internal ``engine'' that produces a wind -- a long energy flow (long
compared to the size of the ``engine'' itself) -- rather than an
explosive ``engine" that produces a fireball whose size is comparable
to the size of the ``engine''.  Not all the energy of the relativistic
shell can be converted to radiation (or even to thermal energy) by
internal shocks \cite{MMM95,KPS97,Mosh97}. The remaining kinetic energy
will most likely dissipate via external shocks that will produce an
``afterglow'' in different wavelength \cite{SaP97a}. This afterglow
was recently discovered, confirming the fireball picture.

At present there is no agreement on the nature of the ``engine'' -
even though binary neutron star mergers \cite{Eichler89} are a
promising candidate. All that can be said with some certainty is that
whatever drives a GRB must satisfy the following general features: It
produces an extremely relativistic energy flow containing $\approx
10^{51}-10^{52}$ergs. The flow is highly variable as most bursts have
a variable temporal structure and it should last for the duration of
the burst (typically a few dozen seconds). It may continue at a lower
level on a time scale of a day or so \cite{KaPS97}.  Finally, it
should be a rare event occurring about once per million years in a
galaxy. The rate is of course higher and the energy is lower if there
is a significant beaming of the gamma-ray emission. In any case the
overall GRB emission in $\gamma$-rays is
$\sim 10^{52}$ergs/10$^6$years/galaxy.

We begin (section \ref{sec:obs}) with a brief review of GRB
observation (see \cite{Fishman95,Greatdebate0,Briggs,Kou0,Hartmann95}
for additional reviews and \cite{Hunts1,Hunts2,Hunts3,Hunts4} for a
more extensive discussion). We then turn to an analysis of the
observational constraints. We analyze the peak intensity distribution
and show how the distance to GRBs can be estimated from this data. We
also discuss the evidence for another cosmological effect:
time-dilation (section \ref{sec:peak}).  We then turn (section
\ref{sec:compact}) to discuss the optical depth or the compactness
problem.  We argue that the only way to overcome this problem is if the sources
are moving at an ultra-relativistic velocity towards us. An essential
ingredient of this model is the notion of a fireball - an optically
thick relativistic expanding electron-positron and photon plasma (for
a different model see however \cite{Pac97}). We discuss fireball
evolution in section \ref{sec:fireball}. Kinematic considerations
which determine the observed time scales from emission emerging from a
relativistic flow provides important clues on the location of the
energy conversion process. We discuss these constraints in section
\ref{sec:Tempstruct} and the energy conversion stage in section
\ref{sec:conv}. We review the recent theories of afterglow formation
\ref{sec:afterglow}. We examine the confrontation of these models with
observations and we discuss some of the quantitative problems.

We then turn to the ``inner engine'' and review the recent suggestions
for cosmological models (section \ref{sec:models}).  As this inner
engine is hidden from direct observation, it is clear that there are
only a few direct constraint that can be put on it.  Among GRB models,
binary neutron star merger \cite{Eichler89} is  unique. It
is the only model that is based on an independently observed
phenomenon \cite{Hulse75}, is capable of releasing the required
amounts of energy \cite{Clark77} within a very short time scale and
takes place at approximately the same rate \cite{NPS,Phinney91,Heuvel_Lorimer}
\footnote{this is assuming that there is no strong cosmic evolution in the 
rate of GRB}.  At present it is not clear if this merger can actually
channel the required energy into a relativistic flow or if it could
produce the very high energy observed in GRB971214.  However, in view
of the special status of this model we discuss its features and the
possible observational confirmation of this model in section
\ref{sec:ns2m}.

GRBs might have important implications to other branches of astronomy.
Relation of GRBs to other astronomical phenomena such as UCHERs,
neutrinos and gravitational radiation are discussed in section
\ref{sec:other}.  The universe and our Galaxy are optically thin to
low energy $\gamma$-rays.  Thus, GRBs constitute a unique cosmological
population that is observed practically uniformly on the sky (there
are small known biases due to CGRO's observation schedule). Most of these 
objects are located at $z\approx 1$ or greater. Thus this population is
farther than any other systematic sample (QSOs are at larger distances
but they suffer from numerous selection effects and there is no all
sky QSOs catalog). GRBs are, therefore, an ideal tool to explore the
Universe.  Already in 1986 Paczy\'nski \cite{Pac86} proposed that
GRBs might be gravitationally lensed. This has led to the suggestion
to employ the statistics of lensed bursts to probe the nature of the
lensing objects and the dark matter of the Universe \cite{Blase92}.
The fact that no lensed bursts where detected so far is sufficient to
rule out a critical density of $10^{6.5} M_\odot$ to $ 10^{8.1}
M_\odot$ black holes \cite{Nemiroff93a}.  Alternatively we may use the
peak-flux distribution to estimate cosmological parameters such as
$\Omega$ and $\Lambda$ \cite{Pi92}.  The angular distribution of GRBs
can be used to determine the very large scale structure of the
Universe \cite{Lamb93a,Piran_Singh97}. The possible direct
measurements of red-shift to some bursts enhances greatly the
potential of these attempts. We conclude in section \ref{sec:cosm} by
summarizing these suggestions.

Over the years several thousand papers concerning GRBs have appeared in the
literature. With the growing interest in GRBs the number  of  GRB  papers 
has been growing at an accelerated rate  recently .
It is, of course, impossible to summarize or even list all this papers here.
I refer the interested reader to the complete GRB bibliography that was
prepared by K. Hurley \cite{Hurley98}.

\section{Observations}
\label{sec:obs}

GRBs are short, non-thermal bursts of low energy $\gamma$-rays.  It is
quite difficult to summarize their basic features.  This difficulty
stems from the enormous variety displayed by the bursts.  I will
review here some features that I believe hold the key to this enigma.
I refer the reader to the proceedings of the Huntsville GRB meetings
\cite{Hunts1,Hunts2,Hunts3,Hunts4} and to other recent reviews
for a more detailed discussion
\cite{Fishman95,Greatdebate0,Briggs,Kou0,Hartmann95}.

\subsection{Duration:}

A ``typical'' GRB (if there is such a thing) lasts about 10sec.
However, observed durations vary by six orders of magnitude, from
several milliseconds \cite{Fishman93} to several thousand seconds
\cite{Klebedasel84}.  About 3\% of the bursts are preceded by a
precursor with a lower peak intensity than the main burst
\cite{Koshut}. Other bursts were followed by low energy X-ray tails
\cite{Laros95}. Several bursts observed by the GINGA detector showed
significant apparently thermal, X-ray emission before and after the
main part of the higher energy emission \cite{Murakami91,Yoshida89}.
These are probably pre-discovery detections of the X-ray afterglow
observed now by BeppoSAX and other X-ray detectors.

The definition of duration is, of course, not unique.  BATSE's team
characterizes it using $T_{90}$ ($T_{50}$) the time needed to
accumulate from 5\% to 95\% (from 25\% to 75\%) of the counts in the
50keV - 300keV band. The shortest BATSE burst had a duration of 5ms
with structure on scale of 0.2ms \cite{Bhat92}. The longest so far,
GRB940217, displayed GeV activity one and a half hours the main burst
\cite{Hurley94a}. The bursts GRB961027a, GRB961027b, GRB961029a and
GRB961029b occurred from the same region in the sky within two days
\cite{Meegan96a} if this ``gang of four'' is considered as a single
very long burst then the longest duration so far is two days!  These
observations may indicate that some sources display a continued
activity (at a variable level) over a period of days \cite{Katz97}.
It is also possible that the observed afterglow is an indication of
a continued activity \cite{KaPS97}.

The distribution of burst durations is bimodal.  BATSE confirmed
earlier hints \cite{Hurley92} that the burst duration distribution can
be divided into two sub-groups according to $T_{90}$: long bursts with
$T_{90}>2$sec and short bursts with $T_{90}<2$sec
\cite{Mazet81,Kou93,Lamb93,MNP,Klebesadel92,Dezaley92}.  The ratio of observed
long bursts to observed short bursts is three to one. This does not
necessarily mean that there are fewer short bursts.  BATSE's triggering
mechanism makes it less sensitive to short bursts than to long
ones. Consequently short bursts are detected to smaller distances
\cite{MNP,CKP,Piran96,Katz96} and we observed a smaller number of short
bursts.

\subsection{Temporal Structure and Variability}
\label{sub-sec:variability}

\begin{figure}%
\begin{center}
\includegraphics[width=10cm]{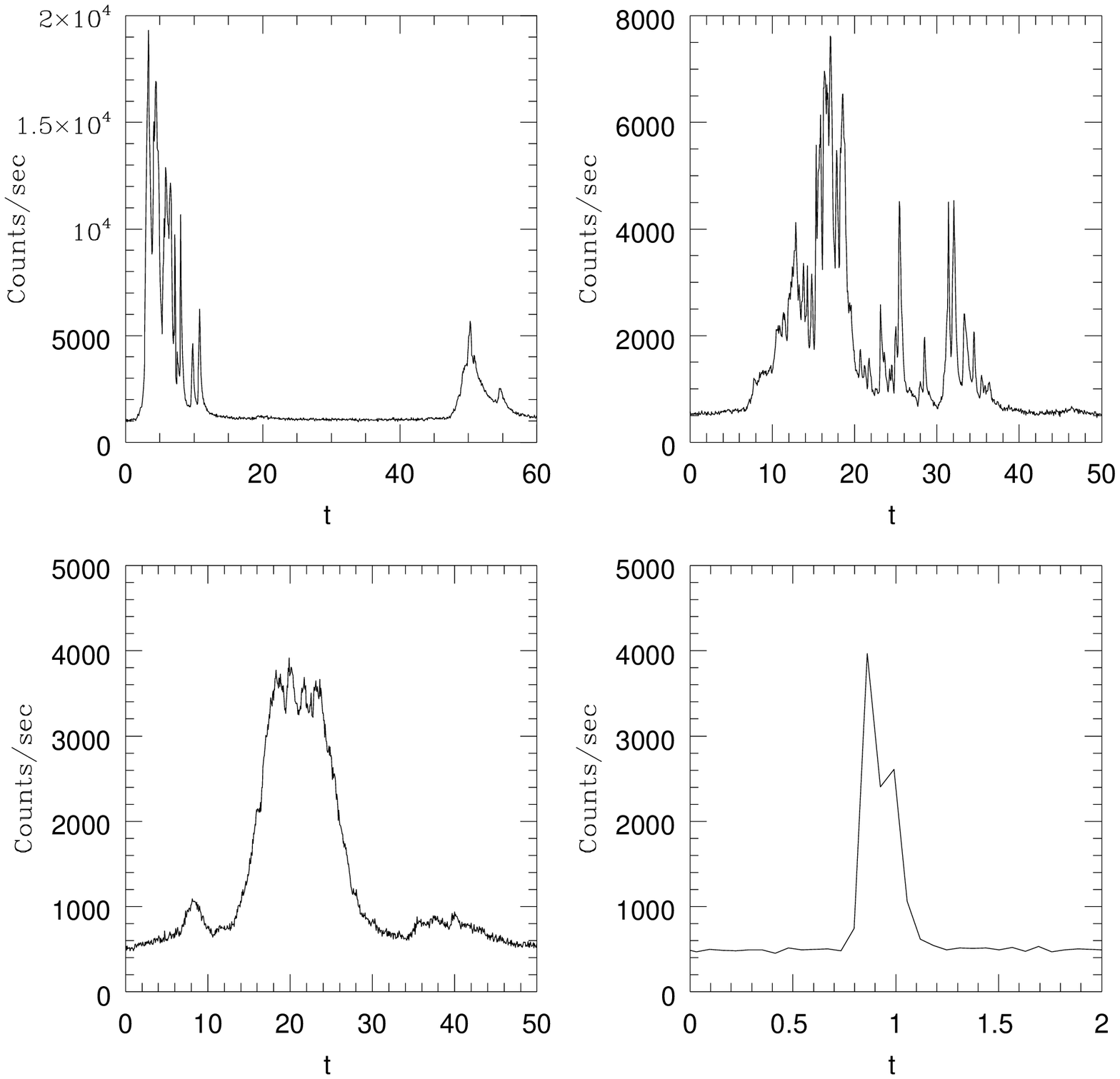}
\caption{\it Total number of counts vs. time for  several  bursts from the BATSE
Catalogue. Note the large diversity of temporal structure observed.}
\label{f:temporal}
\end{center}
\end{figure}

\begin{figure}
\begin{center}
\includegraphics[width=10cm]{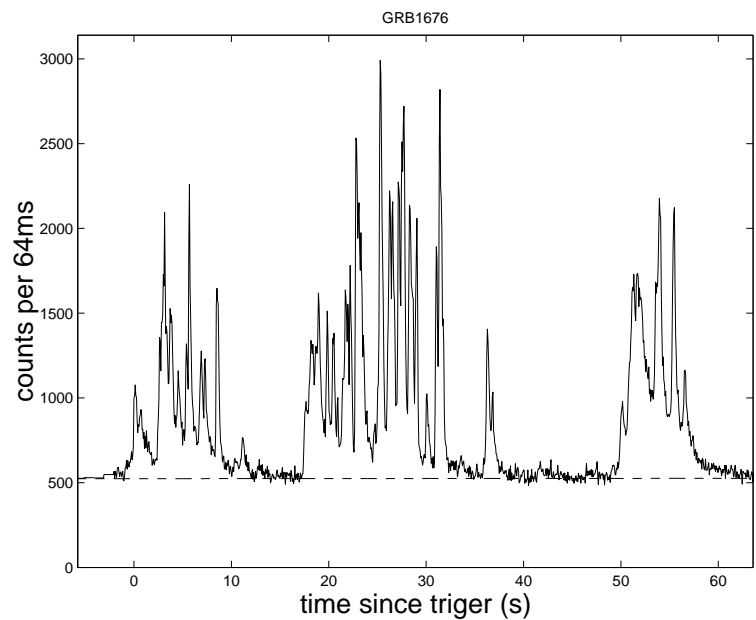}
\caption{\it Counts vs. time for BATSE burst 1676. The bursts lasted $T\sim 60\sec$
and it had peaks of width $\delta T\sim 1\sec $, leading to ${\cal N}
\approx 60$.}
\label{f:temporal_var}
\end{center}
\end{figure}

The bursts have a complicated and irregular time profiles which vary
drastically from one burst to another.  Several time profiles,
selected from the second BATSE catalog, are shown in Fig.
\ref{f:temporal}.  In most bursts, the typical variation takes place on
a time-scale $\delta T$ significantly smaller than the total duration
of the burst, $T$.  In a minority of the bursts there is only one peak
with no substructure and in this case $\delta T \sim T$.  It turns out
that the observed variability provides an interesting clue to the
nature of GRBs. We discuss this in section \ref{sec:Tempstruct}.  We
define the ratio ${\cal N}\equiv T/\delta T$ which is a measure of the
variability. Fig. \ref{f:temporal_var} depicts the total observed
counts (at $E>25$keV) from GRB1676. The bursts lasted $T\sim 100\sec$
and it had peaks of width $\delta T\sim 1\sec $, leading to ${\cal N}
= 100$.

\subsection{Spectrum:}
\label{sub_sec:spectrum}

\begin{figure}
\begin{center}
\includegraphics[width=10cm]{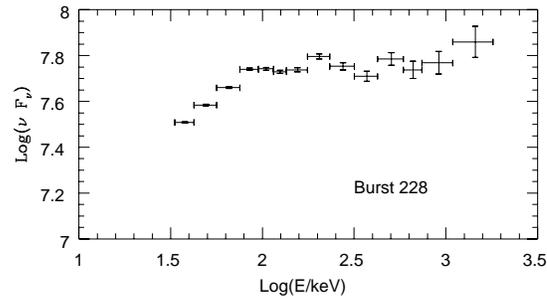}
\caption{\it Observed spectrum of BATSE' burst 228.}
\label{f:spectrum}
\end{center}
\end{figure}

GRBs are characterized by emission in  the few hundred keV ranges with
a non-thermal spectrum (see Fig. \ref{f:spectrum}) X-ray emission is
weaker -- only a few percent of the energy is emitted below 10keV
and prompt emission at lower energies has  not been observed so far. The
current best upper limits on such emission are given by LOTIS.  For
GRB970223 LOTIS finds m$_V > 11$ and provides an upper limit on the
simultaneous optical to gamma-ray fluence ratio of $ < 1.1 \times
10^{-4}$ \cite{Parketl97}.  Most bursts are accompanied, on the other
hand, by a high energy tail which contains a significant amount of
energy -- $E^2 N(E)$ is almost a constant.  GRB940217, for example,
had a high energy tail up to 18 GeV\cite{Hurley94}.  In fact EGRET and
COMPTEL (which are sensitive to higher energy emission but have a
higher threshold and a smaller field of view) observations are
consistent with the possibility that all bursts have high energy tails
\cite{Dingus97,Kippen96}.

An excellent phenomenological fit for the spectrum was introduced by
Band \etall \cite{Band93}:
\begin{equation}
N(\nu) = N_0 \cases { \big({h\nu})^{\alpha} \exp (-{h \nu \over E_0}) & for
$ h\nu < H$ ;\cr
\big[{(\alpha-\beta) E_0 }
\big]^{(\alpha-\beta)} \big({h \nu }\big)^\beta 
& \cr \ \ \ \times \exp (\beta-\alpha), &
for $h \nu > H,$
\cr}
\end{equation}
where $H\equiv(\alpha-\beta)E_0$.
There is no particular theoretical model that predicts this spectral
shape. Still, this function provides an excellent fit to most of the
observed spectra. It is characterized by two power laws joined
smoothly at a break energy $H$.  For most
observed values of $\alpha$ and $\beta$, $\nu F_\nu \propto \nu^2
N(\nu)$ peaks at $E_p = (\alpha+2)E_0 = [(\alpha+2)/(\alpha-\beta)]
H$.  The ``typical'' energy of the observed radiation is $E_p$. That
is this is where the source emits the bulk of its luminosity. $E_p$
defined in this way should not be confused with the hardness ratio
which is commonly used in analyzing BATSE's data, namely the ratio of
photons observed in channel 3 (100-300keV) to those observed in
channel 2 (50-100keV). Sometimes we will use a simple power law 
fit to the spectrum: 
\begin{equation} 
N(E) dE \propto E^{-\alpha} dE . \label{spec}
\end{equation} 
In these cases the power law index will be denoted by $\alpha$. 
A typical spectra index is  
$\alpha \approx 1.8-2$ \cite{Schaefer92b}.

In several cases the spectrum was observed simultaneously by several
instruments.  Burst 9206022, for example, was observed simultaneously
by BATSE, COMPTEL and Ulysses.  The time integrated spectrum on those
detectors, which ranges from 25keV to 10MeV agrees well with a Band
spectrum with: $E_p=457 \pm 30$keV, $\alpha = -0.86\pm 0.15$ and
$\beta = -2.5 \pm 0.07$ \cite{Greiner94}.  Schaefer \etall
\cite{Schaefer_etal98} present a complete spectrum from 2keV to 500MeV
for three bright bursts.

\begin{figure}
\begin{center}
\includegraphics[width=10cm]{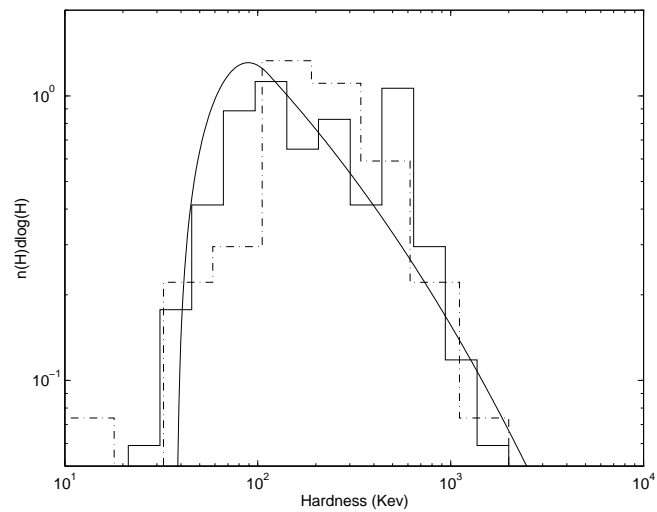}
\caption{\it N(H) -  the number of bursts with hardness , $H$, 
in the Band \etall \cite{Band93} sample (dashed-dotted line) and in
the Cohen \etall sample (solid line) \cite{CNP97} togather with a
theoretical fit of a distribution above $H=120 KeV$ with $\gamma \sim
-0.5$ (a slowly decreasing numbers of GRBs per decade of hardness)}.
\label{f:spectrum_distribution}
\end{center}
\end{figure}

Fig. \ref{f:spectrum_distribution} shows the distribution of observed
values of $H$ in several samples \cite{Band93,Mallozi95,CNP97}. Most
of the bursts are the range $100\,{\rm keV}<H<400\,{\rm keV}$, with a
clear maximum in the distribution around $H\sim 200$keV.  There are
not many soft GRBs - that is, GRBs with peak energy in the tens of keV
range. This low peak energy cutoff is real as soft bursts would have
been easily detected by current detectors.  However it is not known
whether there is a real paucity in hard GRBs and there is an upper
cutoff to the GRB hardness or it just happens that the detection is
easiest in this (few hundred keV) band.  BATSE triggers, for example,
are based mostly on the count rate between 50keV and 300keV. BATSE is,
therefore, less sensitive to harder bursts that emit most of their
energy in the MeV range. Using BATSE's observation alone one cannot
rule out the possibility that there is a population of harder GRBs
that emit equal power in total energy which are not observed because
of this selection effect
\cite{NP95a,CNP97,Petrossian_llyods97,Lingen97}.  More generally, a  harder
burst with the same energy as a soft one emits fewer
photons. Furthermore, the spectrum is generally flat in the high energy
range and it decays quickly at low energies. 
Therefore it is intrinsically more difficult to detect a harder
burst. A study of the SMM data \cite{Harris97} suggests that there is
a deficiency (by at least a factor of 5) of GRBs with hardness above
3MeV, relative to GRBs peaking at $\sim$0.5MeV, but this  data is
consistent with a population of hardness that extends up to 2MeV.

Overall the spectrum is non-thermal. This indicates that the source
must be optically thin. The spectrum deviates from a black body 
 in both the low and the high
energy ends: The X-ray paucity constraint  rules
out optically thick models in which the $\gamma$-rays could be
effectively degraded to X-rays \cite{Imamura}.  The high energy tails
lead to another strong constraint on physical GRB models. These high
energy photons escape freely from the source without producing electron
positron pairs! As we show later, this provides the first and most
important clue on the nature of GRBs.

The low energy part of the spectrum behaves in many cases like a power
law: $F_\nu \propto \nu^\alpha$ with $-\frac{1}{2} < \alpha <
\frac{1}{3} $, \cite{Katz94a,Cohen_etal_97}. This is consistent with the
low energy tail of synchrotron emission from relativistic
electrons -  a distribution of electrons in which
{\it all} the population, not just the upper tail, is
relativistic. This is a direct indication for the existence of
relativistic shocks in GRBs. More than 90\% of the bright bursts
studied by Schaefer \etall \cite{Schaefer_etal98} satisfy this limit. 
However, there may be  bursts whose low
energy tail is steeper \cite{Preece97a}. Such a spectrum cannot be
produced by a simple synchrotron emission model and it is not clear how is 
it produced. 

\subsection{Spectral Evolution}

Observations by earlier detectors as well as by BATSE have shown that
the spectrum varies during the bursts. Different trends were found.
Golenetskii \etall \cite{Golenetskii83} examined two channel data from
five bursts observed by the KONUS experiment on {\it Venera} 13 and 14
and found a correlation between the effective temperature and the
luminosity, implying that the spectral hardness is related to the
luminosity.  Similar results were obtained Mitrofanov et. al.
\cite{Mitrofanov84}. Norris \etall \cite{Norris86} investigated ten
bursts seen by instruments on the SMM ({\it Solar Maximum Mission})
satellite. They found that individual intensity pulses evolve from
hard-to-soft with the hardness peaking earlier than the intensity.
This was supported by more recent BATSE data \cite{Band91}.  Ford
\etall \cite{Ford95} analyzed 37 bright BATSE bursts and found that
the spectral evolution is a mixture of those found by Golenetskii
\etall \cite{Golenetskii83} and by Norris \etall \cite{Norris86}: The
peak energy either rises with or slightly proceeds major intensity
increases and softens for the remainder  of the pulse. For bursts with
multiple peak emission, later spikes tend to be softer than earlier
ones.

A related but not similar trend is shown by the observations that the
bursts are narrower at higher energies with $T(\nu) \propto
\nu^{-0.4}$ \cite{Fenimore95}. As we show in section \ref{sub_sec:synchrotron}
this behavior is consistent with synchrotron emission \cite{SNP96}.

\subsection{Spectral Lines}

Both absorption and emission features have been reported by various
experiments prior to BATSE. Absorption lines in the 20-40keV range
have been observed by several experiments - but never
simultaneously. GINGA has discovered several cases of lines with
harmonic structure \cite{Murakami88,Fenimore88}. These lines were
interpreted as cyclotron lines (reflecting a magnetic field of
$\approx 10^{12}$Gauss) and providing one of the strongest arguments
in favor of the galactic neutron star model.  Emission features near
400keV have been claimed in other bursts \cite{400kev}. These have
been interpreted as red-shifted 511keV annihilation lines with a
corresponding red-shift of $\approx 20$\% due to the gravitational
field on the surface of the Neutron star.  These provided additional
evidence for the galactic neutron star model.

So far BATSE has not found any of the spectral features (absorption or
emission lines) reported by earlier satellites \cite{Palmer94,Band96}. This
can be interpreted as a problem with previous observations (or with
the difficult analysis of the observed spectra) or as an unlucky
coincidence. Given the rate of observed lines in previous experiments
it is possible (at the $\approx 5$\% level) that the two sets of data
are consistent \cite{Band95}.

Recently M\'esz\'aros \& Rees \cite{MR98} suggested that within the
relativitic fireball model the observed spectral lines could have be
blue shifted iron X-ray line.

\subsection{Angular Positions}

BATSE is capable of estimating on its own the direction to a burst. It is
composed of eight detectors that are pointed towards different
directions in the sky. The relative intensity of the counts in the
various detectors allows us to measure the direction to  the burst.
The positional error
of a given burst is the square root of the sum of squares of a
systematic error and a statistical error.  The statistical error
depends on the strength of the burst. It is as large as $20^o$ for a
weak burst, and it is negligible for a strong one. The estimated
systematic error (using a comparison of BATSE positions with IPN
(Inter Planetary Network) localization) is $\approx 1.6^o$
\cite{systematics95}.  A different analysis of this comparison
\cite{Graziani95,systematics97} suggests that this might be slightly
higher, around $3^o$.

The location of a burst is  determined much better using
the difference in arrival time of the burst to several detectors on
different satellites. Detection by two satellites
limits the position to a circle on the sky. Detection by three
determines the position and detection by four or more  
over-determines it. Even in this case the positional error depends on the
strength of the bursts. The stronger the burst, the easier it is to identify
a unique moment of time in the incoming signals. Clearly, the accuracy
of the positional determination is better the longer the distance
between the satellites. The best positions that have been obtained
in this way are with the IPN$^3$ of detectors.  For 12 events the
positional error boxes are of a few arc-minutes \cite{Hurley93b}.

BeppoSAX Wide Field Camera (WFC) that covers about 5\% of the sky
located a few bursts within 3' ($3 \sigma$). BeppoSAX's Narrow Field
Instrument (NFI) obtained the bursts' positions to within 50''. X-ray
observations by ASCA and ROSAT have yielded error boxes of 30'' and
10'' respectively.  Optical identification has led, as usual, to a
localization within 1''.  Finally VLBI radio observation of GRB970508
has yielded a position within 200$\mu$arcsec. The position of at least
one burst is well known.

\subsection{Angular Distribution}

One of the most remarkable findings of BATSE was the observation that
the angular distribution of GRBs' positions on the sky is perfectly
isotropic.  Early studies had shown an isotropic GRB distribution
\cite{Cline75} which have even led to the suggestion that GRBs
are cosmological \cite{vandenBerg83}. In spite of this it was
generally believed, prior to the launch of BATSE, that GRBs are
associated with galactic disk neutron star. It has been expected that
more sensitive detectors would discover an anisotropic distribution
that would reflect the planar structure of the disk of the
galaxy. BATSE's distribution is, within the statistical errors, in
complete agreement with perfect isotropy. For the first 1005 BATSE
bursts the observed dipole and quadrupole (corrected to BATSE sky
exposure) relative to the galaxy are: $\langle \cos\theta \rangle
=0.017\pm 0.018 $ and $\langle
\sin^2 b - 1/3 \rangle =- 0.003\pm 0.009 $. These values are,
respectively, $0.9\sigma$ and $0.3\sigma$ from complete isotropy
\cite{Briggs}.

\subsection{ Quiescent Counterparts and the historical ``No Host" Problem}

One of the main obstacles in resolving the GRB mystery was the lack of
identified counterparts in other wavelengths. This has motivated
numerous attempts to discover GRB counterparts (for a review see
\cite{Schaefer92a,Vrba94}). This is a difficult task - it was not
known what to expect and where and when to look for it.

The search for counterparts is traditionally divided to efforts to
find a flaring (burst), a fading or a quiescent counterpart.  Fading
counterparts - afterglow - have been recently discovered by BeppoSAX
and as expected this discovery has revolutionized GRB studies. This
allowed also the discovery of host galaxies in several cases, which 
will be discussed in the following section
\ref{sub_sec:afterglow}. Soft X-ray flaring (simultaneous with the
GRB) was discovered in several bursts but it is an ambiguous question
whether this should be considered as a part of the GRB itself or is it
a separate component.  Flaring has not been discovered in other
wavelengths yet. Quiescent counter parts were not discovered either.

Most cosmological models suggest that GRBs are in a host galaxy. If so
then deep searches within the small error boxes of some GRBs localized
by the IPN system should reveal the host galaxy.  
until the discovery of GRB afterglow these searches have
yielded only upper limits on the magnitudes of possible hosts. 
 This has lead to what is called the ``No Host"
Problem.  Schaefer \etall conducted searches in the near and far
infrared \cite{Schaefer87} using IRAS, in radio using the VLA
\cite{Schaefer89} and in archival optical photographs
\cite{Schaefer90} and have found only upper limits and no clear
counterpart candidates.  Similar results from multiple wavelength
observations have been obtained by Hurley \etall
\cite{Hurley94a}. Vrba, Hartmann \& Jennings \cite{Vrba95} have
monitored the error boxes of seven bursts for five year.  They did not
find any unusual objects. As for the ``no host'' problem this authors,
as well as Luginbuhl \etall \cite{Luginbuhl95} and Larson, McLean \&
Becklin \cite{Larson96} concluded, using the standard galaxy
luminosity function, that there are enough dim galaxies in the
corresponding GRB error boxes which could be the hosts of cosmological
burst and therefore, there is no ``no host'' problem.

More recently Larson \& McLean \cite{Larson97} monitored in the
infrared nine of the smallest error boxes of burst localized by the
IPN with a typical error boxes of eight arc-min$^2$. They found in all
error boxes at least one bright galaxy with $K\le 15.5$. However, the
error boxes are too large to discern between the host galaxy and
unrelated background galaxies.  Schaefer \etall \cite{Schaefer97},
searched the error boxes of five GRBs using the HST.  Four of these
are smaller boxes with a size of $\sim 1 $ arc-min$^2$. They searched
but did not find any unusual objects with UV excess, variability,
parallax or proper motion.  They have found only faint galaxies. For
the four small error boxes the luminosity upper limits of the host
galaxy are 10-100 times smaller than the luminosity of an $L_*$
galaxy. Band \& Hartmann \cite{Band_hartmann97} concluded that the
error boxes of Larsen \& McLean \cite{Larson97} are too large to
discriminate between the presence or the absence of host galaxies.
However, they find that the absence of host galaxies in the Schaefer 
\etall \cite{Schaefer97} data is significant, at the $2 \cdot 10^{-6}$
level. Suggesting that there are no bright hosts.

This situation has drastically changed and the ``no host'' problem 
has disappeared with afterglow observations.
These observations have allowed for an accurate position determination
and to identification of host galaxies for several GRBs. Most of these
host galaxies are dim with magnitude $24.4 < R< 25.8 $.  This support
the conclusions of the earlier studies that GRBs are not associated
with bright galaxies and definitely not with cores of such galaxies
(ruling out for example AGN type models).
These observations are
consistent with GRBs rate being either a constant or being
proportional to the star formation rate \cite{Hogg_Fruchter98}.
According to this analysis it is not surprising that most hosts are
detected at $R \sim 25$. However, though these two models are
consistent with the current data both predict the existence of host
galaxies brighter than 24 mag, which were not observed so far.
  One could say now that 
the ``no host'' problem has been replaced by the ``no bright host''
problem. But this may not be a promlem but rather an indication on the
nature of the sources.

The three GRBs with measured cosmological redshifts lie in host
galaxies with a strong evidence for star formation.  These galaxies
display prominent emission lines from line associated with
star-formation. In all three cases the strength of those lines is high
for galaxies of comparable magnitude and redshift 
\cite{Djorgovski98a,Djorgovski98b,Bloom98b,Fruchter98,Hogg_Fruchter98}.  
The host of GRB980703, for example, show a  star forming rate of  
$\sim 10 M_\odot$yr$^{-1}$ or higher
with a lower limit of $7M_\odot$yr$^{-1}$ \cite{Djorgovski98b}.
For most GRBs
with afterglow the host galaxy was detected but no emission or
absorption lines were found and no redshift was measured.  This result
is consistent with the hypothesis that all GRBs are associated with
star-forming galaxies.  For those hosts that are at redshift $1.3 < z
< 2.5$ the corresponding emission lines are not observed as for this
redshift range no strong lines are found in the optical spectroscopic
window \cite{Fruchter98}. 

The simplest conclusion of the above observations is that all GRBs are
associated with star forming regions. Still one has to keep in mind
that those GRBs on which this conclusion was based had a strong
optical afterglow, which not all GRBs show.  It is possible that the
conditions associated with star forming regions (such as high
interstellar matter density - or the existance of molecular clouds)
are essential for the appearance of strong optical afterglow and
not for the appearance of the GRB itself.

\subsection{Afterglow}
\label{sub_sec:afterglow}
GRB observations were revolutionized on February 28, 1997 by the
Italian-Dutch satellite BeppoSAX \cite{Piro95} that discovered an
X-ray counterpart to GRB970228 \cite{Costa97a}.  GRB970228 was a
double peaked GRB. The first peak which lasted $\sim 15$sec was hard.
It was followed, 40 seconds later, by a much softer second peak, which
lasted some $\sim 40$sec.  The burst was detected by the GRBM
(Gamma-Ray Burst Monitor) as well as by the WFC (Wide Field Camera).
The WFC, which has a $40^o \times 40^o$ field of view detected soft
X-rays simultaneously with both peaks.  Eight hours latter the NFI
(Narow Field Instrument) was pointed towards the burst's directions
and detected a continuous X-ray emission. The accurate position
determined by BeppoSAX enabled the identification of an optical
afterglow \cite{vanParadijs97a} - a 20 magnitude point source adjacent
to a red nebulae.  HST observations \cite{Sahu97} revealed that the
nebula adjacent to the source is roughly circular with a diameter of
0''.8.  The diameter of the nebula is comparable to the one of
galaxies of similar magnitude found in the Hubble Deep Field,
especially if one takes into account a possible visual extinction in
the direction of GRB970228 of at least one magnitude \cite{Lamb97}.

\begin{figure}
\begin{center}
\includegraphics[width=10cm]{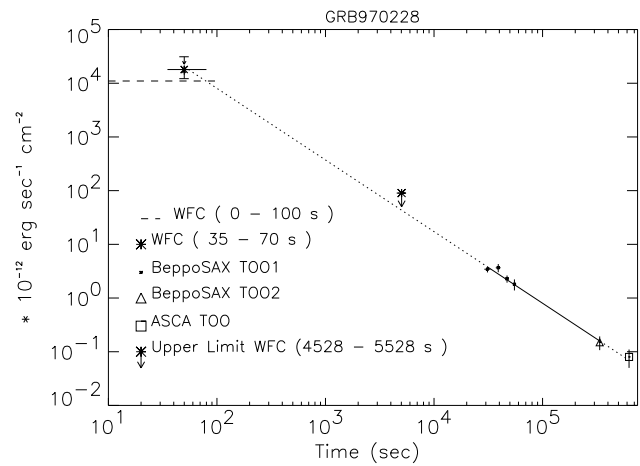}
\caption{\it Decay of the X-ray afterglow from GRB970228, from
  \cite{Costa97a}. Shown is source flux at the 2-10KeV range.
The data is fitted with a power law $t^{-1.32}$.
}
\label{afterglow_xray}
\end{center}
\end{figure}

Following X-ray detections by BeppoSAX \cite{Costa97a,Frontera97}, ROSAT
\cite{Rosat}  and ASCA \cite{ASCA} revealed a decaying X-ray flux
$\propto t^{-1.33\pm0.11}$ (see Fig. \ref{afterglow_xray}).  The
decaying flux can be extrapolated as a power law directly to the X-ray
flux of the second peak (even though this extrapolation requires some
care in determining when is $t=0$).

\begin{figure}
\begin{center}
\includegraphics[width=10cm]{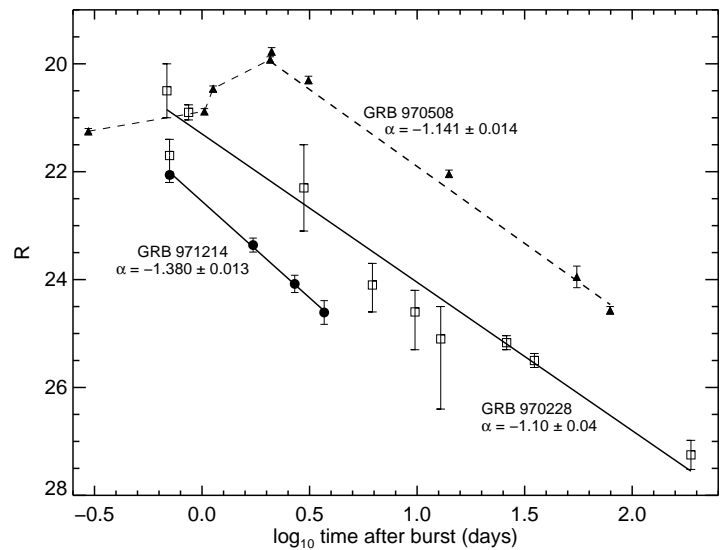}
\caption{\it Decay of the optical  afterglow in GRB070228, GRB9700508 and GRB971214.
A clear power law decay can be seen in all cases.}
\label{afterglow_optical}
\end{center}
\end{figure}

The optical emission also depicts a decaying flux \cite{Galama97a}
(see fig. \ref{afterglow_optical}.
The source could not be observed from late March 97 until early
September 1997. When it was observed again, on Sept. 4th by HST
\cite{Fruchter97,Fruchter98} it was found  that the optical nebulosity
does not decay and the point source shows no proper motion, refuting
earlier suggestions.  The visual magnitude of the nebula on Sept. 4th
was $25.7 \pm 0.25 $ compared with $V= 25.6 \pm 0.25$ on March 26th
and April 7th.  The visual magnitude of the point source on Sept. 4th
was ($ V = 28.0 \pm 0.25$), which is consistent with a decay of the
flux as $t^{-1.14\pm0.05}$ \cite{Fruchter98} .  In spite of extensive
efforts no radio emission was detected and one can set an upper limit
of $\sim 10 \mu$Jy to the radio emission at 8.6 Ghz
\cite{Frail-Kulkarni}.

GRB970508 was detected by both BATSE in $\gamma$-rays
\cite{Kouveliotou97} and BeppoSAX in X-rays \cite{Piro97} on 8 May
1997. The $\gamma$-ray burst lasted for $\sim 15$sec, with a
$\gamma$-ray fluence of $\sim 3 \times 10^{-6}$ergs/cm$^{-2}$.
Variable emission in X-rays, optical
\cite{Bond97,Djorgovski97a,Djorgovski97b,Djorgovski97c,Mignoli97,Chevalier97} 
and radio \cite{Frail97a,Taylor97} followed the $\gamma$-rays. The
spectrum of the optical transient taken by Keck revealed a set of
absorption lines associated with Fe II and Mg II and
 O II emission line with a redshift
$z=0.835$ \cite{Metzger97a}. A second absorption line system with
$z=0.767$ is also seen.   These  lines reveal 
 the existence of an underlying, dim galaxy host. HST images
\cite{Pian98a,Natarajan97} and Keck observations  \cite{Bloom98b}
 show that this host is    very faint ($R
= 25.72 \pm 0.2 $ mag), compact ($\le 1$ arcsec) dwarf galaxy at $z=0.835$
 and nearly coincident on the sky with the transient.

\begin{figure}
\begin{center}
\includegraphics[width=10cm]{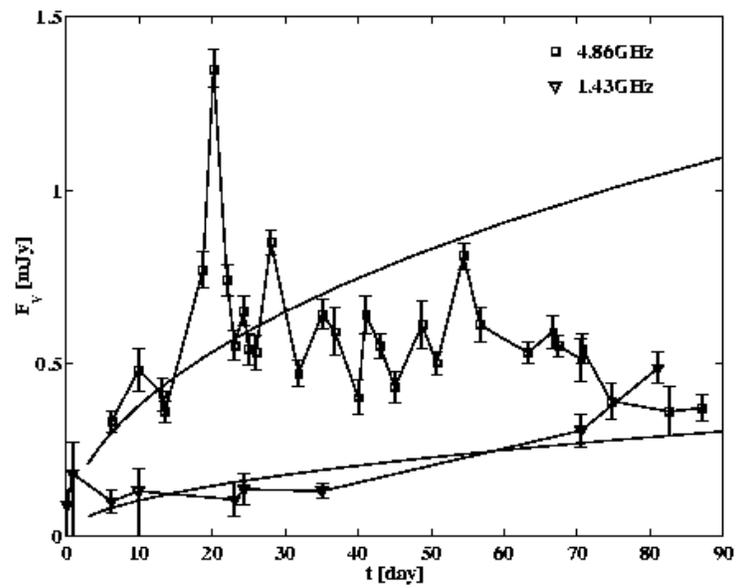}
\caption{\it Light curve of the radio  afterglow from GRB970508, from
\cite{Frail97a}}
\label{afterglow_radio}
\end{center}
\end{figure}

The optical light curve peaks at around 2 days after the burst. 
Assuming isotropic emission (and using
$z=0.835$ and H=100km/sec/Mpc) this peak flux corresponds to a
luminosity of {\it a few} $\times 10^{45}$ergs/sec. The flux decline
shows a continuous power law decay $\propto t^{-1.27\pm 0.02}$
\cite{Galama98b,Castro-Tirado,Pedersen,Sokolov97,Bloom98b}. 
After about 100 days the light curve begun to flatten as the transient
faded and become weaker than the host
\cite{Pedersen,Bloom98a,Castro-Tirado98,Sokolov98a}. 
Integration of this light curve results in an overall
emission of {\it a few } $\times 10^{50}$ ergs in the optical
band. Radio emission was observed, first, one week after the burst
\cite{Frail97a} (see Fig. \ref{afterglow_radio}). This emission showed
intensive oscillations which were interpreted as scintillation
\cite{Goodman97}. The subsequent disappearance of these oscillations
after about three weeks enabled Frail \etall   \cite{Frail97a} to
estimate the size of the fireball at this stage to be $\sim
10^{17}$cm.  This was supported by the indication that the radio
emission was initially optically thick \cite{Frail97a}, which yields a
similar estimate to the size \cite{KP97}.

GRB970828 was a strong GRB that was detected by BATSE on August 28,
1997.  Shortly afterwards RXTE \cite{RXTE1,RXTE2} focused on the
approximate BATSE position and discovered X-ray emission. This X-ray
emission determined the position of the burst to within an elliptical
error box with $5'\times 2'$. However, in spite of enormous effort no
variable optical counterpart brighter than R=23.8 that has changed by
more than 0.2 magnitude was detected \cite{Groot97}.  There was also
no indication of any radio emission.  Similarly X-ray afterglow was
detected from several other GRBs (GRB970615,  GRB970402, GRB970815,  
GRB980519) with no optical or  radio
emission.

Seventeen GRBs have been detected with arcminute positions by July 22,
1998: fourteen by the WFC of BeppoSAX and three by the All-Sky Monitor
(ASM) on board the Rossi X-ray Timing Explorer (RXTE). Of these
seventeen burst, thirteen were followed up within a day in X-rays and
all those resulted in good candidates for X-ray afterglows.  We  will
not  discuss all those here (see table \ref{t:afterglow} for a short
summary of some of the properties). Worth mentioning are however,
GRB971214,  GRB980425 and GRB980703.

GRB971214 was a rather strong burst.  It  was detected on
December 14.9 UT 1997 \cite{Heise97}.  Its optical counterpart was
observed with a magnitude $ 21.2 \pm 0.3$ on the I band by Halpern
\etal, \cite{Halperin97} on Dec.  15.47 UT twelve hours after the
burst. It was observed one day later on Dec. 16.47 with I magnitude
22.6. Kulkarni \etall \cite{Kulkarni98a} obtained a spectrum of the
host galaxy for GRB971214 and found a redshift of z=3.418!  
With a total fluence of $1.09 \times 10^{-5}$ergs
cm$^{-2}$ \cite{Meegan98} this large redshift implies, for isotropic
emission, $\Omega=1$ and $H_0=65$km/sec/Mpc, an energy release of
$\sim 10^{53}$ergs in $\gamma$-rays alone \footnote{This value depends
also on the spectral shape of the burst.}.  The familiar value of $3
\times 10^{53}$ \cite{Kulkarni98a}  is obtained for $\Omega = 0.3$ and 
$H_0=0.55$km/sec/Mpc.

GRB980425 was a moderately weak burst with a peak flux of $3 \pm 0.3
\times 10^{-7}{ \rm ergs \ cm^{-2} \ sec^{-1}}$. It was a single peak
burst with a rise time of 5 seconds and a decay time of about 25
seconds. The burst was detected by BeppoSAX (as well as by BATSE)
whose WFC obtained a position with an error box of $8'$. Inspection of
an image of this error box taken by the New Technology Telescope (NTT)
revealed a type Ic supernova SN1998bw that took place more or less at
the same time as the GRB \cite{Galama98b}.  Since the probability for
a chance association of the SN and the GRB is only $1.1 \times
10^{-4}$ it is likely that this association is real. The host galaxy
of this supernova (ESO 184-G82) has a redshift of $z=0.0085 \pm0.0002$
putting it at a distance of $38 \pm 1$Mpc for H=67km/sec Mpc. The
corresponding $\gamma$-ray energy is $5 \times 10^{47}$ergs. With such a
low luminosity it is inevitable that if the association of this burst
with the supernova is real it must correspond to a new and rare subgroup
of GRBs.

GRB980703 was a very strong burst with an observed gamma-ray fluence
of $(4.59 \pm 0.42) \times 10^{-5}$ ergs cm$^{-2}$ \cite{Kippen98a}.
Keck observations revealed  that the host galaxy has a redshift of 
$z=0.966$. The corresponding energy release (for isotropic emission,
$\Omega=0.2$ and $H_0  = 65$km sec$^{-1}$/Mpc) is $\sim 10^{53}$ ergs
\cite{Djorgovski98b}.

\newcommand{\rb}[1]{\raisebox{1.5ex}[0pt]{#1}}
\begin{center}
\begin{table*}[ht!]
\label{t:afterglow}
\begin{center}
\begin{tabular}{|c||c|c|c|c|c|c|}\hline 
&X-ray & & & $\gamma$-ray fluence & & total energy \\
\rb{burst} & detection & \rb{O} & \rb{R} & in $\rm{[ergs/cm^2]}$ &
 \rb{redshift} & in $\rm{[ergs]}$ \\ \hline \hline
GRB970228 & BeppoSAX & + & - & $\rm{1\times 10^{-5}}$ & - & - \\ \hline 
GRB970508 & BeppoSAX & + & + & $\rm{2\times 10^{-6}}$ & 0.835 &
 $2\times10^{51}$ \\ \hline 
GRB970616 & BeppoSAX & - & - & $\rm{4\times 10^{-5}}$ & - & - \\ \hline 
GRB970815 & RXTE     & - & - & $\rm{1\times 10^{-5}}$ & - & - \\ \hline 
GRB970828 & RXTE & - & - & $\rm{7\times 10^{-5}}$ & - & - \\ \hline 
GRB971214 & RXTE & + & + & $\rm{1\times 10^{-5}}$ & 3.418 &
 $1\times10^{53}$ \\ \hline 
GRB971227 & BeppoSAX & - & - & $\rm{9\times 10^{-7}}$ & - & - \\ \hline 
GRB980326 & BeppoSAX & - & - & $\rm{1\times 10^{-6}}$ & - & - \\ \hline 
GRB980329 & BeppoSAX & + & + & $\rm{5\times 10^{-5}}$ & - & - \\ \hline 
GRB980425 & BeppoSAX & + & + & $\rm{4\times 10^{-6}}$ & 0.0085 &
 $7\times10^{47}$ \\ \hline 
GRB980515 & BeppoSAX & - & - & $\rm{1\times 10^{-6}}$ & - & - \\ \hline 
GRB980519 & BeppoSAX & + & + & $\rm{3\times 10^{-5}}$ & - & - \\ \hline 
GRB980703 & RXTE & + & + & $\rm{5\times 10^{-5}}$ & 0.966 & 
$1\times10^{53}$ \\ \hline 
\end{tabular}\\
\caption{Observational data of several GRBs for which afterglow was
  detected. The two columns O and R indicate whether emission was
  detected in the optical and radio, respectively. The total energy 
  of the burst is estimated through the observed fluence and redshift, 
  assuming spherical emission and a flat $\Omega=1$,
  $\Lambda=0$ universe with $H_0=65\rm{Km/sec/Mpc}$.}
\end{center}
\end{table*}
\end{center}

\subsection{Repetition?}

Quashnock and Lamb \cite{Quashnock93} suggested that there is evidence
from the data in the BATSE 1B catalog for repetition of bursts from
the same source. If true, this would severely constrain most GRB
models. In particular, it would rule out any other model based on a `once in
lifetime' catastrophic event.  This claim has been refuted by several
authors \cite{NP93,Hartmann94} and most notably by the analysis of the
2B data \cite{Meegan95} and 3B data \cite{Tegmark96}.

A unique group of four bursts - ``the gang of four'' -
emerged from the same position on the sky within two days
\cite{Meegan96a}. One of those bursts (the third burst GRB961029a)  was
extremely strong, one of the strongest  observed by BATSE so far.
Consequently it was observed by the IPN network as well, and its
position is known accurately. The other three (GRB961027a,GRB961027b
and GRB961029d) were detected only by BATSE. The precise position of
one burst  is within the $1\sigma$ circles of the three other bursts.
However, two of the bursts are  almost $3\sigma$ away from each other.
Is this a clear cut case of repetition? It is difficult to
assign a unique statistical significance to this question as the
significance depends critically on the a priori hypothesis that one
tests. Furthermore, the time difference between the first and
the last bursts is less than two days. This is only one order of
magnitude longer than the longest burst observed beforehand. It might
still be possible that all those bursts came from
the same source and that they should be considered as one long burst.

\subsection{Correlations with Abell Clusters, Quasars and Supernovae}

Various attempts to search for a correlation between GRBs and other
astronomical objects led to null result.  For example, Blumenthal \&
Hartmann \cite{Nlum_Hart} found no angular correlation between
GRBs and nearby galaxies. They concluded that if GRBs are cosmological
then they must be located at distances larger than 100Mpc. Otherwise,
they would have shown a positive correlation with the galaxy
distribution.

The only exception is the correlation (at 95\% confidence level)
between GRBs at the 3B catalog and Abell clusters
\cite{CKP,KP96}. This correlation has been recently confirmed by
Kompaneetz \& Stern \cite{Kompa_Stern97}.  The correlation is
strongest for a subgroup of strong GRBs whose position is accurately
known. Comparison of the rich clusters auto-correlation with the
cross-correlation found suggests that $\sim 26\pm 15$\% of the
accurate position GRBs sub-sample members are located within
$600\hmpc$. Recently Schartel \etall \cite{Schartel97} found that a
group of 134 GRBs with position error radius smaller than $1.8^o$ are
correlated with radio quiet quasars.  The probability of 
of such
correlation by  chance coincidence is less than 0.3\%.

It should be stressed that this correlation does not imply that there
is a direct association between GRBs and Abell clusters,
such as would have been if GRBs would have emerged from Abell
clusters. All that it means is that GRBs are distributed in space like
the large scale structure of the universe. Since Abell clusters are
the best tracers of this structure they are correlated with GRBs.
Therefore the lack of excess  Abell Clusters in IPN error boxes
(which are much smaller than BATSE's error boxes) \cite{Hurley97} does
not rule out this correlation.

\subsection{$V/V_{max}$, Count and Peak Flux Distributions}

The limiting fluence observed by BATSE is $\approx
10^{-7}$ergs/cm$^2$. The actual fluence of the strongest bursts is
larger by two or three orders of magnitude.  A plot of the number of
bursts vs. the peak flux depicts clearly a paucity of weak bursts.
This is manifested by the low value of $\langle V/V_{max} \rangle$, a
statistic designed to measure the distribution of sources in space
\cite{Schmidt88}.  A sample of the first 601 bursts has $\langle
V/V_{max} \rangle =.328 \pm 0.012$, which is 14$\sigma$ away from the
homogeneous flat space value of $0.5$ \cite{Pen}.  Correspondingly,
the peak count distribution is incompatible with a homogeneous
population of sources in Euclidean space.  It is compatible, however,
with a cosmological distribution (see Fig. \ref{peak_flux}). The
distribution of short bursts has a larger $\langle V/V_{max} \rangle$
and it is compatible with a homogeneous Eucleadian distribution
\cite{CKP,Piran96,Katz96}.

\section{The Distance Scale}
\label{sec:peak}
\subsection{Redshift Measurements.}

The measurements of  redshifts of  several GRB optical counterparts
provide the best and the only direct distance estimates for GRBs.
Unfortunately these measurements are available only for a few
bursts.

\subsection{The Angular Distribution}

Even before these redshift  measurements 
there was a strong evidence that GRBs originate from
cosmological distances. The observed angular distribution is
incompatible with a galactic disk distribution unless the sources are
at distances less than $100$pc.  However, in this case we would expect
that $\langle V/V_{max} \rangle = 0.5$ corresponding to a homogeneous
distribution \cite{Schmidt88} while the observations yield $\langle
V/V_{max} \rangle = 0.33$.

A homogeneous angular distribution could be produced if the GRB originate
from the distant parts of the galactic halo. Since the solar system is
located at $d=8.5$kpc from the galactic center such a population will
necessarily have a galactic dipole of order $d/R$, where $R$ is a
typical distance to a GRB \cite{Katz92}.  
The lack of an observed dipole strongly 
constrain this model.  Such a
distribution of sources is incompatible with the distribution of dark
matter in the halo.  The typical distance to the GRBs must of the
order of 100kpc  to comply with this constraint.  For example, if
one considers an effective distribution that is confined to a shell of
a fixed radius then such a shell would have to be at a distance of
100kpc in order to be compatible with current limits on the dipole
\cite{Hartmann95a}.

\subsection{Interpretation of the Peak Flux Distribution}
\label{sub_sec:peak_flux}
\begin{figure}
\begin{center}
%
\includegraphics[width=10cm]{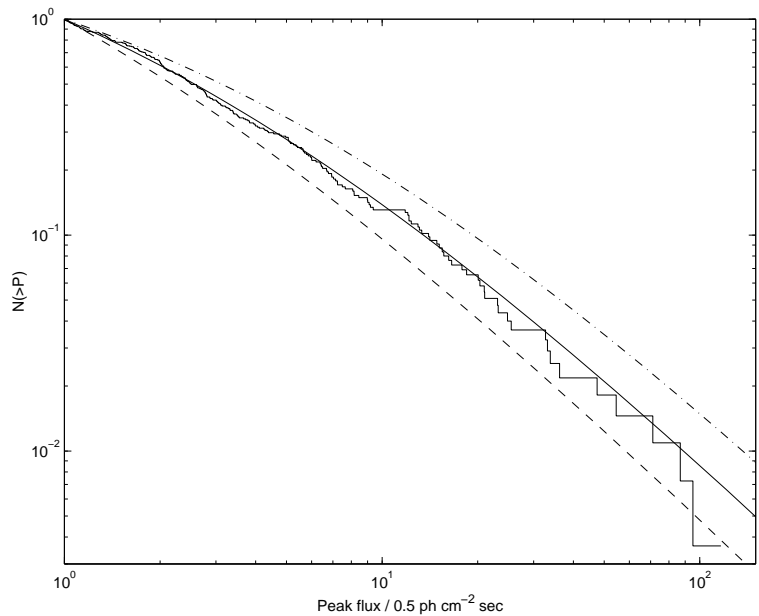}
\caption{\it The observed long burst peak flux distribution
  and three theoretical cosmological distributions with $\O=1$,
  $\Lambda=0$, $\alpha=-1.5$, standard candles and no source
  evolution: $L = 3.4 \cdot 10^{50}$ergs/sec (solid line: best fit),
  $L = 7.2 \cdot 10^{50}$ergs/sec (dashed line: lower 1\% bound), $L =
  1.4 \cdot 10^{50}$ergs/sec (dashed-dotted line: upper 1\% bound) }
\label{peak_flux}
\end{center}
\end{figure}

The counts distribution or the Peak flux distribution of the bursts
observed by BATSE show a paucity of weak burst. A homogeneous count
distribution, in an Eucleadian space should behave like:
$N(C) \propto C^{-3/2}$, where $N(C)$ is the number of bursts with
more than $C$ counts (or counts per second).  
The observed distribution is much flatter
(see Fig. \ref{peak_flux}). This fact is reflected by the low
$\langle V/V_{max} \rangle$ value of the BATSE data: there are
fewer distant sources than expected.

The observed distribution is compatible with a cosmological
distribution of sources. A homogeneous cosmological distribution
displays the observed trend - a paucity of weak bursts relative to the
number expected in a Eucleadian distribution.  In a cosmological
population four factors combine to make distant bursts weaker and by
this to reduce the rate of weak bursts: (i) K correction - the
observed photons are red-shifted.  As the photon number decreases with
energy this reduces the count rate of distant bursts for a detector at
a fixed energy range.  (ii) The cosmological time dilation causes a
decrease (by a factor $1+z$) 
in the rate of arrival of photons. For a detector, like
BATSE, that measures the count rate within a given time window this
reduces the detectability of distant bursts. (iii) 
 The  rate of distant bursts also decreases by a factor $1+z$ and  there are
fewer distant bursts per unit of time (even if the rate at the
comoving frames does not change). 
(iv) Finally, the distant volume
element in a cosmological model is different  than the corresponding
volume element in a Eucleadian space. 
As could be expected, all
these effects are significant only if the typical red-shift to the
sources is of order unity or larger.

The statistics $\langle V/V_{max} \rangle>$ is a weighted average of
the distribution $N(>f)$. Already in 1992 Piran \cite{Pi92} compared
the theoretical estimate of this statistics to the observed one and
concluded that the typical redshift of the bursts observed by BATSE is
$ \zm \sim 1$.  Later Fenimore \etall \cite{Fenimore93a} compared the
sensitivity of PVO (that observes $N(>f) \propto f^{-3/2}$) with the
sensitivity of BATSE and concluded that $\zm(BATSE) \sim 1$ (the maximal $z$
from which bursts are detected by BATSE). This corresponds to a peak
luminosity of $\sim 10^{50}$ ergs/sec.  Other calculations based on
different statistical methods were performed by Horack \& Emslie
\cite{HE}, Loredo \& Wasserman \cite{Loredo1,Loredo2}, Rutledge \etall
\cite{Rutledge95} Cohen \& Piran \cite{CP95} and M\'esz\'aros 
and collaborators \cite{Mes1,Mes2,Mes3,Mes4} and  others.  In
particular Loredo \& Wassermann \cite{Loredo1,Loredo2} give an
extensive discussion of the statistical methodology involved.

Consider a homogeneous cosmological distribution of sources with a peak
luminosity $L$, that may vary from one source to another. It should be
noted that only the luminosity per unit solid angle is accessible by these 
arguments. If there is significant beaming, as inferred  \cite{KP97},
the distribution of total luminosity may be quite different.  The sources
are emitting bursts with a count spectrum: $N(\nu) d\nu = (L/h \bar
\nu) \tilde N(\nu) d\nu$, where $h \bar \nu$, is the average energy.
The observed peak (energy) flux in a fixed energy range,
$[E_{min},E_{max}]$ from a source at a red-shift $z$ is:
\begin{equation}
f(L,z) = {  (1+z)  \over 4 \pi d_l^2 (z)} {L \over h \bar \nu}
\int_{E_{min}}^{E_{max}} \tilde N[\nu (1+z)] h \nu (1+z) h d\nu
\label{f}
\end{equation}
where $d_l(z) $ is the luminosity distance \cite{Weinberg73}.

To estimate the number of bursts with a peak flux larger than $f$,
$N(>f)$, we need the luminosity function, $\psi(L,z)$: the number of
bursts per unit proper (comoving) volume per unit proper time with a
given luminosity at a given red-shift. Using this function we can
write:
\begin{equation}
N(>f) = 4\pi\int_0^\infty   \int_0^{z(f,L)} \psi(L,z)
{d_l^2 \over (1+z)^3} {dr_p(z) \over dz} dz dL
\label{distN}
\end{equation}
where the red-shift, $z(f,L)$, is obtained by inverting Eq. \ref{f}
and $r_p(z)$ is the proper distance to a red-shift $z$.  For a given
theoretical model and a given luminosity function we can calculate the
theoretical distribution $N(f)$ and compare it with the observed one.

A common simple model assumes that $\psi(L,z) = \phi(L)\rho(z)$ -  the
luminosity does not change with time, but the rate of events per unit
volume per unit proper time may change. In this case we have:
\begin{equation}
N(>f) = 4\pi\int_0^\infty   \phi(L) \int_0^{z(f,L)} \rho(z)
{d_l^2 \over (1+z)^3} {dr_p(z) \over dz} dz dL
\label{distf}
\end{equation}

The emitted spectrum, $N(\nu)$, can be estimated from the observed
data.  The simplest shape is a single power law (Eq. \ref{spec}).
with $\alpha=1.5$ or $\alpha = 1.8 $\cite{Schaefer92b}. 
More elaborate studies have used the Band \etall \cite{Band93}
spectrum or even a distribution of such spectra \cite{Rutledge95}.

The cosmic evolution function $\rho(z)$ and the luminosity function
$\phi(L)$ are unknown.  To proceed one has to choose a functional
shape for these functions and characterize it by a few parameters.
Then using maximum likelihood, or some other technique, estimate these
parameters from the data.

A simple  characterization of  $\rho(z)$ is:
\begin{equation}
\rho(z) = \rho_0 (1+z)^{-\beta}. 
\label{evol} 
\end{equation}
Similarly the simplest characterization of the luminosity is as
standard candles:
\begin{equation}
\phi(L)= \delta(L-L_0) ,
\label{std}
\end{equation}
with a single parameter, $L_0$, or equivalently $\zm$, the maximal $z$
from which the source is detected (obtained by inverting Eq. \ref{f}
for $f=f_{min}$ and $L=L_0$).

There are two unknown cosmological parameters: the closure parameter,
$\Omega$, and the cosmological constant $\Lambda$. With the
luminosity function given by Eqs. (5) and (6) we have three unknown
parameters that determine the bursts' distribution: $L_0$, $\rho_0$,
$\beta$. We calculate the likelihood function over this five dimensional
parameter space and find the range of acceptable models (those whose
likelihood function is not less than 1\% of the maximal likelihood).
We then proceed to perform a KS (Kolmogorov-Smirnov) test to check
whether the model with the maximal likelihood is an acceptable fit
to the data.

The likelihood function is practically independent of $\Omega$ in the
range: $ 0.1 <\O < 1$. 
It is also
insensitive to the cosmological constant $\Lambda$ (in the range $0<
\Lambda < 0.9$, in units of the critical density).  This simplifies
the analysis as we are left only with the intrinsic parameters of the
bursts' luminosity function.

\begin{figure}
\begin{center}
\includegraphics[width=10cm]{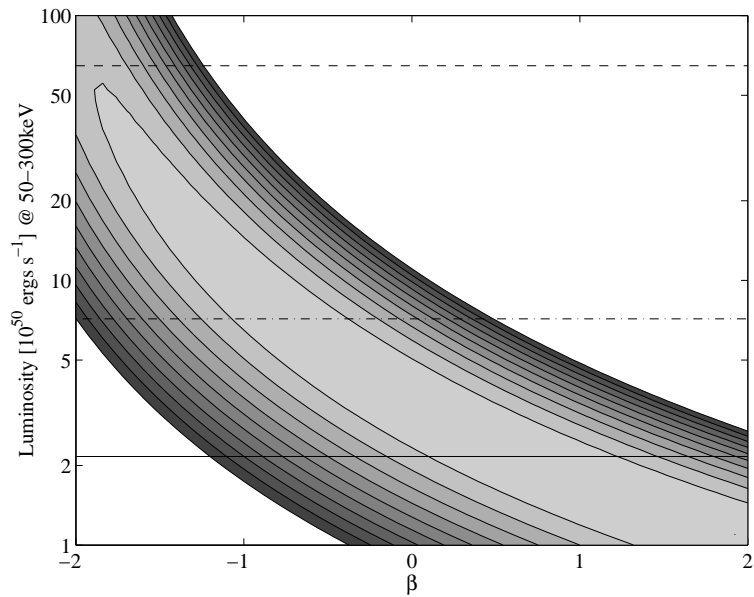}
\caption{\it The likelihood function (levels 33\%, 10\%, 3.3\%
1\% etc..) in the ($\beta,L$) plane for standard candles $\alpha=1.5$,
$\Omega=1$, and evolution given by
$\rho(z)=(1+z)^{-\beta}$. Superimposed on this map is the luminosity
of GRB970508, (solid curve), GRB971214 (dashed curve) and GRB980703
(dahsed-dotted curve). We
have used $h_{75}=1.$} From \cite{CP97a}.
\label{evolution}
\end{center}
\end{figure}

There is an interplay between evolution (change in the bursts' rate)
and luminosity. Fig. \ref{evolution} depicts the likelihood function
in the ($\zm,\beta$) plane for sources with a varying intrinsic rate.
The banana shaped contour lines show that a population whose rate is
independent of $z$ ($\beta=0$) is equivalent to a population with an
increasing number of bursts with cosmological time ($\beta >0$) with a
lower $L_0$ (lower $\zm$).  This tendency saturates at high intrinsic
evolution (large $\beta$), for which the limiting $\zm$ does not go
below $\approx .5$ and at very high $L_0$, for which the limiting
$\beta$ does not decrease below -1.5. This interplay makes it difficult
to constraint the red shift distribution of GRB using the peak flux
distribution alone.
 For completeness we quote here ``typical'' results based on 
standard candles, no evolution and an Einstein-DeSitter cosmology
\cite{CP95}.

Recall that $\langle V/V_{max} \rangle$ of the short bursts
distribution is rather close to the homogeneous Eucleadian value of
0.5. This means that when analyzing the peak flux distribution one
should analyze separately the long and the short bursts \cite{CP95}.
For long bursts (bursts with $t_{90}>2$ sec)  the likelihood
function peaks at $\zm =2.1$ (see Fig.
\ref{peak_flux}) \cite{CP95}. The allowed range at a 1\%
confidence level is: $1.4<\zm < 3.1 $ ($ \zm^{(\alpha=2)} =
1.5^{(+.7)}_{(-.4)}$ for $\alpha=2$).  The maximal red-shift,
$\zm=2.1^{(+1.1)}_{(-0.7)}$, corresponds, with an estimated BATSE
detection efficiency of $\approx 0.3$, to $2.3^{(+1.1)}_{(-0.7)} \cdot
10^{-6}$ events per galaxy per year (for a galaxy density of $10^{-2}
h^3 ~{\rm Mpc}^{-3}$; \cite{Kir}). The rate per galaxy is independent
of $H_0$ and is only weakly dependent on $\O$.  For $\O=1$ and
$\Lambda=0$ the typical energy of a burst with an observed fluence,
$F$, is $7^{(+11)}_{(-4)} \cdot 10^{50} (F/10^{-7}{\rm
ergs/cm}^2)$ergs. The distance to the sources decreases and
correspondingly the rate increases and the energy decreases if the
spectral index is $2$ and not $1.5$.  
These numbers vary slightly if the bursts have a
wide luminosity function.

Short bursts are detected only up to a much nearer distances:
$\zm(short)$ $=0.4^{+1.1}$, again assuming standard candles and no
source evolution. There is no significant lower limit on $\zm$ for
short bursts and their distribution is compatible with a homogeneous
non-cosmological one. The estimate of $\zm(short)$ corresponds to a
comparable rate of $6.3_{(-5.6)} \cdot 10^{-6}$ events per year per
galaxy and a typical energy of $3^{(+39)} \cdot 10^{49} F_{-7}$ ergs
(there are no lower limits on the energy or and no upper limit on the
rate since there is no lower limit on $\zm(short))$.  The fact that
short bursts are detected only at  nearer distances is also
reflected by the higher $\langle V/V_{max} \rangle$ of the population
of these bursts \cite{Katz96}.

Relatively wide luminosity distributions are allowed by the data
\cite{CP95}. For example, the KS test gives a probability of 80\% for
a double peaked luminosity distribution with luminosity ratio of 14.
These results demonstrate that the BATSE data alone allow a
variability of one order of magnitude in the luminosity.

The above considerations should be modified if the rate of GRBs trace
the SFR - the star formation rate
\cite{Totani,Sahuetal97a,Wijersetal98}. The SFR has been determined
recently by two independent studies
\cite{Lilly96,Madauetal96,Connolly97}. The SFR   peaks at
$z\sim 1.25$. This is a strongly evolving non monotonic distribution.
which is drastically different from the power laws considered so far.
Sahu \etall \cite{Sahuetal97a} find that $\rho(z) \propto SFR(z)$
yields $N(>f)$ distribution that is compatible with the observed one
(for $q_0=0.2$, $H_0=50$km/sec$^{-1}$Mpc$^{-1}$) for a narrow
luminosity distribution with $L_\gamma =10^{51}$ergs/sec.   Wijers
\etall \cite{Wijersetal98} find that the implied peak luminosity is
higher $L_\gamma =8.3 \cdot 10^{51}$ergs/sec and it corresponds to a
situation in which the dimmest bursts observed by BATSE originate from
$z \approx 6$!

The direct red-shift measure of GRB970508
\cite{Metzger97a} agrees well with estimates made previously using
peak-flux count statistics (\cite{Fenimore93a,Loredo2,CP95}). The
red-shift of GRB971214, $z=3.418$, and of GRB980703, $z=0.966$, and
the implied luminosities  disagree  with these
estimates. A future detection of additional red-shifts for other bursts
will enable us to estimate directly the luminosity function of GRBs.
It will also enable us to determine the evolution of GRBs. 
Krumholz \etall \cite{Krumholz98} and Hogg \& Fruchter
\cite{Hogg_Fruchter98} find that with a wide luminosity function
both models of a constant GRB rate and a GRB rate following the 
star formation rate are consistent with the peak flux distribution
and with the observed redshift of the three GRBs.

\subsection{Time Dilation}

Norris \etall \cite{Norris94,Norris95} examined 131 long bursts (with a 
duration longer than 1.5s) and found that the dimmest bursts
are longer by a factor of $\approx 2.3$ compared to the bright ones.
With our canonical value of $\zm =2.1$ the bright bursts originate at
$z_{bright} \approx 0.2$.  The corresponding expected ratio due to
cosmological time dilation, $ 2.6$, is in agreement with this
measurement. Fenimore and Bloom \cite{FB94} find, on the other hand,
that when the fact that the burst's duration decreases as a function
of energy as $\Delta t \approx E^{-0.5}$ is included in the analysis,
this time dilation corresponds to $\zm > 6$. This would require a
strong negative intrinsic evolution: $\beta \approx-1.5 \pm 0.3 $.
Alternatively, this might agree with the model in which the GRB rate
follows the SFR \cite{Totani,Wijersetal98} which gives $\zm \approx 6$.
Cohen \& Piran \cite{CP96} suggested a way to perform the time
dilation, spectral and red shift analysis simultaneously.
Unfortunately current data are insufficient for this purpose.

\section{The Compactness Problem and Relativistic Motion.}
\label{sec:compact}

The key to understanding GRBs lies, I believe, in understanding how
GRBs bypass the compactness problem.  This problem was realized very
early on in one form by Ruderman \cite{Ruderman75} and in another way
by Schmidt \cite{Schmidt78}. Both used it to argue that GRBs cannot
originate from cosmological distances.  Now, we understand that GRBs
are cosmological and special relativistic effects enable us to
overcome this constraint.

The simplest way to see the compactness problem is to estimate the
average opacity of the high energy gamma-ray to pair production.
Consider a typical burst with an observed fluence,
$F$. For a source emitting isotropically
at a distance $D$ this fluence corresponds to a total energy release of:
\begin{equation}
E= 4 \pi D^2 F = 10^{50}{\rm ergs} \bigg({ D
\over 3000~{\rm Mpc}}\bigg)^2 \bigg({ F \over 10^{-7} {\rm ergs/cm^2}}
\bigg) .
\label{ene}
\end{equation}

Cosmological effects change this equality by numerical factors of
order unity that are not important for our discussion.  The rapid
temporal variability on a time scale $\delta T
\approx 10~msec$ implies that the sources are compact with a size,
$R_i < c \delta T \approx 3000$~km.  The observed spectrum (see
section \ref{sub_sec:spectrum}) contains a large fraction of high
energy $\gamma$-ray photons.  These photons (with energy $E_1$) could
interact with lower energy photons (with energy $E_2$) and produce
electron-positron pairs via $\gamma
\gamma \rightarrow e^+ e^- $ if $\sqrt{E_1 E_2} > m_ec^2$
(up to an angular factor).  Denote by $f_p$ the fraction of photon
pairs that satisfy this condition. The average optical depth for this process
is \cite{GFR,Carigan_Katz,Piran_Shemi}:
\begin{displaymath}
\tau_{\gamma\gamma} = {f_{p}  \sigma_T F D^2   \over R_i^2 m_e c^2 }\ , 
\end{displaymath}
or
\begin{equation}
\tau_{\gamma\gamma} = 10^{13} f_{p}
\bigg({ F \over 10^{-7} {\rm ergs/cm^2} } \bigg)
\bigg({ D \over 3000~{\rm Mpc}}\bigg)^2
\bigg({ \delta T \over 10~{\rm msec}} \bigg)^{-2} ,  
\label{tau}
\end{equation}
where $\sigma_T$ is the Thompson cross-section.  This optical depth is
very large.  Even if there are no pairs to begin with they will form
rapidly and then these pairs will Compton scatter lower energy
photons, resulting in a huge optical depth for all photons.  However, the
observed non-thermal spectrum indicates with certainty that the
sources must be optically thin!

An alternative calculation is to consider the optical depth of the
highest energy photons (say a GeV photon) to pair production with the
lower energy photons. The observation of GeV  photons shows  that they
are able to escape freely. In other words it means that this optical
depth must be much smaller than unity
\cite{Fenimore93,WoodLoeb}. This consideration leads to
 a slightly stronger but comparable limit on the opacity.

The compactness problem stems from the assumption that the size of 
the  sources emitting the observed radiation is  determined by the 
observed variability time scale. There won't be a problem
if the source emitted the energy in another form and it was converted
to the observed gamma-rays at a large distance, $R_X$, where the
system is optically thin and  $\tau_{\gamma\gamma}(R_X) <1$. A
trivial solution of this kind is based on a weakly interacting
particle, which is converted in flight to electromagnetic radiation.
The only problem with this solution is that there is no known particle
that can play this role (see, however \cite{Loeb}).

\subsection{Relativistic Motion}
\label{ss:motion}

Relativistic effects can fool us and, when ignored, lead to wrong
conclusions. This happened thirty  years ago when rapid variability implied
``impossible'' temperatures in extra-galactic radio sources.  This
puzzle was resolved when it was suggested \cite{Woltjer,Rees67}
that these objects reveal
ultra-relativistic expansion. This was confirmed later by VLBA
measurements of superluminal jets with Lorentz factors of order two 
to ten.  This also happened in the present
case. Consider a source of radiation that is moving towards an
observer at rest with a relativistic velocity characterized by a
Lorentz factor, $\gamma=1/\sqrt{1-v^2/c^2} \gg 1 $.  Photons with an
observed energy $h \nu_{obs}$ have been
blue shifted and their energy at the source was $\approx h
\nu_{obs}/\gamma$.  Since the energy at the source is lower fewer
photons have sufficient energy to produce pairs. Now the observed fraction
$f_p$, of photons that could produce pairs  is not equal to the
fraction of photons that could produce pairs at the source. The latter   
is smaller by a factor $\gamma^{-2\alpha}$ (where $\alpha$
is the high energy spectral index) than the observed fraction. 
At the same time, relativistic
effects allow the radius from which the radiation is emitted, $R_e <
\gamma^2 c
\delta T$ to be larger than the original estimate, $R_e < c \delta T$,
by a factor of $\ga ^2$. We have
\begin{displaymath}
\tau_{\gamma\gamma} =
{f_{p} \over \ga ^{2 \alpha}} {{\sigma_T F D^2} \over {R_e^2 m_e c^2}}\ ,
\end{displaymath}
or
\begin{equation}
\tau_{\gamma\gamma} \approx {10^{13} \over \ga ^{(4+2\alpha)}}
f_{p}
\bigg({ F \over 10^{-7} {\rm ergs/cm^2}} \bigg)
\bigg({ D \over 3000~{\rm Mpc}}\bigg)^2
\bigg({ \delta T \over 10~{\rm msec}} \bigg)^{-2} ,
\label{tauR}
\end{equation}
where the relativistic limit on $R_e$ was included in the second line.
The compactness problem can be resolved if the source is moving
relativistically towards us with a Lorentz factor $\ga >
10^{13/(4+2\alpha)} \approx 10^2$. A more detailed discussion
\cite{Fenimore93,WoodLoeb} gives comparable limits on $\gamma$.  Such
extreme-relativistic motion is larger than the relativistic motion
observed in any other celestial source. Extragalactic super-luminal
jets, for example, have Lorentz factors of $\sim 10$, while the known
galactic relativistic jets \cite{Mirabel} have Lorentz factors of
$\sim 2$ or less.

The potential of relativistic motion to resolve the compactness
problem was realized in the eighties by Goodman \cite{Goo86},
Paczy\'nski \cite{Pac86} and Krolik and Pier \cite{KroPie}. There was,
however, a difference between the first two approaches and the
last one.  Goodman \cite{Goo86} and Paczy\'nski \cite{Pac86}
considered relativistic motion in the dynamical context of fireballs,
in which the relativistic motion is an integral part of the dynamics
of the system.  Krolik and Pier \cite{KroPie} considered, on the other
hand, a kinematical solution, in which the source moves
relativistically and this motion is not necessarily related to the
mechanism that produces the burst.

Is a purely kinematic scenario feasible? In this scenario the source
moves relativistically as a whole.  The radiation is beamed with an
opening angle of $\ga ^{-1}$.  The total energy emitted in the source
frame is smaller by a factor $\ga ^{-3}$ than the isotropic estimate
given in Eq.  (\ref{ene}).  The total energy required, however, is at
least $(M c^2 + 4 \pi F D^2 /\ga ^3) \ga $, where $M$ is the rest mass
of the source (the energy would be larger by an additional amount
$E_{th} \ga $ if an internal energy, $E_{th}$, remains in the source
after the burst has been emitted).  For most scenarios that one can
imagine $Mc^2 \ga \gg (4 \pi /\ga ^2) F D^2$. The kinetic energy is
much larger than the observed energy of the burst and the process is
extremely (energetically) wasteful.  Generally, the total energy
required is so large that the model becomes infeasible.

The kinetic energy could be comparable to the observed energy if it
also powers the observed burst.  This is the most
energetically-economical situation.  It is also the most
conceptually-economical situation, since in this case the $\gamma$-ray
emission and the relativistic motion of the source are related and are
not two independent phenomena.  This will be the case if GRBs result
from the slowing down of ultra relativistic matter. This idea was
suggested by M\'esz\'aros, and Rees \cite{MR1,MR2} in the context of
the slowing down of fireball accelerated material \cite{SP} by the ISM
and by Narayan, et al. \cite{NPP} and independently by Rees and
M\'esz\'aros \cite{MR4} and Paczy\'nski and Xu \cite{PacXu} in the
context of self interaction and internal shocks within the fireball.
It is remarkable that in both cases the introduction of energy
conversion was motivated by the need to resolve the ``Baryonic
Contamination'' problem (which we discuss in the next section).  If
the fireball contains even a small amount of baryons all its energy
will eventually be converted to kinetic energy of those baryons.  A
mechanism was needed to recover this energy back to radiation.
However, it is clear now that the idea is much more general and it is
an essential part of any GRB model regardless of the nature of the
relativistic energy flow and of the specific way it slows down.

Assuming that GRBs result from the slowing down of a relativistic bulk
motion of massive particles, the  rest  mass of the
ultra-relativistic particles is:
\begin{eqnarray}
M = { \theta^2 FD^2 \over \ga \epsilon_c c^2} \approx 
10^{-6} {\rm M}_\odot \epsilon_c^{-1}
\bigg({ \theta^2 \over 4 \pi}\bigg) \\ \nonumber 
\times \bigg({ F \over 10^{-7} {\rm ergs/cm^2}} \bigg) 
\bigg({ D \over 3000~{\rm Mpc}}\bigg)^2
\bigg({ \ga \over 100}\bigg)^{-1} 
\label{mass}
\end{eqnarray}
where $\epsilon_c$ is the conversion efficiency and $\theta$ is the
opening angle of the emitted radiation. 
We see that the allowed mass
is very small. Even though a way was found to convert back the kinetic
energy of the baryons to radiation (via relativistic shocks)
there is still a  ``baryonic contamination'' problem. 
Too much baryonic mass will slow down the flow and it won't be
relativistic.

\subsection{Relativistic Beaming?}

\begin{figure}
\begin{center}
\includegraphics[width=10cm]{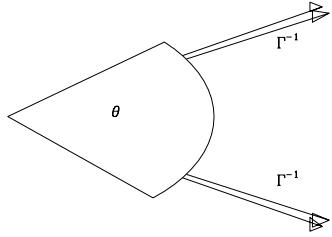}
\caption{\it  Radiation from a relativistic beam with a width
$\theta$. Each observer will detect radiation only from a very narrow
beam with a width $\Gamma^{-1}$.  The overall angular size of the
observed phenomenon can vary, however, with $\Gamma^{-2} < \theta^2 <
4 \pi$.}
\label{beaming}
\end{center}
\end{figure}

Radiation from relativistically moving matter is beamed in the
direction of the motion to within an angle $\ga ^{-1}$. In spite of
this the radiation produced by relativistically moving matter can
spread over a much wider angle.  This depends on the geometry of the
emitting region.  Let $\theta_M$ be the angular size of the
relativistically moving matter that emits the burst.  The beaming
angle $\theta$ will be $\theta_M$ if $\theta_M>\ga ^{-1}$ and $\ga
^{-1}$ otherwise. Thus if $\theta_M = 4 \pi$ - that is if the emitting
matter has been accelerated spherically outwards from a central source
(as will be the case if the source is a spherical fireball) - the
burst will be isotropic even though each observer will observe
radiation coming only from a very small region (see
Fig. \ref{beaming}). The radiation will be beamed into $\ga ^{-1}$
only if the matter has been accelerated along a very narrow beam. The
opening angle can also have any intermediate value if it emerges from
a beam with an opening angle $\theta > \ga ^{-1}$, as will be the case
if the source is an anisotropic fireball \cite{Pi94,NP97} or an
electromagnetic accelerator with a modest beam width.

Beaming requires, of course, an event rate larger by a ratio $4 \pi /
\theta^2$ compared to the observed rate.  Observations of about one
burst per $10^{-6}$ year per galaxy implies one event per hundred
years per galaxy if $\theta \approx \ga^{-1}$ with $\gamma$ given by
the compactness limit of $\sim 100$.

\section{An Overview of the Generic Model}

It is worthwhile to summarize now the essential features of the generic
GRB model that arose from the previous discussion. Compactness has led
us to the requirement of relativistic motion, with a Lorentz factor
$\ga \ge 100$. Ockham's razor and the desire to limit the
total energy have lead us to the idea that the observed gamma-rays
arise in the process of slowing down of a relativistic energy flow, at
a stage that the motion of the emitting particles is  still highly
relativistic.

This leads us to the generic picture mentioned earlier and to the
suggestion that GRBs are composed of a three stage phenomenon: (i) a
compact inner hidden ``engine'' that produces a relativistic energy
flow, (ii) the energy transport stage and (iii) the conversion of this
energy to the observed prompt radiation.  One may add a forth stage
(iv) conversion of the remaining energy to radiation in other
wavelengths and on a longer time scale - the ``afterglow''.

\subsection{Models for The Energy Flow}

The simplest mode of relativistic energy flow is in the form of
kinetic energy of relativistic particles.  A variant that have been
suggested is based on the possibility that a fraction of the energy is
carried by Poynting flux \cite{Thom,Usov94,Usov95,Katz97,MR97a}
although in all models the power must be converted to kinetic energy
somewhere. The energy flow of $\sim 10^{50}$ergs/sec from a compact
object whose size is $\sles 10^7$cm requires a magnetic field of
$10^{15}$Gauss or higher at the source. This large value might be
reached in stellar collapses of highly magnetized stars or amplified
from smaller fields magnethohydrodynamically \cite{Katz97}. Overall
the different models can be characterized by two parameters: the ratio
of the kinetic energy flux to the Poynting flux and the location of
the energy conversion stage ($\approx 10^{12}$ cm for internal
conversion or $\approx 10^{16}$cm for external conversion). This is
summarized in Table \ref{t:energy_transport}.  In the following
section we will focus on the simplest possibility, that is of a
kinetic energy flux.

\begin{center}
\begin{table*}[ht!]
\label{t:energy_transport}
  \begin{center}
\begin {tabular}{|c||c|c|c|}
  \hline & Kinetic Energy & Kinetic Energy and & Poynting Flux \\ &
  Dominated & Poynting Flux & Dominated \\ \hline\hline Internal &
  \cite{NPP,MR4,Katz94} & \cite{Thom} & \cite{Usov94,Katz97,MR97a} \\ 
  conversion & & & \\
  \hline External & \cite{MR1,MR2} & - & \cite{Usov95,Katz97} \\ conversion &
  & & \\ \hline
\end{tabular}
\caption{\it General Scheme for Energy Transport}
\end{center}
\end{table*}
\end{center}

\subsection{Models for The Energy Conversion}

Within the baryonic model the energy transport is in the from of the
kinetic energy of a shell of relativistic particles with a width
$\Delta$. The kinetic energy is converted to ``thermal'' energy of
relativistic particles via shocks.  These particles then release this
energy and produce the observed radiation.  There are two modes of
energy conversion (i) External shocks, which are due to interaction
with an external medium like the ISM. (ii) Internal shocks that arise
due to shocks within the flow when fast moving particles catch up with
slower ones. Similar division to external and internal energy
conversion occurs within other models for the energy flow.

External shocks  arise from the  interaction of the shell with  external
matter.  The typical length scale is the Sedov length, $l \equiv
(E/n_{ism} m_{p}c^2)^{1/3}$. The rest mass energy within a sphere of
radius $l$, equals the energy of the shell. Typically $l\sim
10^{18}$cm.  As we see later (see section \ref{sub_sec:Newtonian})
relativistic external shocks (with a Newtonian reverse shock) convert
a significant fraction of their kinetic energy at
$R_\ga=l/\gamma^{2/3} \approx 10^{15}-10^{16}$cm, where the external
mass encountered equals $\gamma^{-1}$ of the shell's mass.
Relativistic shocks (with a relativistic reverse shock) convert their
energy at $R_\Delta=l^{3/4}\Delta^{1/4} \approx 10^{16}$cm, where the
shock crosses the shell.

Internal shocks occur when one shell overtakes another. If the initial
separation between the shells is $\delta$ and both move with a Lorentz
factor $\gamma$ with a difference of order $\gamma$ these shocks take place
at: $\delta \gamma^2$. A typical value is $10^{12}-10^{14}$cm.

\subsection{Typical Radii}
In table \ref{t:radii} we list the different  radii  that arise in the
fireball evolution.

\begin{center}
\begin{table*}[ht!]

\label{t:radii}
  \begin{center}
\begin {tabular}{|c|c|c|c|}
  \hline
  $R_i$ & Initial Radius & $c \delta t$  & $ \approx 10^7-10^8$cm  \\
  $R_\eta$&Matter dominates &$R_i  \eta$ & $ \approx 10^9$cm \\
  $R_{pair}$& Optically thin to pairs & $
   [ (3 E /4 \pi R_i^3 a)^{1/4}/T_p ] R_i $ &
  $ \approx 10^{10}$cm \\  
  $R_e$& Optically thin & $ ({\sigma_T E/  4 \pi m_p c^2 \eta} )^{1/2}$
  &$\approx 10^{13}$cm \\
  $R_\delta$&Internal collisions &$\delta  \gamma^2$ &
  $\approx 10^{12}-10^{14}$cm\\
  $R_\gamma$ & External Newtonian Shocks& $l \gamma^{-2/3}$&
  $\approx 10^{16}$cm \\
  $R_\Delta$ & External Relativistic shocks&$ l^{3/4} \Delta^{1/4}$ &
  $\approx 10^{16}$cm \\
  $l$ or $L$ &Non relativistic external shock  &$l$ $^{(a)}$ or  $l\ga^{-1/3}$
  $^{(b)}$ & $\approx 10^{17}-10^{18}$cm \\
  l & Sedov Length  &$l=(3E/4 \pi n_{ism} m_p c^2)^{1/3}$ & $\approx 10^{18}$cm\\
  \hline
\end{tabular}

\end{center}
\caption{\it Critical Radii
(a) - adiabatic fireball; (b) - radiative fireball}
\end{table*}
\end{center}

\begin{figure}
\begin{center}
\includegraphics[width=10cm]{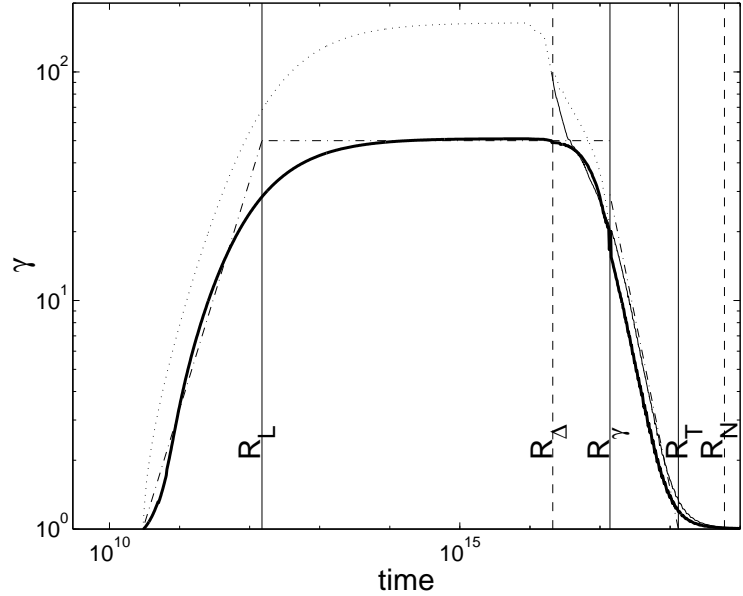}
\caption{\it  Fireball evolution  from its initial
formation at rest to the final Newtonian Sedov solution. The energy
extraction is due to the interaction with the ISM  via a
relativistic forward shock and a Newtonian reverse shock. We have used
for this calculations $\xi=43$, $E_0=10^{52}$ [erg], $\gamma_0=50$
$R_0=3\times10^{10} [cm]$.  Shown are the average value of the Lorentz
factor (thick solid line), the value at the forward shock (thin solid
line), the maximal value (dotted line) and an analytic estimate
(dashed dotted line). From \cite{KPS98}. }
\label{fig:full_NRS}
\end{center}
\end{figure}

\begin{figure}
\begin{center}
\includegraphics[width=10cm]{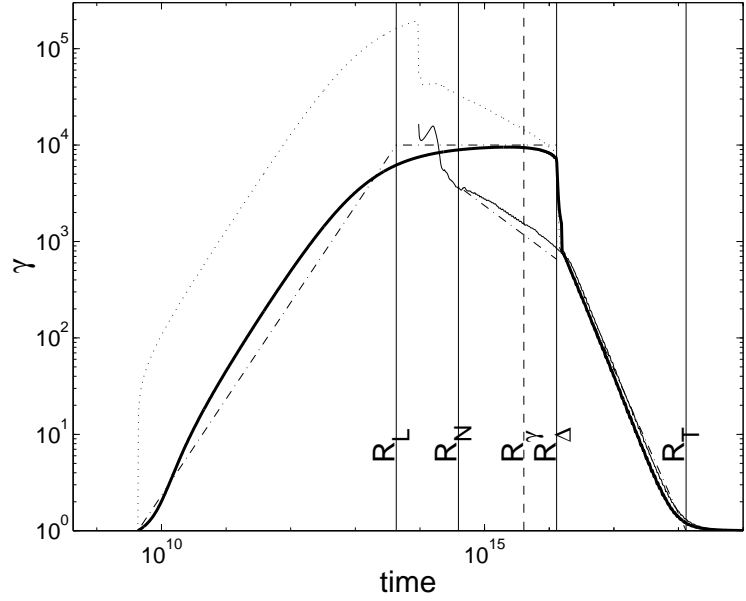}
\caption{\it Fireball  evolufrom its initial
formation at rest to the final Newtonian Sedov solution. The energy
extraction is due to the interaction with the ISM  via relativistic
forward and reverse shocks. The parameters for this computation are:
$\xi=0.1$, $E_0=10^{52}$[erg], $\gamma_0=10^4$,
$R_0=4.3\times10^{9}[cm]$.  Shown are the average value of the Lorentz
factor (thick solid line), the value at the forward shock (thin solid
line), the maximal value (dotted line) and an analytic estimate
(dashed dotted line). From \cite{KPS98}. }
\label{fig:full_RRS}
\end{center}
\end{figure}

Figs. \ref{fig:full_NRS} and \ref{fig:full_RRS} (from \cite{KPS98})
depict a numerical solution of a fireball from its initial configuration
at rest to its final Sedov phase.

\section{Fireballs}
\label{sec:fireball}

Before turning to the question of  how is the kinetic energy of the
relativistic flow  converted to radiation we ask is it possible to
produce the needed flows? More specifically, is it possible to
accelerate particles to relativistic velocities? It is remarkable that
a relativistic particle flow is almost the unavoidable outcome of a
``fireball'' - a large concentration of energy (radiation) in a small
region of space in which there are relatively few baryons. 
The
relativistic fireball model was proposed by Goodman
\cite{Goo86} and by Paczy\'nski \cite{Pac86}. They have shown that the
sudden release of a large quantity of gamma ray photons into a compact
region can lead to an opaque photon--lepton ``fireball'' through the
production of electron--positron pairs.  The term ``fireball" refers
here to an opaque radiation--plasma whose initial energy is
significantly greater than its rest mass.  Goodman \cite{Goo86}
considered the sudden release of a large amount of energy, $E$, in a
small volume, characterized by a radius, $R_i$. Such a situation could
occur in an explosion.  Paczy\'nski \cite{Pac86} considered a steady
radiation and electron-positron plasma wind that emerges from a
compact region of size $R_i$ with an energy, $E$, released on a time
scale significantly larger than $R_i/c$. Such a situation could occur
if there is a continuous source that operates for a while.  As it will
become clear later both configurations display, in spite of the
seemingly large difference between them, the same qualitative
behavior.  Both Goodman \cite{Goo86} and Paczy\'nski
\cite{Pac86} considered a pure radiation fireballs in which
there are no baryons. Later Shemi \& Piran \cite{SP} and Paczy\'nski
\cite{Pac90}  considered the effect of a baryonic load.  They 
showed that under most circumstances the ultimate outcome of a fireball
with a baryonic load will be the transfer of all the energy of the
fireball to the kinetic energy of the baryons. If the baryonic load
is sufficiently small the baryons will be accelerated to a
relativistic velocity with $\gamma \approx E/M$. If it is large
the net result will be a Newtonian flow with $v \simeq \sqrt{{{2E}/{M}}}$.

\subsection{A simple model}

The evolution of a homogeneous fireball can be understood by a simple
analogy to the Early Universe \cite{SP}. Consider, first, a pure
radiation fireball. If the initial temperature is high enough pairs
will form.  Because of the opacity due to pairs, the radiation cannot
escape. The pairs-radiation plasma behaves like a perfect fluid with
an equation of state $p=\rho/3$. The fluid expands under of its own
pressure. As it expands it cools with ${\cal T} \propto R^{-1}$
(${\cal T}$ being the local temperature and $R$ the radius).  The
system resembles quite well a part of a Milne Universe in which
gravity is ignored.  As the temperature drops below the
pair--production threshold the pairs annihilate. When the local
temperature is around $20$keV the number of pairs becomes sufficiently
small, the plasma becomes transparent and the photons escape freely to
infinity.  In the meantime the fireball was accelerated and it is
expanding relativistically outwards.  Energy conservation (as viewed
from the observer frame) requires that the Lorentz factor
that corresponds to this outward motion satisfies $\gamma
\propto R$. 
The escaping photons, whose local energy (relative to the
fireball's rest frame) is $\approx 20$keV are blue shifted. An
observer at rest detects them with a temperature of ${\cal T}_{obs}
\propto \gamma {\cal T}$. Since ${\cal T} \propto R^{-1}$ and $\gamma \propto
R$ we find that the observed temperature, ${\cal T}_{obs}$, approximately
equals $ {\cal T}_0$, the initial temperature.  The observed spectrum, is
however, almost thermal \cite{Goo86} and it is still far from the one
observed in GRBs.

In addition to radiation and $e^+e^-$ pairs, astrophysical fireballs
may also include some baryonic matter which may be injected with the
original radiation or may be present in an atmosphere surrounding the
initial explosion. These baryons can influence the fireball evolution in 
two ways. The electrons associated with this matter increase
the opacity, delaying the escape of radiation.  
Initially, when the local temperature
${\cal T}$ is large, the opacity is dominated by $e^+e^-$ pairs \cite{Goo86}.
This opacity, $\tau_p$, decreases exponentially with decreasing
temperature, and falls to unity when ${\cal T}={\cal T}_p\approx 20$keV.  The
matter opacity, $\tau_b$, on the other hand decreases only as
$R^{-2}$, where $R$ is the radius of the fireball.  If at the point
where $\tau _p =1$, $\tau_{b}$ is still $ > 1$, then the final
transition to $\tau=1$ is delayed and occurs at a cooler temperature.

More importantly, the baryons are accelerated with the rest of the
fireball and convert part of the radiation energy into bulk kinetic
energy.  The expanding fireball has two basic phases: a radiation
dominated phase and a matter dominated phase.  Initially, during the
radiation dominated phase the fluid accelerates with $\ga \propto R$.
The fireball is roughly homogeneous in its local rest frame but due to
the Lorentz contraction its width in the observer frame is $\Delta
\approx R_i$, the initial size of the fireball.
A transition to the matter dominated phase takes place when the
fireball has a size
\begin{equation}
R_\eta = {R_i E \over M c^2} \approx 10^{9}{\rm cm}
R_{i7} E_{52}
({ M / 5\cdot 10^{-6} M_\odot})^{-1}
\end{equation}
and the mean Lorentz factor of the fireball is $\ga \approx E/ M c^2$.
We have defined here $E_{52} \equiv E/10^{52}$ergs and $R_{i7} \equiv
R_i/10^7$cm.  After that, all the energy is in the kinetic energy of
the matter, and the matter coasts asymptotically with a constant
Lorentz factor.

The matter dominated phase is itself further divided into two
sub-phases.  At first, there is a frozen-coasting phase in which the
fireball expands as a shell of fixed radial width in its own local
frame, with a width $\sim \bar\gamma R_i \sim (E/ M c^2) R_i$.
Because of Lorentz contraction the shell appears to an observer with a
width $\Delta \approx R_i$.  Eventually, when the size of the fireball
reaches $R_s = \Delta \ga^2 \approx 10^{11}{\rm cm}(\Delta /10^7 {\rm
cm}) (\ga/100)^2$ variability in $\ga$ within the fireball results in
a spreading of the fireball which enters the coasting-expanding
phase. In this final phase, the width of the shell grows linearly with
the size of the shell, $R$:
\begin{eqnarray}
\Delta(R) \approx R/\ga^2 \approx 10^7 {\rm cm}
\bigg({ R \over 10^{11}{\rm cm}}\bigg)
\bigg({ \ga \over 100}\bigg)^{-2} \\ \nonumber
{\rm for} \ \  R > R_s= 10^{11}{\rm cm}
\bigg({ R_i \over 10^{7}{\rm cm}}\bigg)
\bigg({ \ga \over 100}\bigg)^{2}.
\end{eqnarray}

The initial energy to mass ratio, $\eta = (E/Mc^2) $, determines
the order of these transitions.  There are two critical
values for $\eta$ \cite{SP}:
\begin{equation} \eta_{pair} = ({3 \sigma_T^2 E \sigma T_p^4
\over 4 \pi m_p^2 c^4 R_i} )^{1/2} \approx 3 \cdot 10^{10}
E_{52}^{1/2} R_{i7}^{-1/2}
\end{equation}
and
\begin{equation}
\eta_{b} = ({3 \sigma_T E \over 8 \pi m_p c^2 R_i^2})^{1/3} \approx
10^5 E_{52}^{1/3} R_{i7}^{-2/3}
\end{equation}

These correspond to four different types of fireballs:

\begin{center}
\begin{table*}[ht!]
\label{t:fireballs}
  \begin{center}
\begin {tabular}{|c|c|c|}
  \hline
  Type& $\eta= E/Mc^2$ & $M$ \\
  \hline \hline Pure Radiation & $\eta_{pair}
  < \eta$ & $M< M_{pair} = 10^{-12} M_\odot E_{52}^{1/2} R_{i7}^{1/2}$ \\
  \hline Electrons Opacity & $\eta_{b} < \eta < \eta_{pair} $ &
  $M_{pair}< M < M_b= 2 \cdot 10^{-7} M_\odot E_{52}^{2/3}
  R_{i7}^{2/3}$\\ \hline Relativistic Baryons & $1 < \eta < \eta_{b}
  $ & $M_b < M < 5 \cdot 10^{-3} M_\odot E_{52}$ \\ \hline
  Newtonian & $ \eta < 1$ & $5 \cdot 10^{-4} M_\odot E_{52} < M $\\
  \hline
\end{tabular}
\end{center}
\caption{\it Different Fireballs}
\end{table*}
\end{center}

\begin{itemize}
\item (i) {\bf A Pure Radiation Fireball} ($\eta_{pair} < \eta$):
 The effect of the baryons is negligible and the
 evolution is of a pure photon-lepton fireball.  When the temperature
 reaches ${\cal T}_p$, the pair opacity $\tau_p$ drops to 1 and $\tau _b \ll
 1$.  At this point the fireball is radiation dominated ($E >Mc^2$)
 and most of the energy escapes as radiation.

\item(ii) {\bf Electron Dominated Opacity} ($\eta_b<\eta < \eta_{pair}$): In the late
stages, the opacity is dominated by free electrons associated with the
baryons.  The comoving temperature decreases far below ${\cal T}_p$ before
$\tau$ reaches unity.  However, the fireball continues to be radiation
dominated and most of the energy still escapes as radiation.

\item(iii) {\bf Relativistic Baryonic Fireball} ($1<\eta<\eta_b$):
The fireball becomes matter dominated before it becomes optically
thin.  Most of the initial energy is converted into bulk kinetic
energy of the baryons, with a final Lorentz factor $\ga_f \approx
(E/Mc^2) $. This is the most interesting situation for GRBs.

\item (iv) {\bf Newtonian Fireball} ($\eta<1$): 
This is the Newtonian regime.  The rest energy
exceeds the radiation energy and the expansion never becomes
relativistic. This is the situation, for example in supernova
explosions in which the energy is deposited into a massive envelope.
\end{itemize}

\subsection{ Extreme-Relativistic Scaling Laws.}

The above summary describes the qualitative features of a roughly
homogeneous expanding fireball.  Surprisingly similar scaling laws
exists also for inhomogeneous fireballs \cite{PSN} as well as for 
relativistic winds \cite{Pac86}.  Consider a spherical fireball with
an arbitrary radial distribution of radiation and matter.  Under
optically thick conditions the radiation and the relativistic leptons
(with energy density $e$) and the matter (with baryon mass density
$\rho$) at each radius behave like a single fluid, moving with the
same velocity. The pressure, $p$, and the energy density, $e$, are
related by $p =e/3 $.  We can express the relativistic conservation
equations of baryon number, energy and momentum using characteristic
coordinates: $r$ and $s\equiv t-r$ as \cite{PSN}:
\begin{equation}
{1\over r^2}{\partial \over \partial r}(r^2\rho u)=-{\partial
\over\partial s}\left ({\rho \over \ga+u} \right ),\label{f4}
\end{equation}
\begin{equation} {1\over r^2}{\partial \over \partial
r}(r^2e^{3/4}u)=-{\partial \over\partial s}\left ({e^{3/4} \over
\ga+u}\right ),\label{f5}
\end{equation}
\begin{eqnarray} {1\over r^2}{\partial \over \partial r}\left [r^2\left
(\rho+{4\over 3}e\right )u^2\right ]= -{\partial \over\partial s}\left
[\left (\rho+{4\over 3}e\right ) {u\over \ga +u}\right ]\nonumber \\
+{1\over 3}\left
[{\partial e\over \partial s} -{\partial e\over \partial r}\right
],\label{f6}
\end{eqnarray}
where $u=u^r = \sqrt{\ga^2-1}$, and we use units in which $c=1$ and the
mass of the particles $m=1$.  The derivative $\partial /\partial r$
now refers to constant $s$, i.e.  is calculated along a characteristic
moving outward at the speed of light.  After a short acceleration
phase we expect that the motion of a fluid shell will become highly
relativistic ($\ga \gg 1$).  If we restrict our attention to the
evolution of the fireball from this point on, we may treat $\ga ^{-1}$
as a small parameter and set $\ga \approx u$, which is accurate  to order
$O(\ga ^{-2})$.  Then, under a wide range of conditions the quantities
on the right-hand sides of Eqs. \ref{f4}-\ref{f6} are significantly
smaller than those on the left.  When we neglect the right hand sides
of Eqs. \ref{f4}-\ref{f6} the problem becomes effectively only $r$
dependent. We obtain the following conservation
laws for each fluid shell:
\begin{equation}
r^2 \rho \ga = {\rm const.}, \  r^2e^{3/4}\ga = {\rm const.}, 
\ \ r^2\left (\rho +{4\over 3}e\right )\ga ^2 = {\rm const.}.
\label{f7}
\end{equation}

A scaling solution that is valid in both the radiation-dominated and
matter-dominated regimes, as well as in the transition zone in
between, can be obtained by combining the conserved quantities in
Eq. \ref{f7} appropriately.  Let $t_0$ be the time and $r_0$ be the
radius at which a fluid shell in the fireball first becomes
ultra-relativistic, with $\ga \ \sgreat\ {\rm few}$.  We label various
properties of the shell at this time by a subscript $0$, e.g. $\ga _0$,
$\rho_0$, and $e_0$.  Defining the auxiliary quantity $D$, where
\begin{equation}
{1 \over D} \equiv {\ga_0 \over \ga } + {3\ga_0 \rho_0\over 4e_0 \ga } -
{3\rho_0\over 4e_0},
\label{f10}
\end{equation}
we find that
\begin{equation}
r = r_0 { \ga_0^{1/2} D^{3/2} \over \ga^{1/2}},
\qquad \rho = {\rho_0 \over D^3},
\qquad e = {e_0 \over D^4}.
\label{f11}
\end{equation}
These are parametric relations which give $r,~\rho$, and $e$ of each
fluid shell at any time in terms of the $\ga$ of the shell at that
time.  The parametric solution \ref{f11} describes both the
radiation-dominated and matter-dominated phases of the fireball within
the frozen pulse approximation. That is as long as the fireball does
not spread due to variation in the velocity.

\subsection{The  Radiation-Dominated Phase}

Initially the fireball is radiation dominated.  For $\ga\ll
(e_0/\rho_0)\ga_0$, the first term in Eq \ref{f10} dominates and we
find $D\propto r$, $\ga\propto r$, recovering the radiation-dominated
scaling:
\begin{equation}
\ga\propto r,\qquad \rho\propto r^{-3},\qquad e\propto r^{-4}.
\label{f8}
\end{equation}
The scalings of $\rho$ and $e$ given in Eq. \ref{f8} correspond to
those of a fluid expanding uniformly in the comoving frame.  Indeed,
all three scalings in Eq. \ref{f8} can be derived for a homogeneous
radiation dominated fireball by noting the analogy with an expanding
universe.

Although the fluid is approximately homogeneous in its own frame,
because of Lorentz contraction it appears as a narrow shell in the
observer frame, with a radial width given by $\Delta r \sim r/\ga \sim
{\rm constant}\sim R_i$, the initial radius of the fireball, or the
initial width of the specific shell under discussion when we consider
a continuous wind.  We can now go back to Eqs. \ref{f4}-\ref{f6} and
set $\partial /\partial s \sim \ga /r$.  We then find that the terms
we neglected on the right hand sides of these equations are smaller
than the terms on the left by a factor $\sim 1/\ga$.  Therefore, the
conservation laws \ref{f7} and the scalings \ref{f8} are valid so long
as the radiation-dominated fireball expands ultra-relativistically
with large $\ga$.  The only possible exception is in the very
outermost layers of the fireball where the pressure gradient may be
extremely steep and $\partial /\partial s$ may be $\gg \ga /r$.
Ignoring this minor deviation, we interpret Eq. \ref{f7} and the
constancy of the radial width $\Delta r$ in the observer frame to mean
that the fireball behaves like a pulse of energy with a frozen radial
profile, accelerating outward at almost the speed of light.

\subsection{The Matter-Dominated Phase}

The radiation dominated regime extends out to a radius $r\sim
(e_0/\rho_0) r_0$.  At larger radii, the first and last terms in
Eq. \ref{f10} become comparable and $\ga$ tends to its asymptotic
value of $\ga_f = (4e_0/3\rho_0+1)\ga_0$.  This is the matter
dominated regime.  (The transition occurs when $4e/3=\rho$, which
happens when $\ga=\ga_f/2$.)  In this regime, $D\propto r^{2/3}$,
leading to the scalings:
\begin{equation}
\ga\rightarrow {\rm constant},\qquad \rho\propto r^{-2},
\qquad e\propto r^{-8/3}.
\label{f9}
\end{equation}
The modified scalings of $\rho$ and $e$ arise because the fireball now
moves with a constant radial width in the comoving frame.  (The
steeper fall-off of $e$ with $r$ is because of the work done by the
radiation through tangential expansion.)  Moreover, since $e\ll \rho$,
the radiation has no important dynamical effect on the motion and
produces no significant radial acceleration.  Therefore, $\ga$ remains
constant on streamlines and the fluid coasts with a constant
asymptotic radial velocity.  Of course, since each shell moves with a
velocity that is slightly less than $c$ and that is different from one
shell to the next, the frozen pulse approximation on which
Eq. \ref{f7} is based must ultimately break down at some large radius.

\subsection{Spreading}

At very late times in the matter-dominated phase the frozen pulse
approximation begins to break down.  In this stage the radiation
density $e$ is much smaller than the matter density $\rho$, and the
Lorentz factor, $\ga$, tends to a constant value $\ga_f$ for each
shell.  We may therefore neglect the term $-(1/3)(\partial e /\partial
r)$ in Eq. \ref{f6} and treat $\ga$ and $u$ in Eqs.  \ref{f4}-\ref{f6}
as constants.  We then find that the flow moves strictly along the
characteristic, $\beta _f t-r={\rm constant}$, so that each fluid
shell coasts at a constant radial speed, $\beta _f=u_f/\ga_f$.  We
label the baryonic shells in the fireball by a Lagrangian coordinate
$\tilde R$, moving with a fixed Lorentz factor $\ga_f(\tilde R)$, and
let $t_c$ and $r_c$ represent the time and radius at which the
coasting phase begins, which corresponds essentially to the point at
which the fluid makes the transition from being radiation dominated to
matter dominated.  We then find
\begin{eqnarray}
r(t,\tilde R) - r_c(\tilde R) = {\sqrt{\ga_f^2(\tilde R)-1} \over
\ga_f(\tilde R)} (t-t_c(\tilde R))
\approx \\ \nonumber 
\left [ 1-{1 \over 2 \ga_f^2(\tilde R)}\right ] [t-t_c(r)].
\label{f12}
\end{eqnarray}
The separation between two neighboring shells separated by a
Lagrangian distance $\Delta \tilde R$ varies during the coasting phase
as
\begin{equation}
\left [{d (\partial r /\partial \tilde R) \over dt}\right ] \Delta \tilde R =
\left [{ 1 \over
\gamma_f(\tilde R)^3} {\partial \ga _f \over \partial \tilde R}\right ]
\Delta \tilde R.
\label{f13}
\end{equation}

Thus the width of the pulse at time $t$ is $\Delta r(t) \approx
\Delta r_c + \Delta \gamma_f (t-t_c) / \bar\gamma_f^3
\approx R_i + (t-t_c) /\ga_f^2 $, where $\Delta r_c\sim R_i $
is the width of the fireball when it begins coasting, $\bar\ga_f$ is
the mean $\ga_f$ in the pulse, and $\Delta \gamma_f\sim \bar\ga_f$ is
the spread of $\ga_f$ across the pulse.  From this result we see that
within the matter dominated coasting phase there are two separate
regimes.  So long as $t-t_c < \bar\ga_f^2R_i$, we have a
frozen-coasting phase in which $\Delta r$ is approximately constant
and the frozen pulse approximation is valid.  In this regime the
scalings in Eq. \ref {f9} are satisfied.  However, when $t-t_c
>\bar\ga_f^2 R_i$, the fireball switches to an expanding-coasting
phase where $\Delta r \propto t-t_c$ and the pulse width grows
linearly with time.  In this regime the scaling of $\rho$ reverts to
$\rho\propto r^{-3}$, and, if the radiation is still coupled to the
matter, $e\propto r^{-4}$.

\subsection{Optical Depth}

Independently of the above considerations, at some point during the
expansion, the fireball will become optically thin.  For a pure
fireball this happens when the local temperature drops to about
$20$keV at:
\begin{equation}
R_{pair} \approx 10^{10}~ {\rm cm}~ E_{52}^{1/4} R_{i7}^{-3/4}
\end{equation}

In a matter dominated fireball the optical depth is usually determined
by the ambient electrons. In this case the fireball becomes optically
thin at:
\begin{equation}
\label{optical_depth}
R_e = ({\sigma_T E\over 4 \pi m_p c^2 \eta} )^{1/2}
\approx 6 \cdot 10^{13}~ {\rm cm}~ \sqrt{E_{52}} (\eta/100)^{-1} .
\end{equation}

From this stage on the radiation and the baryons no longer move with
the same velocity and the radiation pressure vanishes, leading to a
breakdown of Eqs. \ref{f4}-\ref{f6}.  Any remaining radiation will
escape freely now. The baryon shells will coast with their own
individual velocities.  If the fireball is already in the matter
dominated coasting phase there will be no change in the propagation of
the baryons. However, if the fireball is in the radiation dominated
phase when it becomes optically thin, then the baryons will switch
immediately to a coasting phase. This transition radius, $R_e$ has
another crucial role in the fireball evolution. It is the minimal
radius in which energy conversion and generation of the observed GRB
can begin.  Photons produced at $R<R_e$ cannot escape.

\subsection{Anisotropic Fireballs}
It is unlikely that a realistic fireball will be spherically
symmetric. In fact strong deviation from spherical symmetry are
expected in the most promising neutron star merger model, in which the
radiation is expected to emerge through funnels along the rotation
axis.  The initial motion of the fireball might be fairly complex but
once $\ga \gg 1$ the motion of each fluid element decouples from the
motion of its neighbors with angular distance larger than $\ga^{-1}$.
This motion can be described by the same asymptotic solution, as if it
is a part of a spherical shell.  
We define the angular range over which different
quantities vary as $\theta$.
Additionally now the motion is not radially outwards and $u^r \ne u$.
We define the spread angle $\phi$ as
$ u^r \equiv u \cos\phi $.  
 The spherical fireball equations hold
locally if:
\begin{equation}
\ga^{-1}  < \theta \ \ \ \ {\bf \rm and }\ \ \ \
 \phi \ll \sqrt{2}/\gamma.
\end{equation}

\section{Temporal Structure and Kinematic Considerations}
\label{sec:Tempstruct}

General kinematic considerations impose constraints on the
temporal structure produced when the energy of a relativistic shell is
converted to radiation.  The enormous variability of the temporal
profiles of GRBs from one burst to another in contrast to the relatively
regular spectral characteristics, was probably the reason that until
recently this aspect of GRBs was largely ignored.  However, it turns
out that the observed temporal structure sets a strong constraint on the
energy conversion models \cite{SaP97a,FenMadNaya}.  GRBs are highly
variable (see section \ref{sub-sec:variability}) and some
configurations cannot produce such temporal profiles.

\subsection{Time-scales}
\label{sec:Timescales}
\begin{figure}
\begin{center}
\includegraphics[width=10cm]{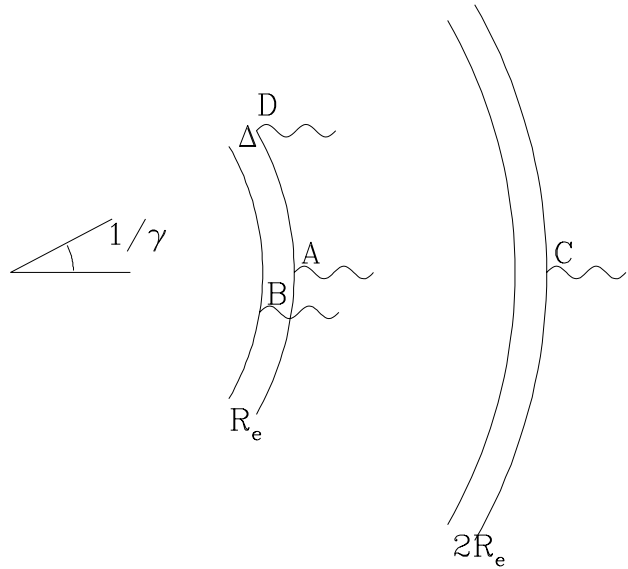}
\caption{\it Different time scales in terms of the arrival time
of four photons: $t_A$,$t_B$, $t_C$, and $t_D$.  $T_{radial}=t_C-t_A$;
$T_{angular} = t_D-t_A$, $\Delta/c = t_B - t_A$.}
\label{fig_timescales}
\end{center}
\end{figure}

Special relativistic effects determine the observed duration of the
burst from a relativistic shell (see Fig. \ref{fig_timescales}).

\begin{itemize}
\item{\bf The Radial Time Scale: $T_{radial}$}:
Consider an infinitely thin relativistic shell with a Lorentz factor
$\ga_E$ (the subscript E is for the emitting region).  Let $R_E$ be a
typical radius characterizing the emitting region (in the observer
frame) such that most of the emission takes place between $R_E$ and $2
R_E$.  The observed duration between the first photon (emitted at
$R_E$) and last one (emitted at $2R_E$) is
\cite{Ruderman75,Katz94}:
\begin{equation}
\label{Tradial}T_{radial}\cong   R_E/2\ga_E ^2c~ .
\end{equation}

\item{\bf The Angular Time Scale: $T_{ang}$}:
Because of relativistic beaming an observer sees up to solid angle of
$\ga_E^{-1}$ from the line of sight. Two photons emitted at the
same time and radius $R_E$, one on the line of sight and the other at
an angle of $\ga_E^{-1}$ away travel different distances to the
observer. The difference lead to a delay in the arrival time by
\cite{Ruderman75,Katz94,FenMadNaya} :
\begin{equation}
\label{Tangular}T_{ang}=R_E/2\ga_E^2c~.
\end{equation}
Clearly this delay is relevant only if the angular width of the
emitting region, $\theta$ is larger than $\ga_E^{-1}$.
\end{itemize}

In addition there are two other time scales that are determined by the
flow of the relativistic particles. These are:
\begin{itemize}
\item{\bf Intrinsic Duration: $\Delta T$}:
The duration of the flow. This is simply the time in which  the source that 
produces the relativistic flow
is active.  $\Delta T = \Delta /c$,
where $\Delta$ is the width of the relativistic wind 
(measured in the observer's rest frame).  For an explosive
source $\Delta \approx R_i$.  However, $\Delta$ could be much larger
for a wind.
The observed duration of the
burst must be longer or equal to $\Delta /c$
\item{\bf Intrinsic Variability $\delta T$}:
The time scale on which the inner source varies and produces 
a subsequent variability with a length scale $\delta =  c\delta T$ in the flow.
Naturally, $\delta T$ sets a lower limit to the variability
time scale observed in any burst.
\end{itemize}
Clearly $\Delta$ and $\delta$ must satisfy:
\begin{equation}
R_{source} \le \delta \le \Delta
\end{equation}
Finally we have to consider the cooling time scale.

\begin{itemize}
\item{\bf The Cooling Time Scale: $T_{cool}$}
This is the difference in arrival time of photons while the shocked
material  cools measured in the rest frame of an observer at rest
at infinity. It is related to the local cooling time, $e /P$ (where $e$
is the internal  energy density and $P$ is the power radiated per unit
volume) in the fluid's rest frame by:
\begin{equation}
T_{cool} = e/P \ga_E
\end{equation}
Note that this differs from the usual time dilation which gives
$\ga_E e/P$. 

For synchrotron cooling there is a unique energy dependence of the
cooling time scale on frequency: $T_{cool} (\nu) \propto  \nu^{-1/2}$
\cite{SNP96} (see Eq. \ref{tausyn2}). If $T_{cool}$ determines the
variability we will have $\delta T (\nu) \propto \nu^{-1/2}$. This is
remarkably close to the observed relation: $\delta T \propto \nu^{-0.4}$
\cite{Fenimore95}.  Quite generally $T_{cool}$ is shorter than the
hydrodynamics time scales \cite{SNP96,SaP97b,Katz97}. However, during the
late stages of an afterglow, $T_{cool}$ becomes the longest time scale
in the system.
\end{itemize}

\subsection{Angular Spreading and  External shocks}

Comparison of Eqs. \ref{Tradial} and \ref{Tangular} (using
$\Delta R_E \sles R_E$) reveals that
$T_{ang} \approx T_{radial}$. As long as the shell's angular width
is larger than $\gamma^{-1}$, any temporal structure that could have
arisen due to irregularities in the properties of the shell or in the
material that it encounters will be spread on a time given by
$T_{ang}$.  This means that $T_{ang}$ is the minimal time
scale for the observed temporal variability: $\delta T \ge
T_{ang}$.

Comparison with the intrinsic time scales yields two cases:
\begin{equation}
T= \cases { T_{ang}=R_E/c\ga_E^2 & if $\Delta < R_E/\ga_E^2$
{\rm \ \ (Type-I )};
\cr
\Delta/c & otherwise {\rm \ \ \ \ \ \ \ (Type-II)}. \cr }
\end{equation}

In Type-I models, the duration of the burst is determined by the
emission radius and the Lorentz factor.
It is independent of $\Delta$.  This type of
models include the standard ``external shock model''
\cite{MR1,Katz94,SaP95} in which the relativistic shell is decelerated
by the ISM, the relativistic magnetic wind model \cite{Usov95} in
which a magnetic Poynting flux runs into the ISM, or the scattering of
star light by a relativistic shell \cite{Shemi93,Shaviv-Dar95}.

In Type-II models, the duration of the burst is determined by the
thickness of the relativistic shell, $\Delta$ (that is by the duration
that the source is active and produces the relativistic wind).  The
angular spreading time (which depends on the the radius of emission) is
shorter and therefore irrelevant. 
These models include the ``internal shock model''
\cite{NPP,MR4,PacXu}, in which different parts of the shell are moving with
different Lorentz factor and therefore collide with one another.  A
magnetic dominated version is given by Thompson \cite{Thom}.

The majority of GRBs have a complex temporal structure (e.g. section
\ref{sub-sec:variability}) with ${\cal N}\equiv T/\delta T$ of order
100.  Consider a Type-I model. Angular spreading means that at any
given moment the observer sees a whole region of angular width
$\ga_E^{-1}$.  Any variability in the emission due to different
conditions in different radii on a time scale smaller than
$T_{ang}$ is erased unless the 
angular size of the emitting region is 
smaller than $\ga_E^{-1}$.  Thus, such a source can
produce only  a smooth single humped burst with ${\cal N}=1$ and no
temporal structure on a time-scale $\delta T < T$. Put in other words
a shell, of a Type-I model, and with an angular width larger than
$\ga_E^{-1}$ cannot produce a variable burst with ${\cal N}
\gg 1$. This is the angular spreading problem.

On the other hand a Type-II model contains a thick shell $\Delta >
R_E/\ga_E^2$ and it can produce a variable burst.  The variability
time scale, is again limited $\delta T>T_{ang}$ but now it can
be shorter than the total duration $T$. The duration of the burst
reflects the time that the ``inner engine'' operates. The variability
reflects the radial inhomogeneity of the shell which was
produced by the source (or the cooling time if it is longer than
$\delta/c$). The observed temporal variability provides an upper limit
to the scale of the radial inhomogeneities in the shell and to the scale
in which the ``inner engine'' varies.  This is a remarkable conclusion
in view of the fact that the fireball hides the ``inner engine''.

\begin{figure}
\begin{center}
\includegraphics[width=10cm]{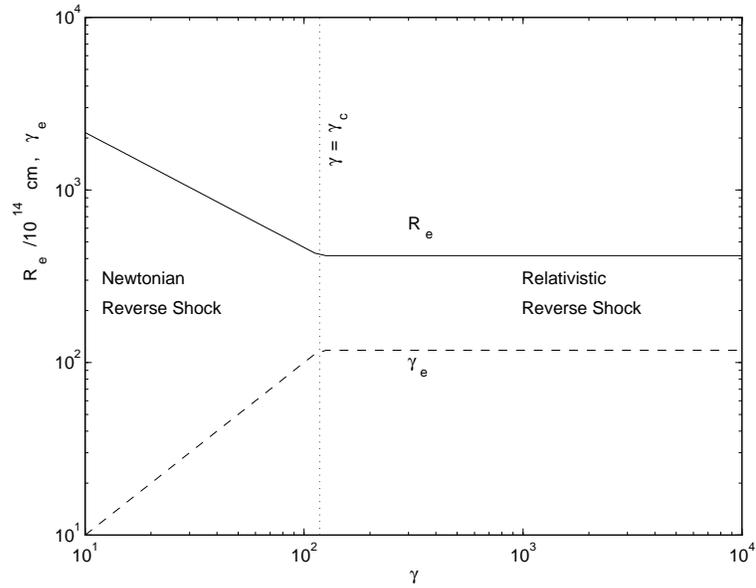}
\caption{\it The deceleration radius $R_e$ and the
Lorentz factor of the shocked shell $\gamma_e$ as functions of the
initial Lorentz factor $\gamma$, for a shell of fixed width $\Delta=3
\times 10^{12}cm$.  For low values of $\gamma$, the shocked material
moves with Lorentz factor $\gamma_e\sim\gamma$. However as $\gamma$
increases the reverse shock becomes relativistic reducing
significantly the Lorentz factor $\gamma_e<\gamma$.  This phenomena
prevents the ``external shock model'' from  being Type-II.}
\label{SaP97a3}
\end{center}
\end{figure}

Can an external shock give rise to a Type-II behavior? This would have
been possible if we could set the parameters of the external shock
model to satisfy $R_E\le 2\ga_E c\delta T$. As discussed in
\ref{sub_sec:Newtonian} the deceleration radius for a thin shell with
an initial Lorentz factor $\gamma$ is given by
\begin{equation}
\label{R_E}
R_E=l \gamma^{-2/3} ,
\end{equation}
and the observed duration is $l\gamma^{-8/3}/c $.  The deceleration is
gradual and the Lorentz factor of the emitting region $\ga_E$ is
similar to the original Lorentz factor of the shell $\gamma$. It seems
that with an arbitrary large Lorentz factor $\gamma$ we can get a
small enough deceleration radius $R_E$. However, Eq. \ref{R_E} is valid
only for thin shells satisfying $\Delta>l\gamma^{-8/3}$ \cite{SaP95}.
As $\gamma$ increases above a critical value
$\gamma\ge\gamma_c=(l/\Delta)^{3/8}$ the shell can no longer be considered
thin. In this situation the reverse shock penetrating the shell becomes
ultra-relativistic and the shocked matter moves with Lorentz factor
$\ga_E=\gamma_c<\gamma$ which is independent
of the initial Lorentz factor of the shell $\gamma$. The deceleration
radius is now given by $R_E=\Delta^{1/4}l^{3/4}$, and it is
independent of the initial Lorentz factor of the shell.  The behavior
of the deceleration radius $R_E$ and observed duration as function of
the shell Lorentz factor $\gamma$ is given in Fig. \ref{SaP97a3} for
a shell of thickness $\Delta=3\times 10^{12}$cm.  This emission radius
$R_E $ is always larger than $\Delta /\ga_E^2$ - thus an external
shock cannot be of type II.

\subsection{Angular Variability and Other Caveats }

In a Type-I model, that is a for a shell satisfying  $\Delta < R_E/\ga_E^2$, 
variability is
possible only if  the
emitting regions are significantly narrower than $\ga_E^{-1}$.  The
source would emit for a total duration $T_{radial}$.  To estimate 
the allowed  opening angle of the emitting
region imagine two points  that emit radiation at the
same (observer) time $t$.  The difference in the arrival time between
two photons emitted at $(R_E,\theta_1)$ and $(R_E,\theta_2)$ at the
same (observer) time $t$ is:
\begin{equation}
\label{DeltaT}\delta T \approx \frac{R_E |\theta_2^2-\theta _1^2 |}{2c} =
\frac{ R_E \bar \theta \left| \theta _2-\theta _1\right|}
{c} = \frac{ R_E \bar \theta \delta \theta} {c} ~,
\end{equation}
where $\theta$ is the angle from the line of sight and 
we have used $\theta _1,\theta _2\ll 1$, $\bar \theta \equiv
(\theta _1+\theta _2)/2$ and $\delta \theta \equiv |\theta_2 -
\theta_1|$. Since an observer sees emitting regions up to an angle
$\ga_E^{-1}$ away from the line of sight $\bar \theta \sim \ga_E
^{-1}$.  The size of the emitting region $r_{s}=R_E \delta \theta  $ is limited by:
\begin{equation}
\label{robj}
r_{s}\le \ga_E c\delta T~.
\end{equation}
The corresponding angular size is:
\begin{equation}
\label{ang_em} \delta \theta \le {\ga_E c\delta T~ \over R_E} =
{1 \over {\cal N} \ga_E} ~.
\end{equation}
Note that Fenimore, Madras and Nayakshin \cite{FenMadNaya} who
examined this issue, considered only emitting regions that are directly
on the line of sight with $\bar \theta \sim \left| \theta _2-\theta
_1\right| $ and obtained a larger $r_s$ which was proportional to
$R_E^{1/2}$. However only a small fraction of the emitting regions
will be exactly on the line of sight.  Most of the emitting regions
will have $\bar \theta \sim
\ga_E ^{-1}$, and thus Eq. \ref{robj} yields the relevant estimate.

\begin{figure}
\begin{center}
\includegraphics[width=7cm]{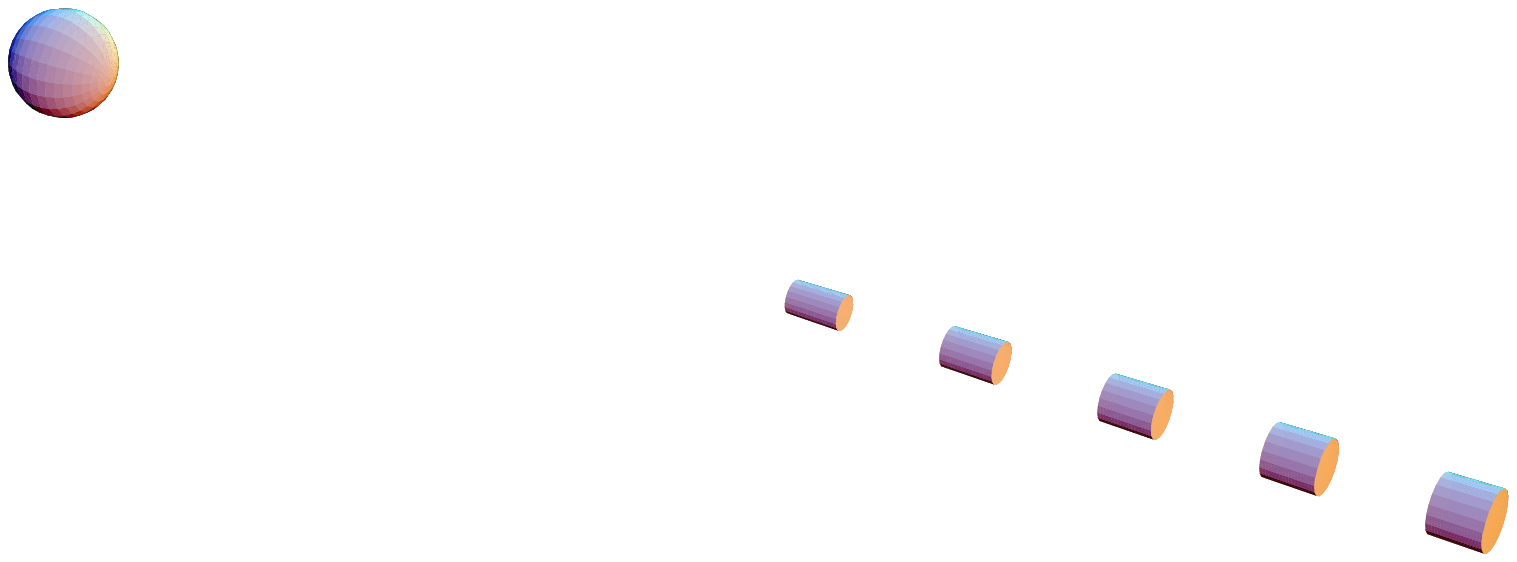}
\includegraphics[width=7cm]{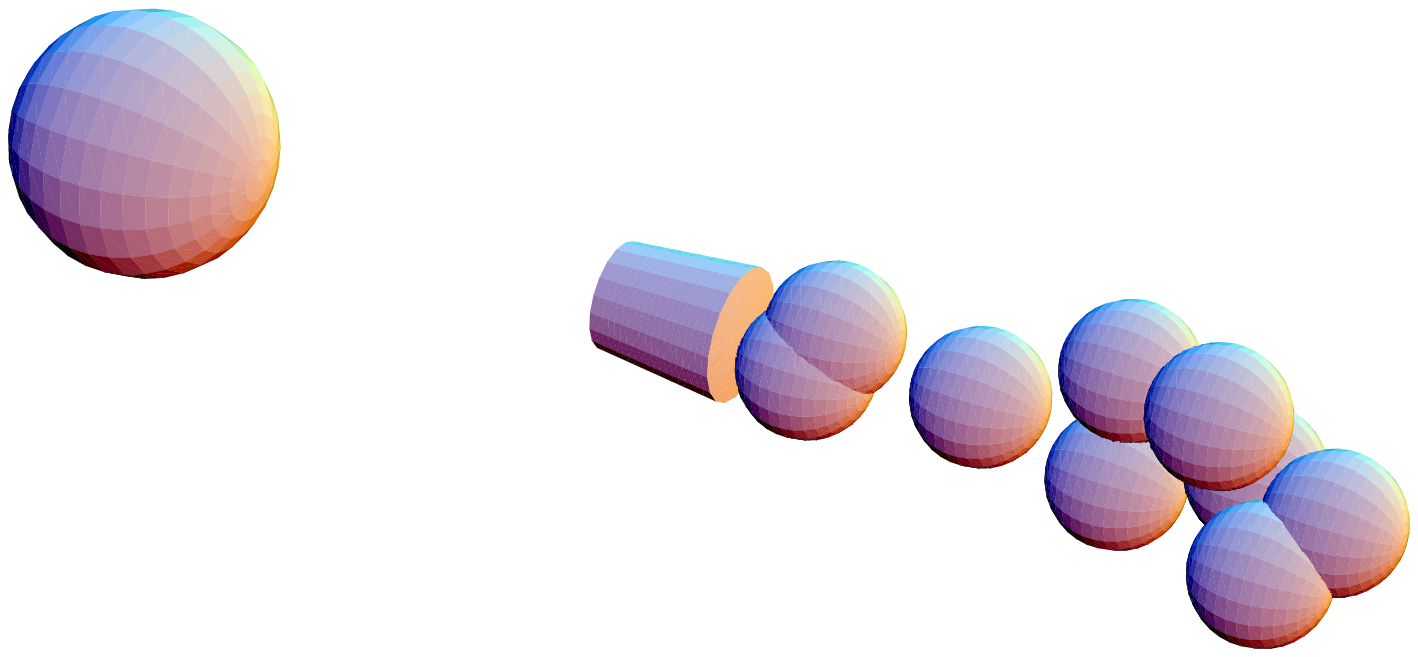}
\caption{\it A very narrow jet of angular size considerably smaller than
$\gamma_e^{-1}$, for which the angular spreading problem does not
exist.  The duration of the burst is determined by the deceleration
distance $\Delta R_e$, while the angular time is assumed small. The
variability could now be explained by either variability in the source
which leads to a pulsed jet (a) or by a uniform jet interacting with
an irregular ISM (b)}
\label{SaP97a1}
\end{center}
\end{figure}

The above discussion suggests that one can produce GRBs with $T
\approx T_{radial} \approx  R_E/c \ga_E^2$ and $\delta T = T/{\cal N}$
if the emitting regions have angular size smaller than $1/{\cal
N}\ga_E \approx 10^{-4}$.  That is, one needs an extremely narrow jet.
Relativistic jets are observed in AGNs and even in some galactic
objects, however, their opening angles are of order of a few  degrees
almost two  orders of magnitude larger.  A narrow jet with such a
small opening angle would be able to produce the observed
variability. Such a jet must be extremely cold (in
its local rest frame); otherwise its internal pressure will cause it to
spread.  It is not clear what could produce such a jet.  Additionally,
for the temporal variability to be produced, either a rapid modulation
of the jet or inhomogeneities in the ISM are needed.  These two
options are presented in Fig. \ref{SaP97a1}.

\begin{figure}
\begin{center}
\includegraphics[width=10cm]{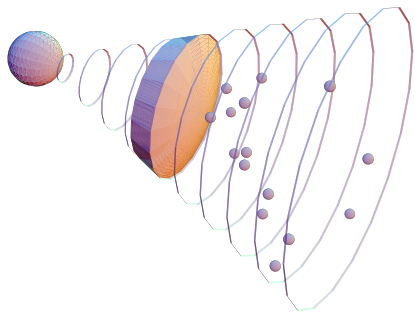}
\caption{\it A shell with angular size
 $\gamma^{-1}$ (the angular size is highly exaggerated). The spherical
 symmetry is broken by the presence of bubbles in the ISM.  The
 relative angular size of the shell and the bubbles is drawn to scale
 assuming that a burst with $N=15$ is to be produced. Consequently
 $N=15$ bubbles are drawn (more bubbles will add up to a smooth
 profile).  The fraction of the shell that will impact these bubbles
 is small leading to high inefficiency. As $N$ increases the
 efficiency problem becomes more severe $\sim N^{-1}$, from
 \cite{SaP97a}}
\label{SaP97a2}
\end{center}
\end{figure}

A  second possibility is that the shell is relatively ``wide'' (wider
than $\ga_E^{-1}$) but the emitting regions are small.  An example
of this situation is schematically described in Fig. \ref{SaP97a2}.
This may occur if either the ISM  or the shell itself are very
irregular.  This situation is, however,  extremely inefficient. The area of the
observed part of the shell is $\pi R_E^2/\ga_E ^2$.  The emitting
regions are much smaller and to comply with the temporal constraint
their area is $\pi r_s^2$.  For high efficiency all the area of the
shell must eventually radiate. The number of emitting regions needed
to cover the shell is at least $(R_E/\ga_E r_s)^2$.  In Type-I
models, the relation $R_E=2c\ga_E ^2T$ holds, and the number of
emitting region required is $4{\cal N}^2$.  But a sum of $4{\cal N}^2$
peaks each of width $1/{\cal N}$ of the total duration does not
produce a complex time structure.  Instead it produces a smooth time
profile with small variations, of order $1/\sqrt{2{\cal N}} \ll 1$, in the
amplitude.

In a highly variable burst there cannot be more than ${\cal N}$
sub-bursts of duration $\delta T=T/{\cal N}$. The corresponding area
covering factor (the fraction of radiating area of the shell) and the
corresponding efficiency is less than $1/4{\cal N}$.  This result is
independent of the nature of the emitting regions: ISM clouds, star
light or fragments of the shell.  This is the case, for example, in
the Shaviv \&  Dar model \cite{Shaviv-Dar95} where a relativistic iron
shell interacts with the starlight of a stellar cluster (a
spherical shell interacting with an external fragmented medium). This
low efficiency poses a series energy crisis for most (if not all)
cosmological models of this kind. In a recent paper Fenimore \etall
\cite{Fenimore98} consider other ways, which are based on low surface covering 
factor, to resolve the angular spreading problems. None seems very promising.

\subsection{Temporal Structure in Internal shocks.}
\label{sub_sec:Internal_time}

\begin{figure}
\begin{center}
\includegraphics[width=10cm]{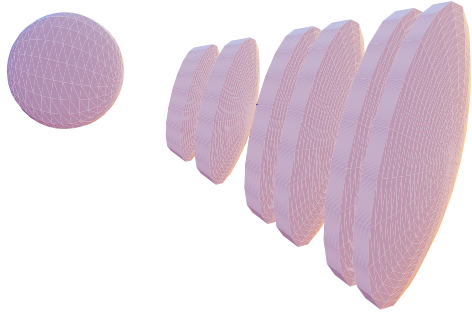}
\caption{\it The internal shock scenario. The source produces multiple shells
as shown in this figures. The shells will have different Lorentz
factors. Faster shells will catch up with slower ones and will
collide, converting some of their kinetic energy to internal energy.
This model is Type-II and naturally produces variable bursts. from
\cite{SaP97a}}
\label{SaP97a4}
\end{center}
\end{figure}

Type-II behavior arises naturally in the internal shock model. 
In this model different shells have different
Lorentz factors. These shells collide with each other and convert some
of their kinetic energy into internal energy and radiate (Fig.
\ref{SaP97a4}).  If the emission radius is sufficiently small angular
spreading will not erase the temporal variability. This requires $R_E
\le 2\ga_E^2 c \delta T$.  This condition is always satisfied as
internal shocks take place at $R_{E}= R_{s} \approx \delta
\gamma^{2}$ with $\gamma_{E}\simeq \gamma$.  Since $\delta < \Delta$
we have $T=\Delta/c> T_{ang}=R_E/2\ga_E ^2c =\delta/c=\delta T$.
Clearly multiple shells are needed to account for the observed temporal
structure. Each shell produces an observed peak of duration $\delta T$
and the whole complex of shells (whose width is $\Delta$) produces a
burst that lasts $T=\Delta/c$. The angular spreading time is
comparable to the temporal variability produced by the ``inner
engine''. They determine the observed temporal structure provided
that they are longer than the  cooling time.

Before concluding that internal shocks can actually produce GRBs we
must address two issues. First, can internal shocks produce the
observed variable structure?  Second, can it be done efficiently? We
will address the first question here and the second one in section
\ref{sub_sec:Internal_shocks} where we discuss energy conversion in
internal shocks.

Mochkovitch, Maitia and Marques \cite{MMM95} and Kobayashi, Piran \&
Sari \cite{KPS97} used a simple model to calculate the temporal
profiles produce by an internal shock. According to this model the
relativistic flow that produces the shocks is characterized by
multiple shells, each of width $l$ and with a separation $L$. The
shells are assigned random Lorentz factors (varying in the range
$[\gamma_{min},\gamma_{max}]$) and random density or energy. One can
follow the motion of the shell and calculate the time of the binary
(two-shell) collisions that take place, until all the shells that
could collide have collided and the remaining flow has a monotonically
decreasing velocity. The energy generated and the emitted radiation in
each binary encounter are then combined to a synthetic sample of a
temporal profile.

The emitted radiation from each binary collision will be observed as a
single pulse characterized by an amplitude and a width.  The amplitude
depends on the amount of energy converted to radiation in a given
collision (see \ref{sub_sec:Internal_shocks}). The time scale depends
on the cooling time, the hydrodynamic time, and the angular spreading
time.  The internal energy is radiated via synchrotron emission and
inverse Compton scattering. In most of cases, the electrons' cooling
time is much shorter than the hydrodynamic time scale
\cite{SNP96,SaP97b,Katz94} so we consider only the latter two.

The hydrodynamic time scale is determined by the time that the shock
crosses the shell, whose width is $d$. In fact there are two shocks: a
forward shock and a reverse shock.  Detailed calculations \cite{KPS97}
reveals that  this time scale
(in the observer's rest frame) is of order of the light crossing time
of the shell, that is: $d/c$.

\begin{figure}
\begin{center}
\includegraphics[width=10cm]{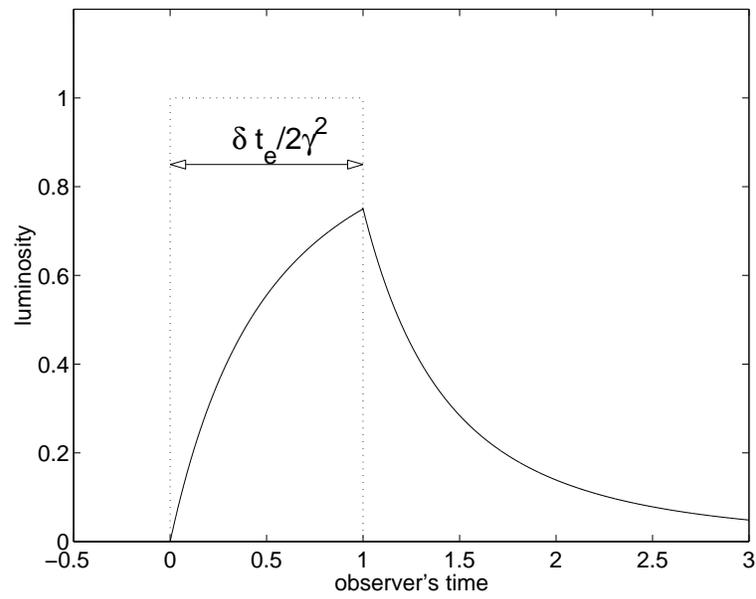}
\caption{\it A peak produced by a collision between two shells. The
luminosity plotted versus the arrival time.  The solid line
corresponds to $R=c\delta t_e$ and the dotted line corresponds to
$R=0$,  from \cite{KPS97}}
\label{KPSf1}
\end{center}
\end{figure}

Let the distance between the shells be $\delta$. A collision takes
place at $\delta /\gamma^2$ and the angular spreading yields an
observed pulse whose width is $\sim \delta /c$.  If  $\delta >d$
the overall pulse width $\delta T$ is determined by angular spreading.
The shape of the pulse become asymmetric with a fast rise and a slower
decline (Fig. \ref{KPSf1}) which GRBs typically show (see section
\ref{sub-sec:variability}).  The amplitude of an individual pulse
depends on the energy produced in the collision, which we calculate
latter in section \ref{sub_sec:Internal_shocks}.

\begin{figure}
\begin{center}
\includegraphics[width=10cm]{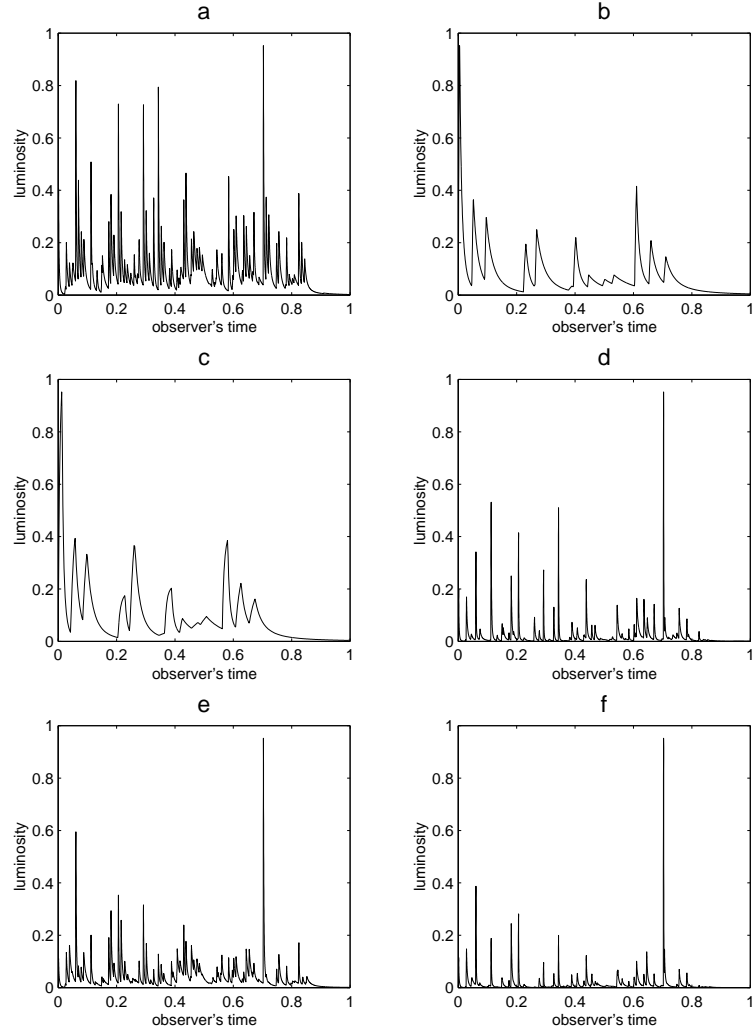}
\caption{\it Luminosity vs. observer time, for different synthetic models
a: $\gamma_{min}=100$,
$\gamma_{max}=1000$, $N=100$, $\eta=-1$ and $L/l=5$
b: $\gamma_{min}=100$, $\gamma_{max}=1000$, $N=100$, $\eta=1$ and $L/l=5$
c: $\gamma_{min}=100$, $\gamma_{max}=1000$, $N=20$ , $\eta=-1$ and
$L/l=5$
d: $\gamma_{min}=100$, $\gamma_{max}=1000$, $N=20$ ,
$\eta=-1$ and $L/l=1$
e: $\gamma_{min}=100$, $\gamma_{max}=1000$,
$N=100$, random energy with  $E_{max}=1000$ and $L/l=5$
 f: $\gamma_{min}=100$, $\gamma_{max}=1000$, $N=100$,
random density with $\rho_{max}=1000$ and $L/l=5$.  From \cite{KPS97}.}
\label{KPSf2}
\end{center}
\end{figure}

Typical synthetic temporal profile are  shown in Fig. \ref{KPSf2}.
Clearly, internal shocks can produce the observed highly variable
temporal profiles, provided that the source of the relativistic
particles, the ``inner engine'', produces a sufficiently variable
wind. Somewhat surprisingly the resulting temporal profile follows, to
a large extent, the shape of the pulse emitted by the source.  Long
bursts require long relativistic winds that last hundred seconds, with
a rapid variability on a time scale of a fraction of a second.  Thus,
unlike previous worries \cite{Piran96,Meszaros95} we find that there
is some direct information on the ``inner engine'' of GRBs: It
produces the observed complicated temporal structure. This severely
constrains numerous models.

\section{Energy Conversion}
\label{sec:conv}

\subsection{Slowing Down of Relativistic Particles}
\label{ss:slowing}

The cross section for a direct electromagnetic or nuclear interaction
between the relativistic particles and the ISM particles is too small to
convert efficiently the kinetic energy to radiation. The fireball
particles can be  slowed down only via some collective interaction such
as a collisionless shock.   Supernova remnants (SNRs) in
which the supernova ejecta is slowed down by the ISM show that
collisionless shocks  form in somewhat similar circumstances.
One can
expect that collisionless shocks  will form here as well \cite{MR1,Katz94}.

GRBs are the relativistic analogues of SNRs. In
both cases the phenomenon results from the conversion of the kinetic
energy of the ejected matter to radiation.  Even the total energy
involved is comparable.  One crucial difference is the amount of
ejected mass.  SNRs involve  a solar mass or more. The corresponding
velocity is several thousands kilometers per second, much less than
the speed of light.  In GRBs, the masses are smaller by several orders
of magnitude and with the same energy the matter attains
ultra-relativistic velocities.  A second crucial difference is that
while SNRs result from the interaction of the ejecta with the ISM,
GRBs result from internal collisions. The interaction of the ejecta
in GRBs with the ISM produces the ``afterglow'' that follows some
GRBs.

The interaction between the SNR ejecta and the ISM takes place on
scales of several pc and it is observed for thousands of years. The
internal interaction of the relativistic matter in GRBs takes place on
a scale of several hundred astronomical units and special relativistic
effects reduce the observed time scale to a fraction of a second.  The
interaction with the ISM that leads to the ``afterglow'' takes place
on a scale of a tenth of a pc. Once more special relativistic
effects reduce the observed time scale to several days.

In the following sections I discuss the slowing down of matter due to
relativistic shocks.  The discussion begins with a general review of
relativistic inelastic collisions and continues with the relativistic
shock conditions. Then I turn to the radiation processes: synchrotron
emission and Inverse Compton. After the general discussion
I apply the general results to internal shocks,  to external shocks
and to the afterglow.

\subsubsection{Relativistic Inelastic Collisions}
\label{sec:inelastic}

Consider a mass (denoted by the subscript $r$) that catches up a
slower one (denoted $s$). The two masses collide and merge to form a
single mass (denoted $m$).  Energy and momentum conservation yield:
\begin{eqnarray}
m_r\gamma_r + m_s \gamma_s = (m_r+m_s+{\cal E}/c^2) \gamma_m 
\\  
m_r \sqrt{\gamma_r^2-1} + m_s \sqrt{\gamma_s^2 -1} =(m_r+m_s+
{\cal E}/c^2) \sqrt{\gamma_m^2 -1}  \nonumber,
\label{ene_mom}
\end{eqnarray}
where ${\cal E}$ is the internal energy generated in the collision (in
the rest frame of the merged mass).

There are two interesting limits. First let $m_{s}$ be at rest: $\ga_s
= 1$. This is the case in external shocks, or in a shock between
relativistic ejecta and a non-relativistic material
that was ejected from the source before it exploded. Eqs.
\ref{ene_mom} reveal that a mass:
\begin{equation}
m_s \approx m_r/\gamma_r \ll m_r
\label{mgamma}
\end{equation}
is needed to yield $\gamma_m \approx \gamma_r /2$ and ${\cal E}
\approx m_r /2$.  The external mass needed to convert half of the
kinetic energy is smaller than the original mass by a factor of
$\gamma_r$ \cite{MR1,Katz94}.

The second case corresponds to an internal collision between shells
that are moving at different relativistic 
velocities: $\gamma_r \sgreat \gamma_s
\gg 1$.   Eqs. \ref{ene_mom} yield:
\begin{equation}
\ga_{m}\simeq \sqrt{\frac{m_{r}\ga_{r}+m_{s}\ga_{s}}{m_{r}/\ga_{r}+m_{s}/
\ga_{s}}},
\label{gammam}
\end{equation}
The internal energy (in the frame of an external observer) of the
merged shell, $E_{int} =\gamma_m{\cal E}$, is the difference of
the kinetic energies before and after the collision:
\begin{equation}
E_{int}=m_{r}c^{2}(\ga_{r}-\ga_{m})+m_{s}c^{2}(\ga_{s}-\ga_{m}).
\end{equation}
The conversion efficiency of kinetic energy into internal
energy  is \cite{KPS97,Katz97}:
\begin{equation}
\epsilon =1-{(m_{r}+m_{s})\gamma _{m} \over (m_{r}\gamma _{r}+m_{s}
\gamma _{s})} .
\label{two-shell-efficiency}
\end{equation}
As can be expected a conversion of a significant fraction of the
initial kinetic energy to internal energy requires that the difference
in velocities between the shells will be significant: $\ga_r \gg \ga_s$
and that the two masses will be comparable $m_r \approx m_s$.

\subsubsection{Shock Conditions}
\label{sss:shock}

Quite often the previous estimates based on the
approximation that the whole shell behaves as a single object are good
enough. However, the time scale between the interaction of different
parts of the shell with the ISM may be relatively long (compared to
the time scale to collect an external mass $M/\gamma$) and in this
case one has to turn to the hydrodynamics of the interaction. This
calculation takes into account the shocks that form during the
collision.  

\begin{figure}
\begin{center}
\includegraphics[width=10cm]{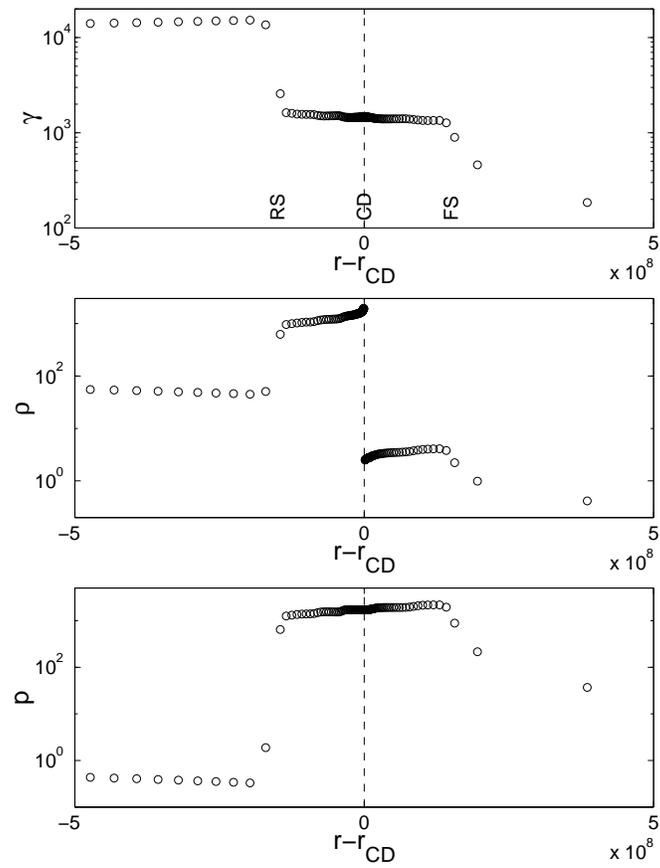}
\caption{\it The Lorentz factor $\gamma$, the
density $\rho$ and the pressure $p$ in the shocks.  There are four
regions: the ISM (region 1), the shocked ISM (region 2), the shocked
shell (region 3) and the unshocked shell (region 4), which are
separated by the forward shock (FS), the contact discontinuity (CD)
and the reverse shock (RS). The initial parameters are the same as in
\ref{fig:full_RRS}. From \cite{KPS98}.}
\label{shock_profile}
\end{center}
\end{figure}

Consider the situation a cold shell (whose internal energy is
negligible compared to the rest mass) that  overtakes another
cold shell or moves into the cold ISM. Generally, Two shocks form: an outgoing
shock that propagates into the ISM or into the external shell, and a
reverse shock that propagates into the inner shell, with a contact
discontinuity between the shocked material (see
Fig. \ref{shock_profile}).  Two quantities determine the shocks'
structure: $\ga$, the Lorentz factor of the motion of the inner shell
(denoted 4) relative to the outer one - or the ISM (denoted 1) and the
ratio between the particle number densities in these regions, $f
\equiv n_4/n_1$.

There are three interesting cases:
{\bf (i)} Ultra-relativistic shock ($\ga \gg 1$) with $f > \ga^2$.  This
happens during the early phase of an external shock or during the very
late external shock evolution when there is only a single shock. We
call this configuration ``Newtonian'' because the reverse
shock is non-relativistic (or mildly relativistic). In this case the
energy conversion takes place in the forward shock:
Let $\gamma_2$ be the Lorentz factor of the motion of the shocked
fluid relative to the rest frame of the fluid at 1 (an external
observer for interaction with the ISM and the outer shell in case of
internal collision) and let $\bar \gamma_3$ be the Lorentz factor of
the motion of this fluid  relative to the rest
frame of the relativistic shell (4).
\begin{equation}
\ga_2  \approx \ga \ \ \ ; \ \ \ \bar \gamma_{3} \approx 1,
\label{nr1}
\end{equation}
The particle and energy densities $(n, e)$ in the shocked regions
satisfy:
\begin{equation}
n_2 \approx 4 \ga n_1,
\ \  ;  \ \  e \equiv e_2  = 4 \ga^2 n_1 m_p c^2
\ \ ; \ \ n_3 = 7 n_4,
\ \  ;  \ \   e_3 = e .
\label{nr3}
\end{equation}

{\bf (ii)} Later during the propagation of the shell the density ratio
decreases and $f < \ga^2$.  Both the forward and the reverse shocks
are relativistic.  The shock equations between regions 1 and 2 yield:
\cite{BLmc1,BLmc2,Pi94}:
\begin{equation}
\gamma_2  =
f^{1/4} \ga^{1/2} /\sqrt 2 \ \ ; \ \ n_2 = 4 \gamma_2 n_1 \ \ ; \ \ e
\equiv e_2 = 4 \gamma^2_2 n_1 m_p c^2 ,
\label{cond12}
\end{equation}
Similar relations hold for the reverse shock:
\begin{equation}
\bar \gamma_3 = f^{-1/4} \ga^{1/2} /\sqrt 2 \ \
;  \ \ n_3 = 4 \bar \gamma_{3} n_4.
\label{cond34}
\end{equation}
In addition we have $e_3=e$ and $\bar
\gamma_3\cong (\ga/\ga _2+\ga_2/\ga)/2$ which follow from
the equality of pressures and velocity on the contact discontinuity.
Comparable amounts of energy are converted to thermal energy in both
shocks when both shocks are relativistic.  But only a negligible
amount of energy is converted to thermal energy in the reverse shock
if it is Newtonian \cite{SaP95}.

{\bf (iii)} Internal shocks are characterized by $f \approx 1$ - both
shells have similar densities, and by a Lorentz factor of order of a
few ($2 < \ga < 10$) describing the relative motion of the shells. 
In this case, for an adiabatic index (4/3) we have:
\begin{eqnarray}
\ga_2 =\sqrt{(\ga^2+1)/2} \approx max [ 1, \sqrt{\ga/2}] \ \ ; \\ \nonumber 
n_2 = (4 \ga_2 +3 ) n_1 \approx 4 \ga_2 n_1
\ \ ; \ \ e_2  =  \ga_2 n_2 m_p c^2
\label{internal_conditions}
\end{eqnarray}
Both shocks are mildly relativistic and their strength depends on the 
relative Lorentz factors of the two shells.

\subsubsection{Lorentz Factors in Different Emitting Regions}

Before concluding this section and turning to the radiation mechanisms
we summarize briefly the different relativistic motions
encountered when considering different emitting regions.  The
relativistic shocks are characterized by $\ga_{sh}$ that describes
the shock's velocity as well as the ``thermal'' motion of the shocked
particles. It is measured relative to a rest frame in which the
unshocked material is at rest. The Lorentz factor of the forward shock
is usually different from the Lorentz factor of the reverse shock.
The emitting region - the shocked material - moves relativistically
relative to an observer at rest at infinity. This is
characterized by a Lorentz factor, $\ga_E$.  Table
\ref{Lorentz_factors} summarizes the different values of $\ga_{sh}$
and $\ga_E$ for external and internal shocks and for the afterglow.

\begin{center}
\begin{table*}[ht!]
\label{Lorentz_factors}
\begin{center}
\begin {tabular}{|c|c|c|c|c|}
\hline
\multicolumn{3}{|c|} {Shock type}        & $\ga_E$  & $\ga_{sh}$ \\ \hline\hline

& & Forward    & $\ga$  &  $\ga$    \\ \cline{3-5}
& \raisebox{1.5ex}[0pt]{Newtonian}  & Reverse    & $\ga$  &   1       \\ \cline{2-5}
\raisebox{1.5ex}{External}& & Forward & $\ga \xi^{3/4}$ & $\ga \xi^{3/4}$  \\ \cline{3-5}
&\raisebox{1.5ex}[0pt]{Relativistic}& Reverse  &  $\ga \xi^{3/4}$& $\xi^{-3/4}$ \\
\hline\hline
\multicolumn{3}{|c|} {Internal}        & $\ga$  &
$\sqrt{\ga_{int}/2}\sim$ a few \\ \hline\hline
\multicolumn{3}{|c|} {Afterglow}       &  $\ga(t)$ & $\ga(t)$  \\ \hline\hline
\end{tabular}
\end{center}
\caption{\it Lorentz Factors}
\end{table*}
\end{center}

\subsection{Synchrotron Emission from Relativistic Shocks}
\label{sub_sec:synch}
\subsubsection{General Considerations}
\label{sub_sec:syn_general}
The most likely radiation process in GRBs is  synchrotron emission
\cite{MLR,Katz94,Pi94,SNP96}. The observed low energy spectra
provide an indication that this is indeed the case
\cite{Cohen_etal_97,Schaefer97a}. 
  
The parameters that determine synchrotron emission are the magnetic
field strength, $B$, and the electrons' energy distribution
(characterized by the minimal Lorentz factor $\ga_{e,min}$ and the
index of the expected power-law electron energy distribution $p$).
These parameters should be determined from the microscopic physical
processes that take place in the shocks. However, it is difficult to
estimate them from first principles.  Instead we define two
dimensionless parameters, $\epsilon_B$ and $\epsilon_e$, that
incorporate our ignorance and uncertainties \cite{Pi94,SNP96}.

The dimensionless parameter $\epsilon_B$ measures the ratio of the
magnetic field energy density to the total thermal energy $e$:
\begin{equation}
\label{EpsilonB} \epsilon_B\equiv {U_B\over e} = {B^2\over8\pi e},
\end{equation}
so that, after substituting the shock conditions we have:
\begin{equation}
B= \sqrt{32\pi} c \epsilon_B^{1/2} \gamma_{sh} m_p^{1/2} n_{1}^{1/2}.
\label{Bvalue}
\end{equation}
There have been different attempts to estimate $\epsilon_B$
\cite{Thom,MLR,Usov95}. We keep it as a free parameter. Additionally we
assume that the magnetic field is  randomly oriented in space.

The second parameter, $\epsilon_e$, measures the fraction of the total
thermal energy $e$ which goes into random motions of the electrons:
\begin{equation}
\epsilon _e\equiv {U_e\over e}.
\end{equation}

\subsubsection{The Electron Distribution}

We call consider a  ``typical'' electron as  one that has  the 
average $\gamma_e$ of the electrons distribution: 
\begin{equation}
\langle\gamma_e\rangle={m_{\rho}\over m_{e}}
\epsilon_{e}\gamma_{sh}\;.
\label{typical}
\end{equation}

Collisionless acceleration of electrons can be efficient if they
are tightly coupled to the protons and the magnetic field by mean
of plasma waves \cite{Kirk94}.
Since the electrons receive their random motions through
shock-heating, we assume (following the treatment of non-relativistic
shocks) that they develop a power law distribution of Lorentz factors:
\begin{equation}
N(\gamma_e )\sim \gamma_e^{-p}\ \ \ {\rm for}\ \gamma_e
>\gamma _{e,min}\;.
\label{e_distribution}
\end{equation}
We require $p>2$ so that the energy does not diverge at large
$\gamma_e$.  Since the shocks are relativistic we assume that all the
electrons participate in the power-law, not just a small fraction in
the tail of the distribution as in the Newtonian case.  An indication
that this assumption is correct is given by the lower energy spectrum
observed in some GRBs \cite{Cohen_etal_97,Schaefer97a}.  The minimum
Lorentz factor, $\gamma _{e,min}$, of the distribution is related to
$\epsilon _e$ and to the total energy $e\sim\gamma_{sh}nm_pc^2$:
\begin{equation}
\gamma_{e,min}= {m_p \over m_e} {p-2\over p-1}\epsilon_e\gamma_{sh}=
 {p-2\over p-1} 
\langle\gamma_e\rangle.
\label{gemin}
\end{equation}
where $\gamma_{sh}$ is the relative Lorentz factor across the 
corresponding shock.

The energy index $p$ can be fixed by requiring that the model should
be able to explain the high energy spectra of GRBs.  If we assume that
most of the radiation observed in the soft gamma-rays is due to
synchrotron cooling, then it is straightforward to relate $p$ to the
power-law index of the observed spectra of GRBs, $\beta$. The mean
spectral index of GRBs at high photon energies $\beta\approx-2.25$,
\cite{Band93} corresponds to $p\approx 2.5$.  
This agrees, as we see later (\ref{sub_sec:GRB970508}) with the value
inferred from afterglow observations ($p \sim 2.25$).  We assume this
value of $p$ in what follows. The corresponding ratio that appears in
Eq. \ref{gemin} $(p-2)/(p-1)$ equals $1/3$ and we have $\gamma_{e,min}
= 610 \ga_{sh}$.

The shock acceleration mechanisms cannot accelerate the electrons to 
arbitrary
high energy. For the  maximal  electron's energy,  with a corresponding 
$\ga_{e,max}$,  the acceleration time equals to the  cooling time.
The acceleration time is determined by the Larmor radius $R_L$ and the
Alfv\'en velocity $v_A$ \cite{Hillas84}: 
\begin{equation}
t_{acc} = {c R_L \over v^2_A} \ .
\label{tacc}
\end{equation}
This time scale should be compared with the synchrotron cooling time
$\ga_e m_e c^2 /P_{syn}$ (in the local frame).  Using $v_A^2 \approx
\epsilon_B^2$, Eq. \ref{Bvalue} to estimate $B$ and
Eq. \ref{syn_power} below to estimate $P_{syn}$ one finds:
\begin{equation}
\ga_{e,max}= \sqrt{{ 24 \pi \epsilon_B^2  e \over B \sigma_T \ga_{sh} }}
= 3.7 \times 10^8 {\epsilon_B^{3/2} \over \ga_{sh} n_1^{1/2}} \  .
\end{equation}{cooling}  
This value is quite large and generally  it does not effect the observed
spectrum in the soft gamma ray range. 

\subsubsection{Synchrotron Frequency and Synchrotron Power}

The typical energy of synchrotron photons as well as the synchrotron
cooling time depend on the Lorentz factor $\gamma_e$ of the
relativistic electron under consideration and on the strength of the
magnetic field  (see e.g.  \cite{Rybicki_Lightman79}). 
Since the emitting material  moves with a Lorentz factor
$\gamma_E$ the photons are blue shifted.
The characteristic photon energy in the observer  frame is
given by:
\begin{equation}
\label{syn_obs}
(h\nu_{syn})_{obs}=\frac{\hbar q_eB}{m_ec}\gamma _e^2\gamma_E .
\end{equation}

The power emitted by a single electron due to synchrotron radiation
in the local frame is:
\begin{equation}
P_{syn} = \frac 4 3 \sigma_T c U_B \gamma_e^2 \ ,
\label{syn_power}
\end{equation}
where $\sigma_T$ is the Thomson cross section.  The cooling time of
the electron in the fluid frame is then $\gamma_e m_e c^2/P$. The
observed cooling time $t_{syn}$ is shorter by a factor of $\gamma_E$:
\begin{equation}
\label{cooling}
t_{syn}(\gamma_e) =\frac{3 m_e c}{4\sigma _T U_B\gamma_e\gamma_E}.
\end{equation}

Substituting the value of $ \gamma_e$ from equation
\ref{syn_obs} into the cooling rate Eq. \ref{cooling} we
obtain the cooling time scale as a function of the observed photon
energy:
\begin{equation}
\label{tausyn2}
t_{syn} (\nu) = \frac{3}{\sigma_T}  \sqrt{\frac{ 2 \pi c \hbar m_e q_e}
{B^{3} \ga_E  \nu}} 
\end{equation}

Since $\ga_e$
does not appear explicitly in this equation $t_{syn}$ at a given observed
frequency is independent of the electrons' energy distribution within
the shock. This is provided, of course, that there are electrons with
the required $\ga_e$ so that there will be emission in the frequency
considered.  As long as there is such an electron the cooling time is
``universal''.   This equation shows a characteristic scaling
of $t_{syn} (\nu) \propto \nu^{-1/2}$. This is not very different from
the observed relation $\delta T \propto \nu^{-0.4}$ \cite{Fenimore95}.
However, it is not clear if the cooling time and not another time
scale determined the temporal profile.

The cooling time calculated above sets a lower limit to the
variability time scale of a GRB since the burst cannot possibly
contain spikes that are shorter than its cooling time.  Observations
of GRBs typically show asymmetric spikes in the intensity variation,
where a peak generally has a fast rise and a slower exponential
decline (FRED).  A plausible explanation of this observation is that
the shock heating of the electrons happens rapidly (though
episodically), and that the rise time of a spike is related to the
heating time.  The decay time is then set by the cooling, so that the
widths of spikes directly measure the cooling time.

\subsubsection{The Integrated Synchrotron Spectrum}
\label{sub_sec:syn_spectra}

The instantaneous synchrotron spectrum of a single electron with an
initial energy $\ga_e m_e c^2$ is a power law with $F_\nu \propto
\nu^{1/3}$ up to $\nu_{syn}(\ga_e)$ and an exponential decay above it.
If the electron is energetic it will cool rapidly until it will reach
$\ga_{e,c}$. This is the Lorentz factor of an electron that cools on a
hydrodynamic time scale.  For a rapidly cooling electron we have to
consider the time integrated spectrum above $h_{syn}(\ga_{e,c})$:
$F_\nu \propto \nu^{-1/2}$ from $\nu_{syn}(\ga_{e,c})$ up to
$\nu_{syn}(\ga_e)$.

To calculate the overall spectrum due to all the electrons we need to
integrate over $\gamma_e$.  Our discussion here follows \cite{SPN98a}.
We consider a power-law electron distribution with a power index $p$
and a minimal Lorentz factor $\ga_{e,min}$ (see
Eq. \ref{e_distribution}). Overall we expect a broken power law
spectrum with a break frequency around the synchrotron frequency of
the lowest energy electrons $\nu_{syn}(\ga_{e,min})$.  These power
law indices depend on the cooling rate. The most energetic electrons will
always be cooling rapidly (independently of the behavior of the ``typical
electron''). Thus the highest spectrum is always satisfy: 
\begin{equation}
F_\nu = N[(\gamma(\nu)]
m_e c^2 (\gamma(\nu) d\gamma /d\nu \propto \nu^{-p/2}.
\label{fastfnu}
\end{equation}
similarly the low energy electrons will always be slow cooling and
thus the lowest part of the spectrum will behave like $F_\nu \propto
\nu^{1/3}$.

For slow cooling we have the instantaneous spectrum: $F_\nu \propto
\nu^{1/3}$ for the lower part of the spectrum. For the upper part we
have
\begin{equation}
F_\nu = N[(\gamma(\nu)]
P[( \gamma(\nu)] d\gamma /d\nu \propto \nu^{-(p-1)/2} ,
\end{equation}
where $\gamma(\nu)$ is the Lorentz factor for which the synchrotron
frequency equals $\nu$. The most energetic electrons will cool rapidly
even when the overall system is in slow cooling.  These electrons emit
practically all their energy $m_e c^2 \gamma$, at their synchrotron
frequency.  Thus the uppermost part of the spectrum will satisfy:

For fast cooling we have $F_\nu \propto \nu^{-1/2}$ for the lower part
and $F_\nu \propto \nu^{-p/2}$ for the upper part.  Here at the lower
end  the least energetic electrons will be cooling slowly even when
the typical electron is cooling rapidly.  Thus we will have $f_\nu
\propto \nu^{1/3}$ in the lowest part of the spectrum.

The critical parameter that determines if the electrons are cooling
fast or slow is  $\ga_{e,c}$, the Lorentz factor of an electron that
cools on a hydrodynamic time scale. 
To estimate $\ga_{e,c}$ we compare $t_{syn}$
(Eq. \ref{cooling}) with $t_{hyd}$, the hydrodynamic time scale (in
the observer's rest frame):
\begin{equation}
\ga_{e,c} = {{3m_{e} c }\over{4 \sigma _{T}  U_{B}\ga_{E}t_{hyd}}}
\label{ga_c}
\end{equation}
{\it Fast cooling } occurs if $\ga_{e,c} < \ga_{e,min}$.  All the
electrons cool rapidly and the electrons' distribution effectively
extends down to $\ga_{e,c}$. If $\ga_{e,c} > \ga_{e,min}$ only the
high energy tail of the distribution (those electrons above
$\ga_{e,c}$) cool and the system is in the {\it slow cooling}  regime.

For the GRB itself we must impose the condition of  fast cooling:
the relativistic shocks must emit their energy effectively - otherwise
there will be a serious inefficiency problem. Additionally we won't be able
to explain the variability if the cooling time is too long. 
The electrons must cool
rapidly and release all their energy. In this case 
$\gamma_{e,min} >\gamma_{e,c}$ \cite{SNP96} and all the  electrons  cool
down roughly to $\gamma_{e,c}$.  The observed flux, $F_\nu$, is
given by:
\begin{equation}\label{spectrumfast}
F_\nu \propto \cases{ ( \nu / \nu_c )^{1/3} F_{\nu,max}, & $\nu_c>\nu$, \cr (
\nu / \nu_c )^{-1/2} F_{\nu,max}, & $\nu_m>\nu>\nu_c$, \cr ( \nu_m /
\nu_c )^{-1/2} ( \nu / \nu_m)^{-p/2} F_{\nu,max}, & $\nu>\nu_m$, \cr }
\end{equation}
where $\nu_m \equiv \nu_{syn}(\gamma_{e,min}), \nu_{c} \equiv
\nu_{syn}(\gamma_{e,c})$ and $F_{\nu,max}$ is the observed peak flux.

It is  most likely that during the latter stages of an
external shock (that is within the afterglow phase - provided  that it
arises  due to external shocks) there will be a transition from fast to slow
cooling \cite{MR97,Waxman97a,MesReesWei97,Waxman97b,KP97}.
When $\gamma_c>\gamma_{e,min}$, only those electrons with
$\gamma_e>\gamma_c$ can cool. We call this slow cooling, because the
electrons with $\gamma_e\sim\gamma_{e,min}$, which form the bulk of
the population, do not cool. Integration over the
electron distribution gives in this case:
\begin{equation}
\label{spectrumslow}
F_\nu \propto \cases{  (\nu/\nu_m)^{1/3} F_{\nu,max},
            & $\nu_m>\nu$, \cr
(\nu/\nu_m)^{-(p-1)/2} F_{\nu,max},
            & $\nu_c>\nu>\nu_m$, \cr
\left( \nu_c/\nu_m \right)^{-(p-1)/2}
\left( \nu/\nu_c \right)^{-p/2} F_{\nu,max},
            & $\nu>\nu_c$. \cr
}
\end{equation}

For fast cooling $\nu_c < \nu_m$. We find that the peak flux is at
$\nu_c$ while the peak energy emitted (which corresponds to the peak
of $\nu F_\nu$) is at $\nu_m$.  For slow cooling
the situation reverses $ \nu_m < \nu_c $. The peak flux is at $\nu_m$
while the peak energy emitted is at $\nu_c$.

Typical spectra corresponding to fast and slow cooling are shown in
Fig. \ref{fig:syn_spec}. The light curve depends on the
hydrodynamic evolution, which in turn determines the time dependence
of $\nu_m,\nu_c$ and $F_{\nu,max}$.

For fast cooling the power emitted is simply the power given
to the electrons, that is $\epsilon_e $ times the power generated by
the shock:
\begin{equation}
P_{fast} = \epsilon_e {dE \over dt} .
\label{Pfast-cool}
\end{equation}
For slow cooling the emitted power is determined by the
ability of the electrons to radiate their energy:
\begin{equation}
P_{slow} = N_e P_{syn} (\ga_{e,min})
\label{Pslow-cool}
\end{equation}
where, $N_e$ is the number of electrons in the emitting region and
$P_{syn} (\ga_{e,min}) $, the synchrotron power of an  electron with 
$\ga_{e,min}$, is given by Eq. \ref{syn_power}.

\subsection{Synchrotron Self Absorption}
\label{sub_sec:SA}

An important effect that we have ignored so far is the possibility of
self absorption. This is irrelevant during the GRB itself. One of
the essential features of the GRB spectrum is that it is produced in an
optically thin region. However, self absorption may appear at late
time and typically in radio emission
\cite{Katz94,Waxman97b,KP97,Wijers_Galama98,jgranot98b}.
When it appears it will cause a steep cutoff of the low energy spectrum,
either as the commonly known $\nu^{5/2}$ or as $\nu^2$.

To estimate the self absorption frequency we need the 
optical depth along the line of sight. A simple approximation is:
$\alpha'_{\nu'}R_l/\gamma_l$ where 
$\alpha'_{\nu'}$ is the absorption coefficient \cite{Rybicki_Lightman79}:
\begin{equation}
\label{alpha_nu} 
\alpha'_{\nu'} = {(p+2) \over 8 \pi m_e \nu'^2}\int^{\infty}_{\gamma_{min}} 
d\gamma_e P'_{\nu',e}(\gamma_e){N(\gamma_e) \over \gamma_e} \  \  .
\end{equation}
The self absorption frequency $\nu_a$ 
satisfies: $\alpha'_{\nu'_0} R/\gamma=1$. It can be estimates
only once we have a model for the hydrodynamics and how do $R$ 
and $\gamma$ change  with time \cite{Wijers_Galama98,jgranot98b}.

The spectrum below the 
the self-absorption frequency 
depends on the electron distribution. One obtaine the well known
\cite{Rybicki_Lightman79}, 
$\nu^{5/2}$ slop  when the synchrotron frequency of the electron emitting
the self absorbed radiation   is inside
the self absorption range. One obtains a slop of 
$\nu^2$2 if there is  self absorption, but the radiation in that range is
due to the low energy tail of electrons radiating effectively 
at higher energies. For this latter case, which is more appropriate for
GRB afterglow  we find that \cite{Katz94,KP97}:
\begin{equation}
F_\nu \propto \nu^2  [k_B T_e / (\gamma m_p c^2)] {R^2},
\label{sa_spec}
\end{equation}
where $R$ is the radius of the radiating shell and the 
factor $k_B T_e / (\gamma m_p c^2)$ describes the degree of
electron equipartition in the plasma shock-heated to an internal energy per
particle $\gamma m_p c^2$ and moving with Lorentz factor $\gamma$.

\subsection{Inverse Compton  Emission}
\label{sub_sec:IC}

Inverse Compton (IC) scattering may modify our analysis in several
ways.  IC can influence the spectrum even if the system is optically
thin (as it must be) to Compton scattering (see e.g.
\cite{Rybicki_Lightman79}).  
In view of the high energies involved we assume that only one IC scattering
takes place. After this scattering the photon's energy is so high that 
in the electron's rest frame it
is above the Klein-Nishina energy and in this case the decrease in the Compton
cross
section makes this scattering unlikely.
The effect of IC depends on the Comptonization
parameter $Y=\gamma^2 \tau_e$.  For fast cooling one
can  show \cite{SNP96} that  $Y$
satisfies:
\begin{eqnarray}
Y= {\epsilon _e/\epsilon _B}~~~     & \ \ {\rm  if }\ \ & \epsilon_e
\ll \epsilon _B\\ \nonumber 
Y= \sqrt{\epsilon _e/\epsilon_B} & \ \ {\rm  if }\ \ & \epsilon_e
\gg \epsilon _B . \nonumber
\end{eqnarray}
IC is unimportant if $Y<1$ and in this case it can be ignored.

If $Y>1$, which corresponds to $\epsilon_e > \epsilon_B$ and to
$Y=\sqrt{\epsilon_e/\epsilon_B}$ then a large fraction of the low
energy synchrotron radiation will be up scattered by IC and a large
fraction of the energy will be emitted via the IC processes. If those
IC up scattered photons will be in the observed energy band then the
observed radiation will be IC and not synchrotron photons. Those IC
photons might  be too energetic, that is their energy may be
beyond the observed energy 
range. In this case  IC will not influence the observed
spectra directly. However, as IC will take a significant fraction of the
energy of the cooling electrons it will influence the observations in two ways:
it will shorten the cooling time (the emitting electrons will be
cooled by both synchrotron and IC process).  Second, assuming that
the observed $\ga$-ray photons results from synchrotron emission,
IC will influence
the overall energy budget and reduce the efficiency of the production
of the observed radiation. We turn now to each of this cases.

Consider, first, the situation in which $Y>1$ and the IC photons are
in the observed range so that some of the observed radiation may be
due to IC rather than synchrotron emission. This is an interesting
possibility since one might expect that the IC process will ease the
requirement of rather large magnetic fields that is imposed by the
synchrotron process. We show here that, somewhat surprisingly, this 
cannot be the case.

An IC scattering boosts the energy of the photon by a factor
$\gamma^2_e$.  Typical  IC photons will be observed at the energy:
\begin{equation}
\label{IC_obs}
(h\nu_{IC})_{obs}=\frac{\hbar q_eB}{m_ec}\gamma _e^4 \gamma_E = 12
{\rm Mev} { B_{1G} }
\left({\ga_{E100}}\right )
\left[ {\ga_e \over (m_p/m_e)} \right]^4.
\end{equation}
where $B_{1G}=B/1Gauss$  and $\ga_{E100} \equiv \ga_{E/100}$.
The Lorentz factor of electrons radiating synchrotron photons which
are IC scattered on electrons with the same Lorentz factor and have
energy $h \nu$ in the observed range is the square root of the $\gamma_e$
required to produce synchrotron radiation in the same frequency. The required
value for $\ga_e$ is rather low relative to what one may expect in an
external shock (in which $\ga_{e,ext} \sim \epsilon_e (m_p/m_e)
\ga_{sh}$). In internal shocks we expect  lower values ($\ga_{e,int} \sim
\epsilon_e (m_p/m_e) $) but in this case the equipartition magnetic field
is much stronger (of the order of few thousand Gauss, or higher). Thus
IC might produce the observed photons in internal shocks if
$\epsilon_B$ is rather small (of order $10^{-5}$).

These electrons are cooled both by synchrotron and by IC.  The latter
is more efficient and the cooling is enhanced by the Compton parameter
$Y$.  The cooling time scale is: 
$$ t_{IC}={\frac{6 \pi c^{3/4}
\sqrt{\epsilon_B} \hbar^{1/4} m_e^{3/4}{q_e}^{1/4}} {B^{7/4}
\sqrt{\epsilon_e} (h \nu)^{1/4} \ga_E^{3/4} \sigma_T}} = $$
\begin{equation}
\label{tIC}
8 \cdot 10^{4} {\rm sec} \sqrt{{\epsilon_B \over \epsilon_e}}
{B_{1G}}^{-7/4} 
\left( {\ga_{E100}} \right )^{-3/4}
\left( {{h \nu} / {100 {\rm keV}}} \right )^{-1/4}
\end{equation}

As we see in the following discussion for external shocks,
$t_{IC}(100keV)$, the IC cooling time if the IC radiation is in the
observed range (soft gamma-rays) is too long, while for internal
shocks $t_{IC}(100keV)$ is marginal. However, even if IC does not
produce the observed $\ga$-ray photons it still influences the process
if $Y> 1$.  It will speed up the cooling of the emitting regions and
shorten the cooling time, $t_{syn}$ estimated earlier
(Eq. \ref{tausyn2}) by a factor of $Y$.  Additionally IC also reduces
the efficiency by the same factor, and the efficiency becomes
extremely low as described below.

\subsection{Radiative Efficiency}
\label{sub_sec:radiative_efficiency}

The efficiency of a burst depends on three factors: First only the
electrons' energy is available. This yields a factor
$\epsilon_e$. Second, if $\epsilon_B < \epsilon_{e}$ there is an
additional factor of ${\rm min} [1, \sqrt{\epsilon_B/\epsilon_e}]$ if
the IC radiation is not observed.  Third, there is a specific Lorentz
factor, $\hat \gamma_e$, of an electron which emits synchrotron (or
IC) radiation in the 100keV energy band.  Therefore, only the energy
radiated by electrons with $ \gamma _{e} \ge \hat \gamma_e $ is
observed as soft $\ga$-rays. Assuming a power law electron
distribution with an index $p=2.5$ (see \cite{SNP96}) this gives a
factor of $(\gamma_{e,min}/\hat \gamma_e )^{1/2}$ (which is valid of
course provided that $\gamma_{e,min} <\hat \gamma_e$). The total
efficiency is the multiplication of those three factors and it is
given by:
\begin{equation}
\epsilon=\epsilon_e {\rm min}[1,\sqrt{\epsilon_B/\epsilon_e}]
(\gamma_{e,min}/\hat \gamma_e )^{1/2}
\label{rad_eff}
\end{equation}

The efficiency  depends first of all on the electrons' energy density
and to a lesser extend on the magnetic energy density. Both should be close
to equipartition in order that the efficiency will be large.
Additionally, in order that there will be photons in the 
100keV range 
$\ga_{e,min}$ should be smaller than  $\hat\ga$. However
efficient production of soft $\ga$-rays requires that $\ga_{e,min}$
will  not be too  small compared with
$\hat\ga$. This estimate is, of course, different if the 
observed $\gamma$-rays are produced by  inverse Compton scattering.

\subsection{Internal Shocks}
\label{sub_sec:Internal_shocks}
Internal shocks are the leading mechanism for energy conversion and
production of the observed gamma-ray radiation. We discuss, in this
section, the energy conversion process, the typical radiation frequency
and its efficiency.

\subsubsection{Parameters for Internal Shocks}
\label{sub_sec:Internal_parameters}

Internal shocks take place when an inner shell overtakes a slower
outer shell. Consider a fast inner shell with a Lorentz factor $\ga_r$
that collides with a slower shell whose Lorentz factor is $\ga_s$.  If $\ga_r
\sgreat  \ga_s \sim \ga$ then the inner shell will overtake the outer one at:
\begin{equation}
R_\delta \approx \ga^2 \delta \approx 10^{14}~ {\rm cm}~
\delta_{10} \ga_{100}^2
\label{rs}
\end{equation}
where $\delta $ is the initial separation between the shells in the
observer's rest frame and $\d10=\delta / 10^{10}~{\rm cm }$ and
$\ga_{100}=\ga /100$.  Clearly internal shocks are relevant only if
they appear before the external shock that is produced as the 
shell sweeps up the ISM.  We show in section \ref{sub_sec:Newtonian} that the
necessary condition for internal shocks to occur before the external
shock is: 
\begin{equation}
\xi^{3/2} >  \zeta \ .
\label{isrelevance}
\end{equation}
where $\xi$ and $\zeta$ are two
dimensionless parameters. 
The   parameter, $\xi$, characterizes the interaction of the flow
with the external medium and it is defined in Eq. \ref{xi}
(see section \ref{sub_sec:Newtonian}).
The second  parameter, $\zeta$, characterizes
the variability of the flow:
\begin{equation}
\zeta \equiv
{\delta \over \Delta} \le 1 .
\label{zetadef}
\end{equation}
We have seen in section
\ref{sub_sec:Internal_time} that for internal shocks 
the  duration of the burst $T \approx
\Delta/c $ and the duration of individual spikes $\delta T \approx
\delta /c$. The observed ratio ${\cal N}$ defined in section
\ref{sub-sec:variability} must equal $1/\zeta$ and  this sets
$\zeta \approx 0.01$.

\begin{figure}
\begin{center}
\includegraphics[width=10cm]{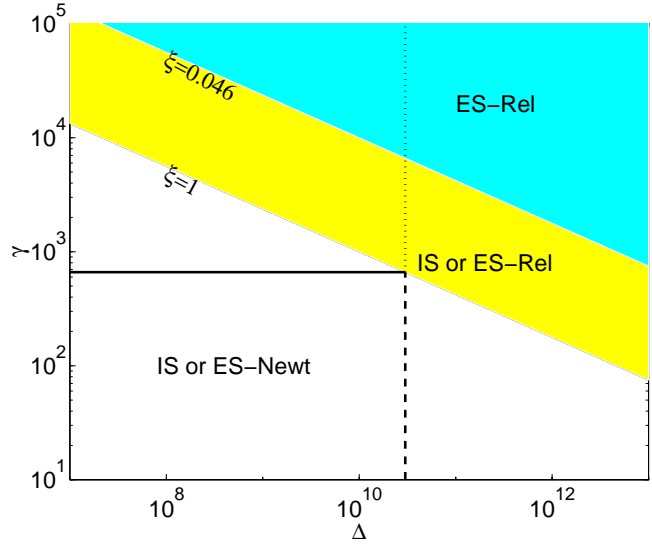}
\caption{\it Different scenarios in the $\Delta$ (in cm) - $\gamma$ plane
for $\zeta\equiv \delta/\Delta=0.01$. Relativistic ES occur for large
$\Delta$ and large $\gamma$ - upper right -above the $\xi=1$ line
(dark gray and light gray regions). Newtonian ES occur below $\xi=1$ -
lower left - white region.  IS occur, if there are sufficient
variation in $\gamma$ below the $\xi=\zeta^{2/3}$ line (light gray and
white regions).  The equal duration $T=1$s curve is shown for
Newtonian ES (solid line) a relativistic ES (dotted line) and IS
(dashed line).  Note that a relativistic ES and an internal shock with
the same parameters have the same overall duration $T$ but different
temporal substructure depending on $\delta$. From \cite{SaP97b}.}
\label{SaP97bf2}
\end{center}
\end{figure}

The overall duration of a burst produced by internal shocks
equals  $\Delta/c$. Thus,
 whereas external shocks
require an extremely large value of $\ga$   to produce a very
short burst, internal shocks can produce a short burst with
a modest value of the Lorentz factor $\gamma $. This eases somewhat 
the baryon purity constraints on
source models. The condition \ref{isrelevance} can be turned into a
condition that $\gamma$ is sufficiently small:
\begin{equation}
\label{xige1} \gamma \le 2800 ~ {\zeta_{0.01}}^{-1/2}
{T_{10}}^{-3/8} l_{18}^{3/8} ,
\end{equation}
where we have used $T=\Delta /c$ and we have defined
$T_{10} = {T/10{\rm s}}$ and $\zeta_{0.01}=\zeta/0.01$.  It 
follows that internal shocks
take place in relatively ``low'' $\gamma$ regime. Fig. \ref{SaP97bf2}
depicts the regimes in the physical parameter space ($\Delta, \gamma$)
in which various  shocks are possible. It also depicts an example of
a $T=\Delta /c =1$s line.

Too low a value of the Lorentz factor leads to a large optical depth in
the internal shocks region.  Using Eq. \ref{optical_depth} for $R_e$, at
which the optical depth for Compton scattering of the photons on the
shell's electrons  equals one, Eq.  \ref{rs} for $R_\delta$ and the
condition $R_e \le R_\delta$ we find:
\begin{equation} \label{mingamma}
\ga  \ge \left(  {\frac{E \sigma_T }{4 c^2 \delta^2  m_p \pi}} \right)^{1/5} =
130~{T_{10}}^{-2/5} {\zeta_{0.01}}^{-2/5} E_{52}^{1/5} \ .
\end{equation}

In addition, the radius of emission should be large enough so that the
optical depth for $\gamma \gamma \rightarrow e^+e^-$ will be less than
unity ($\tau_{\gamma\gamma} < 1$). There are several ways to consider
this constraint.  The strongest constraint is obtained if one demands
that the optical depth of an observed high energy, e.g. $100$MeV photon
will be less than unity \cite{Fenimore93,WoodLoeb}. Following these
calculations and using Eq. \ref{rs} to express $R_\delta$ we find:
\begin{equation}
\label{mingammagamma}
\gamma> 570{\left({\zeta_{0.01} T_{10} }\right)}^{-1/4}~~.
\end{equation}
This constraint, which is due to the $\gamma\gamma$ interaction, is
generally  more important than the constraint due to Compton scattering:
that is $\tau_{\gamma\gamma} >
\tau_e$.

\begin{figure}
\begin{center}
\includegraphics[width=10cm]{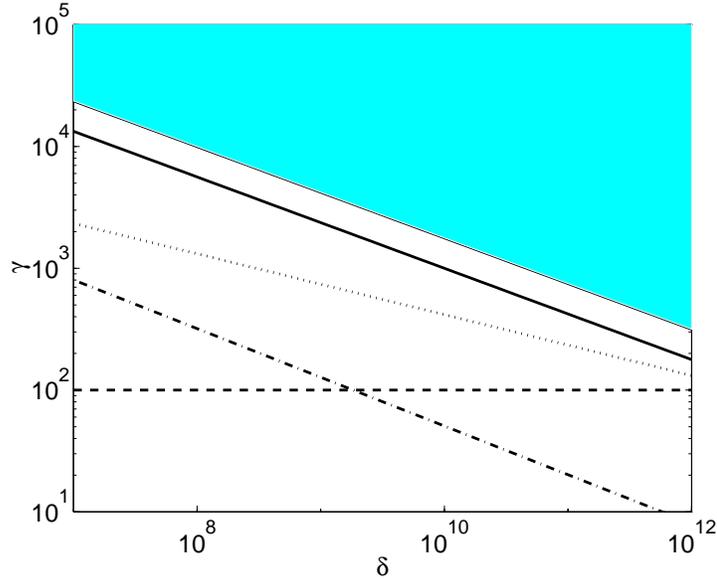}
\caption{\it Allowed regions for internal shocks in the $\delta$ (in cm),
$\gamma$ plane.  Note that the horizontal $\delta$ axis also
corresponds to the typical peak duration, $\delta t$ multiplied  by
$c$. Internal shocks are impossible in the upper right (light gray)
region.  The lower boundary of this region depends on $\zeta \equiv
\delta/\Delta$ and are marked by two solid curves, the lower one for $\zeta=1$
and the upper one for $\zeta=0.01$. Also shown are
$\tau_{\gamma\gamma}=1$ for an observed spectrum with no upper bound
(dotted line), $\tau_{\gamma\gamma}=1$ for an observed spectrum with
an upper bound of $100$MeV (dashed line) and $\tau_e =1$
(dashed-dotted).  The optically thin internal shock region is above
the $\tau=1$ curves and below the $\xi=\zeta^{2/3}$ (solid) lines.
From \cite{SaP97b}. }
\label{SaP97bf3}
\end{center}
\end{figure}

Eq.  \ref{xige1}, and the more restrictive Eq. \ref{mingammagamma}
constrains $\gamma$ to a relatively narrow range:
\begin{equation}
\label{gammainternal}  570~\zeta_{0.01}^{-1/4}
{T_{10}}^{-1/4}
\le \gamma_E \le 2800 ~  \zeta_{0.01}^{-1/2} T_{10}^{-3/8} l_{18}^{3/8} .
\end{equation}
This can be translated to a rather narrow range of emission radii:
\begin{equation}
10^{15} {\rm cm} ~\zeta_{0.01}^{1/2}
T_{10}^{1/2}
\le R_\delta \le 2.5 \cdot 10^{16} {\rm cm} ~
T_{10}^{1/4} l_{18}^{3/4} .
\label{Rsrange}
\end{equation}
In Fig. \ref{SaP97bf3}, we plot the allowed regions in the $\gamma$
and $\delta$ parameter space.
Using the less restrictive $\tau_{e}$ limit \ref{mingamma} we find:
\begin{eqnarray}
5 \times 10^{13} {\rm cm} ~ \zeta _{0.01}^{1/5}
~T_{10}^{1/5} E_{52}^{2/5}\le & R_\delta & \le 2.5 \cdot 10^{16} {\rm cm} ~
~T_{10}^{1/4} l^{3/4} \nonumber   \\ \nonumber 
130 ~  \zeta _{0.01}^{-2/5}
~T_{10}^{-2/5} E_{52}^{1/5}\le & \gamma_{E} & \le 2800 ~ \zeta _{0.01}^{-1/2}
~T_{10}^{3/8} .
        \label{Comptonscattering}
\end{eqnarray}

Three main conclusions emerge from the discussion so far. First, if
the spectrum of the observed photons extends beyond 100MeV (as was the
case in the bursts detected by EGRET \cite{Kippen96}) and if those
high energy photons are emitted in the same region as the low energy
ones then the condition on the pair production, $\tau_{\gamma\gamma}$,
Eq. \ref{mingammagamma} is stronger than the condition on Compton
scattering Eq. \ref{Comptonscattering}. This increases the required
Lorentz factors. Second, the Compton scattering limit (which is
independent of the observed high energy tail of the spectrum) poses
also a lower limit on $\gamma$.  However, this is usually less
restrictive then the $\tau_{\ga \ga}$ limit. Finally, one sees in
Fig. \ref{SaP97bf3} that optically thin internal shocks are produced
only in a narrow region in the $(\delta,\gamma)$ plane.  The region is
quite small if the stronger pair production limit holds. In this case
there is no single value of $\gamma$ that can produce peaks over the
whole range of observed durations.  The allowed region is larger if we
use the weaker limits on the opacity. But even with this limit there
is no single value of $\gamma$ that produces peaks with all
durations. The IS scenario suggests that bursts with very narrow peaks
should not have  very high energy tails and that very short bursts may
have a softer  spectrum.

\subsubsection{Physical Conditions and Emission from Internal Shocks}
\label{sub_sec:Internal_physical}

Provided that the different parts of the shell have comparable Lorentz
factors differing by factor of $\sim 2$, the internal shocks are mildly
relativistic. The protons' thermal Lorentz factor will be of order of
unity, and the shocked regions will still move highly relativistically
towards the observer with approximately the initial Lorentz factor
$\gamma$.  In front of the shocks the particle density of the shell is
given by the total number of baryons $E/\gamma m_p c^2$ divided by the
co-moving volume of the shell at the radius $R_\delta$ which is $4 \pi
R_\delta^2 \Delta \gamma$. The particle density behind the shock is higher
by a factor of $7$ which is the limiting compression possible by
Newtonian shocks (assuming an adiabatic index of relativistic gas,
i.e., $4/3$).   We  estimate the pre-shock density of the
particles in the shells as: $[E/(\ga m_p c^2)]/(4 \pi (\delta \ga^2 )^2
\ga \Delta )$.  We introduce $\ga_{int}$ as the Lorentz factor of the internal
shock. As this shock is relativistic (but not extremely relativistic)
$\ga_{int}$ is of order of a few.  Using Eq. \ref{internal_conditions}
for the particle density $n$ and the thermal energy density $e$ behind
the shocks we find:
\begin{eqnarray}
\label{hydroshocks}
n_{int} \approx {4 E (\ga_{int}/2)^{1/2}\over 4 \pi \gamma^6 c^2 m_p
\delta^2 \Delta} = \\ \nonumber 
2 \times 10^{10}~ {\rm cm}^{-3} ~ E_{52} ({\ga_{int}}/{2})^{1/2}
{\g100} ^{-6} {\D12} ^{-1} {\d10}^{-2} ,
\end{eqnarray}
\begin{equation}
e_{int} = (\ga_{int}/2)^{1/2} n_{int} m_p c^2 .
\end{equation}
We have defined here $\D12=\Delta/ 10^{12}{\rm cm}$. Using
Eq. \ref{Bvalue} we find:
\begin{equation}
B_{int} = 6 \cdot 10^5~ {\rm Gauss}~ \epsilon_B^{1/2} E^{1/2}_{52}
{\g100} ^{-3} {\D12} ^{-1/2} {\d10 } ^{-1}( {\ga_{int}}/{2}) ^{1/2} .
\label{Bint}
\end{equation}

Using Eqs. \ref{Bvalue}, \ref{typical},
\ref{syn_obs} and \ref{hydroshocks} we can
estimate the typical synchrotron frequency from an internal shock.
This is the synchrotron frequency of an electron with a ``typical''
Lorentz factor:
\begin{eqnarray}
\label{int_syn_obs}
(h\nu_{syn})_{obs}|_{\langle \ga_e \rangle}
=\frac{\hbar q_eB}{m_ec}\gamma _e^2\gamma
=
220~ {\rm keV}~  E^{1/2}_{52} \epsilon_B^{1/2} \\ \nonumber 
\times {\zeta}_{0.01}(\ga_{int}/{2})^{1/2} {\g100} ^{-2} {\D12} ^{-3/2}
[{\ga_e}/{(m_p/m_e)}] ^2 .
\end{eqnarray}
The corresponding observed synchrotron cooling time is:
\begin{equation} 
t_{syn}|_{\langle \ga_e \rangle} =
1.3 \cdot 10^{-6}~{\rm sec}~ \epsilon_B^{-1}
\d10^{2} \D12 \g100^{5} {E_{52}}^{-1} ({\ga_{int}}/{2})^{-1}.
\end{equation}

Using Eq. \ref{gemin} we can express $\ga_{e,min}$ in terms of
$\ga_{int}$ to estimate the minimal synchrotron frequency:
\begin{eqnarray}
\label{int_syn_obs2}
(h\nu_{syn})_{obs}|_{\ga_{e,min}}=
24~ {\rm keV}~  E^{1/2}_{52} \epsilon_B^{1/2} \epsilon_e^2
\\ \nonumber 
\times {\zeta}_{0.01}^{-1}( {\ga_{int}}/{2})^{5/2} {\g100} ^{-2} {\D12 } ^{-3/2} .
\end{eqnarray}
The energy emitted by a ``typical electron'' is around 220keV.  The
energy emitted by a ``minimal energy'' electron is about one order of
magnitude lower than the typical observed energy of $\sim
100$keV. This should correspond to the break energy of the spectrum.
This result seems in a good agreement with the observations.  But this
estimate might be misleading as both $\epsilon_B$ and $\epsilon_e$
might be significantly lower than unity.  Still these values of $(h
\nu_{syn})_{obs}$ are remarkably close to the observations.  One might
hope that this might explain the observed lower cutoff of the GRB
spectrum. Note that a lower value of $\epsilon_B$ or $\epsilon_e$ might be
compensated by a higher value of $\ga_{int}$. This is advantageous as
shocks with higher $\ga_{int}$ are more efficient (see section
\ref{sub_sec:efficiency_internal}).

The synchrotron cooling time at a given frequency (in the observer's
frame) is given by:
\begin{eqnarray} t_{syn} (h \nu) =
2 \times 10^{-6}~{\rm sec}~ \epsilon_B^{-3/4}
\left({h\nu_{obs}\over100~{\rm keV}}\right)^{-1/2} \\ \nonumber 
\times \d10 ^{3/2}\D12^{3/4}\g100^{4} E_{52}^{3/4} ({\ga_{int}}/{2})^{-3/4}.
\end{eqnarray}
We recover  the general trend $t_{syn} \propto (h \nu)^{-1/2}$ of 
synchrotron emission. However if (as we expect quite generally) 
this cooling time is  much
shorter than $T_{ang}$  it does not determine the 
width of the observed peaks. It will correspond to the observed
time scales if, for example, $\epsilon_B$ is small. But then the
``typical'' photon energy will be far below the observed range.
Therefore, it is not clear this relation can explain the observed
dependence of the width of the bursts on the observed energy.

\subsubsection{Inverse Compton in Internal Shocks}

The calculations of section \ref{sub_sec:IC} suggest that the typical
Inverse Compton (IC) (actually synchrotron - self Compton) radiation
from internal shocks will be at energy higher by a factor $\ga_e^2$
then the typical synchrotron frequency. Since synchrotron emission is
in the keV range and $\ga_{e,min} \approx m_p/m_e$, the expected IC
emission should be in the GeV or even TeV range. This radiation might
contribute to the prompt very high energy emission that accompanies
some of the GRBs \cite{Kippen96}.

However, if the magnetic field is extremely low: $\epsilon_B \sim
10^{-12}$ then we would expect the IC photons to be in the observed
$\sim 100$keV region:
\begin{eqnarray}
h \nu_{IC-int} = 800~ {\rm keV}~ ({\epsilon_B / 10^{-12}})^{1/2} \\ \nonumber 
\times \d10^{-1} \D12^{-1/2} \g100^{-2} [\ga_e / (mp/me)]^{4}
E_{52}^{1/2} [{\ga_{int}}/{2}]^{1/2}.
\end{eqnarray}
Using Eqs. \ref{tIC} and \ref{Bint} we find that the cooling time for
synchrotron-self Compton in this case is:
\begin{eqnarray}
\tau_{IC_int} = 1~ {\rm sec}~
\epsilon_e^{-1/2} ({\epsilon_B / 10^{-12}})^{-1/2} \\ \nonumber 
\times {\d10 }^{2} {\D12 } \g100^{5}
[\ga_e / (mp/me)]^{-1} E_{52}^{-1} ({\ga_{int}}/{2})^{-1} .
\end{eqnarray}
This is marginal. It is too large for some bursts and possibly
adequate for others. It could possibly  be adjusted by a proper
choice of the parameters.  It is more likely that if Inverse Compton is
important then it contributes to the very high (GeV or even TeV) signal
that accompanies the lower energy GRB (see also \cite{Pila_Loeb97}).

\subsubsection{Efficiency in Internal Shocks}
\label{sub_sec:efficiency_internal}
The elementary unit in the internal shock model (see section
\ref{sub_sec:Internal_time}) is a a binary (two shells) encounter
between a rapid shell (denoted by the subscript $r$) that catches up a
slower one (denoted $s$).  The two shells merge to form a single shell
(denoted $m$). The system behaves like an inelastic collision between
two masses $m_{r}$ and $m_{s}$.

The efficiency of a single collision between two shells was calculated earlier
in section \ref{sec:inelastic}. For multiple collisions the efficiency
depends on the nature of the random distribution. It is highest if the
energy is distributed equally among the different shells.  This can be
explained analytically. Consider a situation in which the mass of the
shell, $m_i$ is correlated with the (random) Lorentz factor, $\gamma_i$
as $m_i \propto \gamma_i^{\eta}$.  Let all the shells collide and merge
and only then emit the thermal energy as radiation. Using conservation
of energy and momentum we can calculate the overall efficiency:
\begin{equation}
\epsilon =1-\Sigma \gamma _{i}^{\eta }/\sqrt{\Sigma \gamma _{i}^{\eta
-1}\Sigma \gamma _{i}^{\eta +1}}.
\end{equation}
Averaging over the random variables $\gamma _{i}$, and assuming a
large number of shells $N\to \infty $ we obtain:
\begin{eqnarray}
\langle \epsilon \rangle \sim 1-\frac{(\gamma _{max}/\gamma _{min})^{\eta
+1}-1}{\eta +1} \times \\ \nonumber 
\sqrt{\frac{\eta (\eta +2)}{[(\gamma _{max}/\gamma
_{min})^{\eta }-1][(\gamma _{max}/\gamma _{min})^{\eta +2}-1]}}.
\label{analitic}
\end{eqnarray}
This formula explains qualitatively the numerical results: the
efficiency is maximal when the energy is distributed equally among
the different shells (which corresponds to $\eta = -1$).

In a realistic situation we expect that the internal energy will be
emitted after each collision, and not after all the shells have
merged. In this case there is no simple analytical formula. However,
numerical calculations show that the efficiency of this process is low
(less than $2\%$) if the initial spread in $\gamma $ is only a factor
of two \cite{MMM95}. However the efficiency could be much higher
\cite{KPS97}. The most efficient case is when the shells have a
comparable energy but very different Lorentz factors. In this case
($\eta =-1$, and spread of Lorentz factor $\gamma _{max}/\gamma
_{min}>10^{3}$) the efficiency is as high as $40\%$. For a moderate
spread of Lorentz factor $\gamma _{max}/\gamma _{min}=10$, with $\eta
=-1$, the efficiency is $20\%$.

The efficiency discussed so far is the efficiency of conversion of
kinetic energy to internal energy. One should multiply this by the
radiative efficiency, discussed in \ref{sub_sec:radiative_efficiency}
(Eq. \ref{rad_eff}) to obtain the overall efficiency of the
process. The resulting values may be rather small and this indicates that
some sort of beaming may be required in most GRB models in order not
to come up with an unreasonable energy requirement.

\subsubsection{Summary - Internal Shocks}

Internal shocks provide the best way to explain the observed temporal
structure in GRBs. These shocks, that take place at distances of $\sim
10^{15}$cm from the center, convert two to twenty percent of the
kinetic energy of the flow to thermal energy.  Under reasonable
conditions the typical synchrotron frequency of the relativistic
electrons in the internal shocks is around 100keV, more or less in the
observed region.

Internal shocks require a variable flow.  The situation in which an
inner shell is faster than an outer shell is unstable
\cite{WaxmanPiran}. The instability develops before the shocks form
and it may affect the energy conversion process. The full implications of
this instability are not understood yet.

Internal shocks can extract at most half of the shell's energy
\cite{MMM95,KPS97,Katz97}. Highly relativistic flow with a kinetic 
energy and a Lorentz factor comparable to the original one remains
after the internal shocks. Sari \& Piran \cite{SaP97a} pointed out
that if the shell is surrounded by ISM and collisionless shock occurs
the relativistic shell will dissipate by ``external shocks'' as
well. This predicts an additional smooth burst, with a comparable or
possibly greater energy.  This is most probably the source of the
observed ``afterglow'' seen in some counterparts to GRBs which we
discuss later.  This leads to the Internal-External scenario
\cite{SaP97c,SaP97a,KP97a} in which the GRB itself is produced by an Internal
shock, while the ``afterglow'' that was observed to follows some GRBs
is produced by an external shock.

The main concern with the internal shock model is its low efficiency
of conversion of kinetic energy to $\ga$-rays. This could be of order
twenty percent under favorable conditions and significantly lower
otherwise. If we assume that the ``inner engine'' is powered by a
gravitational binding energy of a compact object (see section
\ref{subsec:innerengine}) a low efficiency may require beaming to overcome an
overall energy crisis.

\subsection{Shocks with the ISM - External shocks}
\label{sss:ism}

We turn now to the interaction of a relativistic shell with the
ISM. We have seen in section \ref{sub_sec:Internal_time} that external
shocks cannot produce bursts with a complicated temporal structure.
Still it is worthwhile to explore this situation.  First, there are
some smooth bursts that might be produced in this way. Second, one
needs to understand the evolution of external shocks in order to see
why they cannot satisfy the condition $R_E/ \gamma^2 \le \Delta$.
Third, it is possible that in some bursts emission is observed from both
internal and external shocks \cite{Sari97a}. Finally, as we see in the
following section \ref{sec:afterglow} the observed afterglow is most
likely produced by external shocks.

\subsubsection{Newtonian vs. Relativistic Reverse Shocks}
\label{sub_sec:Newtonian}
The interaction between a relativistic flow and an external medium
depends, like in SNRs, on the Sedov length, $l \equiv (E/n_{ism} m_p
c^2)^{1/3}$. The ISM rest mass energy within a volume $l^3$ equals the energy
of the GRB: $E$.  For a canonical cosmological burst with $E
\approx 10^{52}$ergs and a typical ISM density $n_{ism} =
1$particle/cm$^3$ we have $ l \approx 10^{18}$cm.  A second length
scale that appears in the problem is $\Delta$, the width of the
relativistic shell in the observer's rest frame.

There are two possible types of external shocks \cite{SaP95}. They are
characterized according to the nature of the reverse shock: Newtonian
Reverse Shock (NRS) vs.  Relativistic Reverse Shock (RRS).  If the
reverse shock is relativistic (RRS) then it reduces significantly the
kinetic energy of each layer that it crosses.  Each layer within the
shell loses its energy independently from the rest of the shock.  The
energy conversion process is over once the reverse shock crosses the
shell (see Fig. \ref{fig:full_RRS}).  A Newtonian or even mildly
relativistic reverse shock (NRS) is comparatively weak.  Such a shock
reduces the energy of the layer that it crosses by a relatively small
amount.  Significant energy conversion takes place only after the
shock has crossed the shell several time after it has been reflected
as a rarefraction wave from the inner edge (see
Fig. \ref{fig:full_NRS}).  The shell behaves practically like a single
object and it loses its energy only by the time that it accumulates an
external mass equal to $M/\gamma$.

The question which scenario is taking place depends on the parameters of
the shell relative to the parameters of the ISM. As we see shortly it
depends on a single dimensionless parameter $\xi$ constructed from
$l$, $\Delta$ and $\ga$:
\cite{SaP95}:
\begin{equation}
\label{xi}
\xi \equiv (l/ \Delta )^{1/2} \ga^{-4/3} \ .
\end{equation}

As the shell propagates outwards it is initially very dense and the
density ratio between the shell and the ISM, $f\equiv n_4/n_1$, is
extremely large (more specifically $f> \gamma^2$).  The reverse shock
is initially Newtonian (see Eq. \ref{nr1}). Such a shock converts only
a small fraction of the kinetic energy to thermal energy.  As the
shell propagates the density ratio, $f$, decreases (like $R^{-2}$ if
the width of the shell is constant and like $R^{-3}$ if the shell is
spreading).  Eventually the reverse shock becomes relativistic at
$R_N$ where $f=\gamma^2$.  The question where is the kinetic energy
converted depends on whether the reverse shock reaches the inner edge
of the shell before or after it becomes relativistic.

There are four different radii that should be considered. 
The following estimates assume a spherically symmetric shell, or that
$E$ and $M$  are energy and rest mass divided by the fraction of a 
sphere into which they are launched.
The reverse shock becomes relativistic
at $R_N$, where $f=n_4/n_1 = 1$:  
\begin{equation}
R_N = l^{3/2} /\Delta^{1/2} \gamma^2
\end{equation}
Using the expression for the velocity of the reverse shock into the
shell (Eq. \ref{cond34}) we find that the reverse
shock reaches the inner edge of the shell at $R_\Delta$ \cite{SaP95}:
\begin{equation}
R_\Delta = l^{3/4} \Delta^{1/4} \ .
\label{Rdelta}
\end{equation}
A third radius is $R_\gamma$, where the shell collects an ISM mass of
$M/\ga$ \cite{MR1,Katz94}. For NRS this is where an effective energy
release occurs:
\begin{eqnarray}
R_\ga = {l \over \gamma^{2/3}} =
\bigg({E  \over n_{ism} m_p c^2 \gamma^2} \bigg)^{1/3} = \\ \nonumber 
5.4 \times 10^{16}~{\rm cm }~ E_{52}^{1/3} n_{1}^{-1/3} \g100^{-2/3} ,
\label{rm}
\end{eqnarray}
where we defined $n_1= n_{ism}/ 1~ {\rm particle/ cm}^3$.  Finally we
have $R_\delta = \delta \gamma^2$, (see Eq. \ref{rs}).  The
different radii are related by the dimensionless parameter $\xi$, and
this determines the character of the shock:
\begin{equation}
\label{order0}
R_\delta/\zeta =     R_\Delta /\xi^{3/2}=  R_\ga \xi^{2} = R_N /\xi^{3}
\end{equation}

If $\xi > 1$ then:
\begin{equation}
\label{order1}
R_\delta < R_\Delta  <   R_\ga < R_N .
\end{equation}
The reverse shock remains Newtonian or at best mildly relativistic
during the whole energy extraction process.  The reverse shock reaches
the inner edge of the shock at $R_\Delta$ while it is still Newtonian.
At this stage a reflected rarefraction wave begins to move
forwards. This wave is, in turn, reflected from the contact
discontinuity, between the shell's material and the ISM material, and
another reverse shock begins. The overall outcome of these waves is
that in this case the shell acts as a single fluid element of mass $M
\approx E /\ga c^2$ that is interacting collectively with the ISM.  It
follows from Eq. \ref{mgamma} that an external mass $m=M/\ga$ is
required to reduce $\ga$ to $\ga/2$ and to convert half of the kinetic
energy to thermal energy.  Energy conversion takes place at $R_\ga$
 Comparison of $R_\ga$ with $R_e$ (equation
\ref{optical_depth}) shows that the optical depth is much smaller
than unity.

\begin{figure}
\begin{center}
\includegraphics[width=20cm]{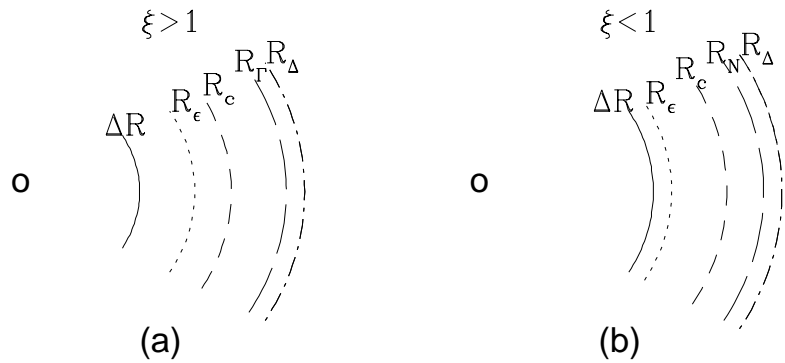}
\caption{\it (a) Schematic description of the different radii
for the case $\xi>1$. The different distances are marked
on a logarithmic scale. Beginning from the inside we have
$\Delta R$, the initial size of the shell, $R_\eta$, the radius
in which a fireball becomes matter-dominated (see the following
discussion), $R_c$, the radius where inner shells overtake
each other and collide, $R_\Delta$, where the reverse shock reaches
the inner boundary of the shell, and $R_\Gamma$, where the kinetic
energy of the shell is converted into thermal energy.
(b) Same as (a) for $\xi < 1$. $R_\Gamma$ does not appear here
since it is not relevant. $R_N$ marks the place where the
reverse shock becomes relativistic. From \cite{Piran97}}
\label{hydro_conditions}
\end{center}
\end{figure}

If the shell propagates with a constant width then $R_N/\xi = R_\ga
=\sqrt \xi R_\Delta$ (see Fig. \ref{hydro_conditions}) and for
$\xi>1$ the reverse shock remains Newtonian during the energy
extraction period.  If there are significant variations in the
particles velocity within the shell it will spread during the
expansion.  If the typical variation in $\ga$ is of the same order as
$\ga$ then the shell width increases like $R/\ga^2$. Thus $\Delta $
changes with time in such a manner that at each moment the current
width, $\Delta (t)$, satisfies $\Delta (t)
\sim {\rm max}[ \Delta (0),R/\gamma^2]$. This delays the time
that the reverse shock reaches the inner edge of the shell and
increases $R_\Delta$.  It also reduces the shell's density which, in
turn, reduces $f$ and leads to a decrease in $R_N$. The overall result
is a triple coincidence $R_N \approx R_\ga \approx R_\Delta$ with a
mildly relativistic reverse shock and a significant energy conversion
in the reverse shock as well.  This means that due to spreading a
shell which begins with a value of $\xi>1$ adjusts itself so as to
satisfy $\xi=1$.

For $\xi \ge 1$ we find that $T_{radial} \sim T_{ang}\sim R_\ga
/\ga^2 > \Delta$. Therefore, NRS can produce only smooth bursts.  The
bursts' duration is determined by the slowing down time of the shell.
In section \ref{sec:Tempstruct} we have shown that only one time scale
is possible in this case.  Given the typical radius of energy
conversion, $R_\ga$ this time scale is:
\begin{equation}
\delta T \approx T_{obs} \approx R_\ga /(\gamma_E^2 c) \approx
R_\ga /(\ga^2 c) \approx
170~{\rm sec} ~ E_{52}^{1/3} n_1^{-1/3} \g100^{-8/3} ,
\label{tobs1}
\end{equation}

If $\ga$ or $\Delta$ are larger then $\xi<1$. In this case the order
is reversed:
\begin{equation}
\label{order}
R_N  <   R_\ga < R_\Delta  .
\end{equation}
The reverse shock becomes relativistic very early (see
Fig. \ref{hydro_conditions}).  Since $\ga_{sh} = \ga_2 \ll \ga$ the
relativistic reverse shock converts very efficiently the kinetic
energy of the shell to thermal energy. Each layer of the shell that is
shocked loses effectively all its kinetic energy at once and the time
scale of converting the shell's kinetic energy to thermal energy is
the shell crossing time. The kinetic energy is consumed at $R_\Delta$,
where the reverse shock reaches the inner edge of the shell.  Using
Eq. \ref{Rdelta} for $R_\Delta$ and Eq. \ref{cond12} we find that at
$R_\Delta$
\begin{equation}
\ga_E = \ga_2 = (l/\Delta)^{3/8} .
\end{equation}
Note that $\ga_E$ is independent of $\ga$.  The observed radial or
angular time scales are:
\begin{equation}
T_{radial} \approx T_{ang} \approx
R_\Delta / \gamma_E^2 \approx \Delta /c = 30~{\rm sec}~\D12   \ .
\end{equation}
Thus even for RRS we find that $\delta T \sim T$ and there is only one
time scale.  This time scale depends only on $\Delta$ and it is
independent of $\ga$! Spreading does not affect this estimate since
for $\xi<1$ spreading does not occur before the energy extraction. 

In the following discussions we focus on the RRS case and we express
all results in terms of the parameter $\xi$.  By setting $\xi<1$ in
the expressions we obtain results corresponding to RRS, and by
choosing $\xi=1$ in the same expressions we obtain the spreading NRS
limit. We shall not discuss the case of non-spreading NRS ($\xi\gg1$),
since spreading will always bring these shells to the mildly
relativistic limit ($\xi\sim1$).  Therefore, in this way, the same
formulae are valid for both the RRS and NRS limits.

If $\xi>1$ it follows from Eq. \ref{order1} that internal shocks will
take place before external shocks. If $\xi < 1$ then the condition for
internal shocks $R_\delta < R_\Delta$ becomes Eq. \ref{isrelevance}:
$\xi^{3/2} > \zeta $ .
As we have seen earlier (see section
\ref{sub_sec:Internal_parameters}) this sets an upper limit on
$\gamma$ for internal shocks.

\subsubsection{Physical Conditions in External Shocks}

The interaction between the outward moving shell and the ISM takes
place in the form of two shocks: a forward shock that propagates into
the ISM and a reverse shock that propagates into the relativistic
shell. This results in four distinct regions: the ISM at rest (denoted
by the subscript 1 when we consider properties in this region), the
shocked ISM material which has passed through the forward shock
(subscript 2 or f), the shocked shell material which has passed
through the reverse shock (3 or r), and the unshocked material in the
shell (4). See Fig. \ref{shock_profile}.  The nature of the emitted
radiation and the efficiency of the cooling processes depend on the
conditions in the shocked regions 2 and 3.  Both regions have the same
energy density $e$. The particle densities $n_2$ and $n_3$ are,
however, different and hence the effective ``temperatures,'' i.e. the
mean Lorentz factors of the random motions of the shocked protons and
electrons, are different.

The bulk of the kinetic energy of the shell is converted to thermal
energy via the two shocks at around the time the shell has expanded to
the radius $R_\Delta$. At this radius, the conditions at the forward
shock are as follows,
\begin{equation}
\label{hydroforward}
\gamma _2  =  \gamma \xi ^{3/4},   \ \ \
n_2  =  4\gamma _2n_{1},  \ \ \
e_2  =  4\gamma _2^2n_{1}m_pc^2,
\end{equation}
while at the reverse shock we have
\begin{equation}
\label{hydroreverse}
\bar \gamma _3  =  \xi^{-3/4},   \ \ \
\gamma_3        =  \gamma\xi^{3/4}, \ \ \
n_3             =  4\xi ^{9/4}\gamma ^2n_{1}, \ \ \
e_3            =  e_2.
\end{equation}

Substitution of $\ga_{sh}=\ga_2 = \ga \xi^{3/4}$ in Eq. \ref{Bvalue} yields:
\begin{equation}
B= \sqrt{32 \pi} c \epsilon_B^{1/2} \ga \xi^{3/4} m_p^{1/2} n_{1}^{1/2}
=(40~{\rm G})~\epsilon_B^{1/2}\xi^{3/4}
{\g100} n_{1}^{1/2}.
\end{equation}
If the magnetic field in region 2 behind the forward shock is obtained
purely by shock compression of the ISM field, the field would be very
weak, with $\epsilon_B \ll 1$.  Such low fields are incompatible with
observations of GRBs.  We therefore consider the possibility that
there may be some kind of a turbulent instability which may bring the
magnetic field to approximate equipartition.  In the case of the
reverse shock, magnetic fields of considerable strength might be
present in the pre-shock shell material if the original exploding
fireball was magnetic. The exact nature of magnetic field evolution
during fireball expansion depends on several assumptions. Thompson
\cite{Thom} found that the magnetic field will remain in equipartition
if it started off originally in equipartition.  M\'esz\'aros, Laguna
\& Rees \cite{MLR} on the other hand estimated that if the magnetic
field was initially in equipartition then it would be below
equipartition by a factor of $10^{-5}$ by the time the shell expands
to $R_\Delta$. It is uncertain which, if either, is right.  As in the
forward shock, an instability could boost the field back to
equipartition.  Thus, while both shocks may have $\epsilon_B\ll 1$
with pure flux freezing, both could achieve $\epsilon_B\rightarrow1$
in the presence of instabilities.  In principle, $\epsilon_B$ could be
different for the two shocks, but we limit ourselves to the same
$\epsilon_B$ in both shocks.

In both regions 2 and 3 the electrons have a power law distribution
with a  minimal Lorentz factor $\gamma_{e,min}$ given by Eq.
\ref{gemin} with the corresponding Lorentz factors for the forward and
the reverse shock.

\subsubsection{Synchrotron Cooling in External Shocks}
\label{sub_sec:synchrotron}
The typical energy of synchrotron photons as well as the synchrotron
cooling time depend on the Lorentz factor $\gamma_e$ of the
relativistic electrons under consideration and on the strength of the
magnetic field. 
Using Eq. \ref{gemin} for $\gamma_{e,min} $
we find the characteristic synchrotron  energy 
for the forward shock:
\begin{equation}
\label{hnu_gemin}(h\nu_{syn})_{obs}|_{\ga_{e,min}}=
160~ {\rm keV}~ \epsilon_B^{1/2} \epsilon_e^2
({\ga_2/  100})^4 n_1^{1/2} ,
\end{equation}
and
\begin{equation}
\label{cooling_gemin}t_{syn}|_{\ga_{e,min}}=
0.085~ {\rm sec} ~\epsilon_B^{-1} \epsilon_e^{-1}
({\ga_2/  100})^{-4} n_1^{-1} .
\end{equation}
The characteristic frequency and the corresponding cooling time for
the ``typical'' electron are larger by a factor of $[(p-2)/(p-1)]^2$
and shorter by a factor of $[(p-2)/(p-1)]^2$, correspondingly.

These photons seems to be right in the observed soft
gamma-ray range. However, one should recall that the frequency
calculated in Eq. \ref{hnu_gemin} depends on the forth power of
$\ga_2$. An increase of the canonical $\ga_2$ by a factor of 3
(that is $\ga_2 =
300$ instead of  $\ga_2 = 100$)  will yield a 
``typical'' synchrotron emission at the 16MeV instead of 160keV.
The Lorentz factor of a ``typical electron'' in the reverse shock is
lower by a factor $\xi^{3/2}$.  Therefore the observed energy is lower
by a factor $\xi^3$ while the cooling time scale is longer by a factor
$\xi^{-3/4}$.

Alternatively we can check the conditions in order that there are
electrons with a Lorentz factor $\hat \gamma _e$ that be emitting soft
gamma-rays with energies $\sim 100$keV.  Using Eq. \ref{syn_obs} we
calculate $\hat \gamma _e$:
\begin{eqnarray} \label{gamma_hat}
\hat \gamma _e=\left( \frac{m_ech\nu_{obs}}
{\hbar q_e\gamma _2B}\right) ^{1/2} = 
5 \times10^4 ~\epsilon_B^{-1/4}\\ \nonumber 
\times \left({h\nu_{obs}\over100~{\rm keV}}\right)^{1/2}
{\g100 }^{-1} \xi^{3/4} n_1^{-1/4}.
\end{eqnarray}
Electrons with $\gamma_e=\hat\gamma_e$ are
available in the shocked material if 
$\gamma_{e,min} <\hat \gamma_e$. This  corresponds to the condition
\begin{equation}
\label{maxeer}
\epsilon _{e~|r}<80~\epsilon _b^{-1/4}
\left( {h\nu_{obs}\over100~{\rm keV}}\right)^{1/2}
\g100^{-1} n_1^{-1/4}
\end{equation}
in the reverse shock, and the condition
\begin{equation}
\label{maxeef} \epsilon _{e~|_f}<0.8 ~\epsilon _B^{-1/4}
\left({h\nu_{obs}\over100~{\rm keV}}\right)^{1/2}
\g100^{-2} \xi^{-3/2} n_1^{-1/4}
\end{equation}
in the forward shock.  Since by definition $\epsilon_e\leq1$, we see
that the reverse shock always has electrons with the right Lorentz
factors to produce soft gamma-ray synchrotron photons.  However, the
situation is marginal in the case of the forward shock.  If $\ga> 100$
and if the heating of the electrons is efficient, i.e. if
$\epsilon_{e~|_f}\sim1$, then most of the electrons may be too
energetic.  Of course, as an electron cools, it radiates at
progressively softer energies.  Therefore, even if $\gamma_{\rm min}$
is initially too large for the synchrotron radiation to be in soft
gamma-rays, the same electrons would at a later time have
$\gamma_e\sim\hat\gamma_e$ and become visible.  However, the energy
remaining in the electrons at the later time will also be lower (by a
factor $\hat\gamma/\gamma_{\rm min}$), which means that the burst will
be inefficient.  For simplicity, we ignore this radiation.

Substituting the value of $\hat \gamma_e$ from equation
\ref{gamma_hat} into the cooling rate Eq. \ref{cooling} we
obtain the cooling time scale as a function of the observed photon
energy to be
\begin{equation}
\label{tausyn3}t _{syn}(h \nu) 
=1.4\times 10^{-2}~{\rm sec}~ \epsilon_B^{-3/4}
\left({h\nu_{obs}\over100~{\rm keV}}\right)^{-1/2}
\g100^{-2}n_1^{-3/4}.
\end{equation}
Eq. \ref{tausyn3} is valid for both the forward and reverse
shock, and is moreover independent of whether the reverse shock is
relativistic or Newtonian.

The cooling time calculated above sets a lower limit to the
variability time scale of a GRB since the burst cannot possibly
contain spikes that are shorter than its cooling time.  However,
it is unlikely that this cooling time actually determines the observed
time scales. 
\subsection{The Internal - External Scenario}
\label{sub_sec:intextscen}

Internal shocks can convert only a fraction of the total energy to
radiation \cite{MMM95,KPS97,Katz97}.  After the flow has produced a GRB via
internal shocks it will interact via an external shock with the
surrounding medium \cite{SaP97a}. This will produce the afterglow -
a signal that will follow the GRB. The idea of an afterglow in other
wavelengths was suggested earlier \cite{PacRho93,Katz94,MR97} but it
was suggested as a follow up of the, then standard, external shock
scenario. In this case the afterglow would have been a direct
continuation of the GRB activity and its properties would have scaled
directly to the properties of the GRB.

According to internal-external models (internal shocks for the GRB and
external shocks for the afterglow) different mechanisms produce the
GRB and the afterglow.  Therefore the afterglow should not be scaled
directly to the properties of the GRB.  This was in fact seen in the
recent afterglow observations \cite{KP97,KP97a}.  In all models of
external shocks the observed time satisfy $t \propto R /\gamma_e^2$
and the typical frequency satisfies $\nu \propto \gamma_e^4$. Since
most of the emission takes place at practically the same radius and
all that we see is the variation of the Lorentz factor we expect quite
generally \cite{KP97}: $\nu \propto t^{2 \pm \iota} $.  The small
parameter $\iota$ reflects the variation of the radius and it depends
on the specific assumptions made in the model. We would expect that
$t_x /t_\gamma \sim 5$ and $t_{opt}/t_\gamma \sim 300$.  The
observations of GRB970508 show that $(t_{opt}/t_\gamma)_{observed}
\approx 10^4$. This is in a clear disagreement with the single
external shock model for both the GRB and the afterglow.

Under quite general conditions the initial typical synchrotron energy
for either the forward or the reverse external shock may fall in the
soft GRB band. In this case the initial stage of the afterglow might
overlap the $\ga$-ray emission from the internal shock \cite{Sari97a}.
The result will be superposition of a rapidly varying signal on top of
a long smooth and softening pulse. This possibility should be explored
in greater detail.

\section{Afterglow}
\label{sec:afterglow}
It is generally believed that the observed afterglow results from
slowing down of a relativistic shell on the external ISM.  The
afterglow is produced, in this case, by an external shock. A second
alternative is of ``continuous emission''.  The ``inner engine'' that
powers the GRB continues to emit energy for much longer duration with
a lower amplitude \cite{KaPS97} and may produce the earlier part
(first day or two in GRB970228 and GRB970508) of the afterglow. It is
most likely that both processes take place to some extent
\cite{KP97a}.  We discuss in this section theoretical models for the
production of the afterglow focusing on the external shock model.

\subsection{Hydrodynamics of a Slowing Down Relativistic Shell}

Within the external shock model there are several possible physical
assumptions that one can make.  The ``standard'' model assumes
adiabatic hydrodynamics (energy losses are negligible and do not
influence the hydrodynamics), slow cooling (the electrons radiate a
small fraction of the energy that is generated by the shock) and
synchrotron emission
\cite{PacRho93,Katz94,MR97,Wiejers_MR97,Waxman97a,Waxman97b,MesReesWei97}.
However there are other possibilities. First, the electrons' energy
might be radiated rapidly. In this case the radiation process is fast
and the observed flux is determined by the rate of energy generation
by the shock. If the electrons carry a significant fraction of the
total internal energy fast cooling will influence the hydrodynamics
which will not  be adiabatic any more. In this case  we have a radiative
solution \cite{KP97,Vietri97} which differs in its basic scaling
laws from the adiabatic one. The different possibilities are summarized in
Table \ref{t:afterglow_models}

\begin{center}

\begin{table*}[ht!]
\label{t:afterglow_models}
  \begin{center}
\begin {tabular}{|c||c|c|}
  \hline &Adiabatic Hydrodynamics & Radiative Hydrodynamics \\ \hline
  \hline Slow Cooling & Arbitrary $\epsilon_e$ & Impossible \\ \hline
  Fast Cooling & $\epsilon_e< 1$ & $\epsilon_e \approx 1$ \\ \hline
\end{tabular}
\end{center}
\caption{\it Afterglow Models}
\end{table*}
\end{center}

\subsubsection{A Simple Collisional Model}
\label{sub_sec:simplecolmod}

We consider first a simple model for the slowing down of the shell.
In this model the slowing down is described by a series of
infinitesimal inelastic collisions between the shell and 
infinitesimal external masses.  We assume a homogeneous shell described
by its rest frame energy $M$ (rest mass and thermal energy) and its
Lorentz factor $\gamma $. Initially, $E_0=M_0 c^2 \ga_0$.  The shell
collides with the surrounding matter. We denote the mass of the ISM
that has already collided with the shell by $m(R)$. As the shell
propagates it sweeps up more ISM mass.  Additional ISM mass elements,
$dm$, which are at rest collides inelastically with the shell.

Energy and momentum conservation yield:
\begin{equation}
{\frac{d\gamma }{{\gamma ^{2}-1}}}=-{\frac{dm}{M}}  \label{dgamma} ,
\end{equation}
and
\begin{equation}
dE=(\gamma -1)dm ,
\end{equation}
where $dE$ is the thermal energy produced in this collision.  We
define $\epsilon $ as the fraction of the shock generated thermal
energy (relative to the observer frame) that is radiated. The
incremental total mass satisfies:
\begin{equation}
dM=(1-\epsilon )dE+dm=[(1-\epsilon )\gamma +\epsilon ]dm  \label{dM} .
\end{equation}
These equations yields  analytic relations between the Lorentz factor
and the total mass of the shell:
\begin{equation}
\frac{(\gamma -1)(\gamma +1)^{1-2\epsilon }}{(\gamma _{0}-1)(\gamma
_{0}+1)^{1-2\epsilon }}=(M/M_{0})^{-2}  \label{Mgamma} ,
\end{equation}
and between $m(R)$ (and therefore $R$) and $\gamma$.
\begin{eqnarray}
\frac{m(R)}{M_{0}}
=-(\gamma _{0}-1)^{1/2}(\gamma _{0}+1)^{1/2-\epsilon }
\\ \nonumber 
\times \int_{\gamma_{0}}^{\gamma }(\gamma' -1)^{-3/2}(\gamma' +1)^{-3/2+\epsilon }d\gamma' .
\label{mgamma1}
\end{eqnarray}
These relations completely describe the hydrodynamical evolution of
the shell.

Two  basic features can be seen directly from Eq. \ref{mgamma1}.
First, we can estimate the ISM mass $m$ that should be swept to get
significant deceleration.  Solving Eq. \ref{mgamma1} with an upper
limit $\gamma _{0}/2$ and using $\gamma _{0}\gg 1$ we obtain the well
known result: a mass $m \cong M_{0}/(2\gamma _{0})$ is required to
reach $\gamma =\gamma _{0}/2$.  Apparently this result is independent
of the cooling parameter $\epsilon $.

A second simple result can be obtained in the limit that $\gamma _{0}
\gg \gamma \gg 1$:
\begin{equation}
m(R)=\frac{M_{0}}{(2-\epsilon )\gamma _{0}}\left( \frac{\gamma }{\gamma _{0}}%
\right) ^{-2+\epsilon },
\end{equation}
so that $\gamma \propto R^{-3/(2-\epsilon )}$. For $\epsilon =0$ this
yields the well known adiabatic result:
\begin{equation}
{4 \pi \over 3} R^3 n m_p c^2  \gamma^2= E_0 ,
\label{g_adiabatic}
\end{equation}
and $\gamma \propto R^{-3/2}$
\cite{BLmc1,Katz94,Wiejers_MR97,Waxman97a,Waxman97b}.   
For $\epsilon =1$ this yields the completely radiative result:
\begin{equation}
{4 \pi \over 3} R^3 n m_p c^2  \gamma \gamma_0  = E_0,
\label{g_radiative}
\end{equation}
and $\gamma \propto R^{-3}$ \cite{BLmc1,Vietri97,KP97}.

For comparison with observations we have to calculate the observed
time that corresponds to different radii and Lorentz factors.
The well known formula
\begin{equation}
t_{obs} = {R \over 2 \ga^2 c}
\end{equation}
is valid only for emission along the line of sight
from a  shell that propagates
with a constant velocity. Sari \cite{Sari97a} pointed out that as the
shell decelerates this formula should be used only in a differential
sense:
\begin{equation}
d t_{obs} = {dR \over 2 \ga^2 c} .
\label{dtobs}
\end{equation}
Eq. \ref{dtobs} should be combined with the relation \ref{g_adiabatic}
or \ref{g_radiative} and integrated to get the actual relation between
observed time and emission radius. For an adiabatic expansion, for
example, this yields: $t_{obs} = R/16 \ga^2 c$ \cite{Sari97a}.
Eq. \ref{dtobs} is valid only along the line of sight. The situation
is complicated further if we recall that the emission reaches the
observe from an angle of order $\ga^{-1}$ around the line of sight.
Averaging on all angles yields another numerical factor
\cite{Sari97b,Waxman97c,Meszaros97} and altogether we get
\begin{equation}
t_{obs} \approx { R \over c_{\gamma} \ga_2 c},
\label{tobs}
\end{equation}
where the  value of the 
numerical factor, $c_{ga}$, depends on the details of the solution
and it varies between  $\sim3$ and $\sim 7$.
Using Eqs. \ref{tobs} and \ref{g_adiabatic} or \ref{g_radiative} we obtain
the following relations between $R$, $\ga$ and $t$:
\begin{equation}
\label{R_t}
R(t) \cong \cases {
(3Et/\pi m_p n c)^{1/4}, & ad, \cr
(4ct/L)^{1/7} L, & rad,}
\end{equation}
\begin{equation}
\label{g_t}
\gamma(t) \cong \cases {
(3E/256\pi n m_p c^5 t^3)^{1/8}, & ad, \cr
(4ct/L)^{-3/7}, & rad,}
\end{equation}
where $L \equiv ({{3E}/  {4 \pi n m_p c^{2}\ga}})^{1/3}$ is the radius
where the external mass equals the mass of the shell.

One can proceed and use the relation between $R$ and $\gamma$ and
$t_{obs}$ (Eqs. \ref{R_t} and \ref{g_t}) to estimate the physical
conditions at the shocked material using Eqs. \ref{nr3}. Then one can
estimate the emitted radiation from this shock using Eqs.
\ref{syn_obs} and \ref{syn_power}. However, before doing so we explore
the Blandford-McKee self similar solution \cite{BLmc1}, which
describes more precisely the adiabatic expansion. This solution is
inhomogeneous with a well determined radial profile.  The matter at
the front of the shell moves faster than the average speed. This
influences the estimates of the radiation emitted from the shell.

\subsubsection{The Blandford-McKee  Self-Similar Solution}
\label{sub_sec:BM_solution}

Blandford \& McKee \cite{BLmc1} discovered a self-similar solution
that describes the adiabatic slowing down of an extremely relativistic
shell propagating into the ISM.  Using several simplifications and
some algebraic manipulations we rewrite the Blandford-McKee solution
as
\cite{Sari97a}:
\begin{eqnarray}
\label{selfsimilar}
n(r,t)       &=& 4 n    \gamma(t)   \left[ 1+16\gamma(t)^2(1-r/R) \right]^{-5/4}, \cr
\gamma(r,t)  &=&        \gamma(t)   \left[ 1+16\gamma(t)^2(1-r/R) \right]^{-1/2}, \cr
e(r,t)       &=&4nm_pc^2\gamma(t)^2 \left[ 1+16\gamma(t)^2(1-r/R) \right]^{-17/12} ,
\end{eqnarray}
where $n(r,t)$, $e(r,t)$ and $\gamma(r,t)$ are, respectively, the
density, energy density and Lorentz factor of the material behind the
shock (not to be confused with the ISM density $n$) and
$\gamma(t)=\gamma(R(t))$ is the Lorentz factor of material just behind
the shock.  $n(r,t)$ and $e(r,t)$ are measured in the fluid's rest
frame while $\gamma(r,t)$ is relative to an observer at rest.  The
total energy in this adiabatic flow equals $E=E_0$, the initial
energy.  The scaling laws of $R(t)$ and $\gamma(t)$ that follow from
these profiles and from the condition that the total energy in the
flow equals $E$ is:
\begin{eqnarray}
\label{scalings}
R(t)&=&\left( {17 E t \over \pi m_p n c} \right)^{1/4} =3.2\times
10^{16}~ {\rm\ cm}~ E_{52}^{1/4}n_1^{-1/4} t_s^{1/4} , \cr
\gamma(t)&=&\frac 1 4 \left( {17 E \over \pi n m_p c^5 t^3} \right)^{1/8}
          =260 E_{52}^{1/8} n_1^{-1/8} t_s^{-3/8} .
\end{eqnarray}
The scalings \ref{scalings} are consistent with the scalings \ref{R_t}
and \ref{g_t} which were derived using conservation of energy and
momentum. They provide the exact numerical factor that cannot be
calculated by the simple analysis of section
\ref{sub_sec:simplecolmod}. These equations can serve as a starting
point for a detailed radiation emission calculation and a comparison
with observations.

The Blandford-McKee solution is adiabatic and as such it does
not allow for any energy losses. 
With some simplifying assumptions it is possible to derive a
self-similar radiative solution in which an arbitrary fraction of
the energy generated by the shock is radiated away \cite{CPS98}. 

\subsection{Phases  in a Relativistic   Decelerating  Shell}

There are several phases in the deceleration of a relativistic shell:
fast cooling (with either radiative or adiabatic hydrodynamics) is
followed by slow cooling (with adiabatic hydrodynamics). Then if the
shell is non spherical its evolution changes and a phase of sideways
expansion and much faster slow down begins when the Lorentz factor
reaches $\theta^{-1}$ \cite{Rhodas97}. Finally the shell becomes
Newtonian when enough mass is collected and $\ga \approx 1$. In the
following we estimate the time scale for the different transitions. We
define $\ga_{e,min} \equiv \tge \epsilon_e (m_p / m_e) \ga$ and
$t_{obs} = (1+z) {R / 4 c_t c \ga^2}$ such that the factors $\tge$ and
$c_t$ reflect some of the uncertainties in the model. The canonical
values of these factors are: $\tge \approx 0.5$ and $c_t \approx 1$.

The deceleration begins in a fast cooling phase.  If $\epsilon_e$ is
close to unity than this cooling phase will also be radiative.  The
first transition is from fast to slow cooling.  
There are several different  ways to estimate this transition. 
One can compare the  cooling time scale  to the hydrodynamic time scale;
alternatively one can calculate  the fast cooling rate (given by the
rate of energy generation by the shell)
and compare it to the  slow cooling rate (given by the emissivity of 
the relativistic electrons). We have chosen here to calculate this
time as the time when the ``typical electron'' cools - that is
when $\nu_c = \nu_m$:
\begin{equation}
\label{tfc}
t_{fs} =\cases{
210 ~{\rm days}~ \epsilon_B^2 \epsilon_e^2 E_{52} n_1 ,
       & ad, \cr
4.6 ~{\rm days}~
\epsilon_B^{7/5} \epsilon_e^{7/5} E_{52}^{4/5}\gamma_{100}^{-4/5}
n_1^{3/5} ~,
       & rad.}
\end{equation}
All methods of estimating $t_{fs}$ give the same dependence on
the parameters. However, the numerical factor is quite sensitive to 
the definition of this transition.

If the solution is initially radiative the transition from fast to
slow cooling and from a radiative hydrodynamics to adiabatic
hydrodynamics takes place at:
\begin{eqnarray}
t_{rad-ad} = 1.3 ~{ \rm days} E_{52}^{4/5} n_1^{3/5}
\epsilon_e^{7/5}\epsilon_B^{7/5} ((1+z)/2)^{12/5} \\ \nonumber
(\tge/0.5)^{14/5} c_t^{-12/5} (\ga_0/100)^{-4/5} .
\end{eqnarray}
During a radiative evolution the energy in the shock decreases with
time. The energy that appears in Eqs.  \ref{R_t} in the radiative
scalings is the initial energy. When a radiative shock switches to
adiabatic evolution, it is necessary to use the reduced energy to
calculate the subsequent adiabatic evolution. The energy $E_{f,52}$
which one should use in the adiabatic regime is related to the initial
$E_{i,52}$ of the fireball by
\begin{equation}
\label{finalE}
E_{f,52}=0.022\epsilon_B^{-3/5}\epsilon_e^{-3/5}
E_{i,52}^{4/5}\gamma_{100}^{-4/5}n_1^{-2/5}.
\end{equation}

If the shell is not spherical and it has an opening angle: $\theta$,
then the evolution will change when $\ga \sim \theta^{-1}$
\cite{Rhodas97}. Earlier on the jet expands  too rapidly to expand sideways
and it evolves as if it is a part of a spherical shell.  After this
stage the jet expands sideways and it accumulates much more mass and
slows down much faster. This transition will take place, quite
generally, during the adiabatic phase at:
\begin{equation}
t_\theta \approx 0.5 {\rm days} E_{52}^{1/3} n_1^{-1/3} ((1+z)/2)(\theta
/0.1)^{8/3}
 c_t^{-1}
\end{equation}

The shell eventually becomes non relativistic. This happens at: $R
\approx l = (4 E_0/4 \pi n_{1} m_p c^2)^{1/3}$ for an adiabatic
solution.  This corresponds to a transition at:
\begin{equation}
t_{NR,ad}\approx l/c \approx 300 ~{\rm\ days}~E_{52}^{1/3} n_1^{-1/3}  .
\end{equation}
A radiative shell loses energy faster and it becomes non relativistic
at $R=L=l/\ga_0^{1/3}= (4 E_0/4 \pi n_{ism} m_p c^2
\ga_0)^{1/3}$. This will take place at:
\begin{equation}
t_{NR,rad}\approx 65~{\rm\ days}~ E_{52}^{1/3} n_1^{-1/3}
(\ga_0/100)^{-1/3} .
\end{equation}
However, the earlier estimate of the transition from fast to slow
cooling suggests that the shell cannot remain radiative for such a long
time.

\subsection{Synchrotron Emission from a Relativistic Decelerating  Shell}

We proceed now to estimate the expected instantaneous spectrum and
light curve from a relativistic decelerating shell. The task is fairly
simple at this stage as all the ground rules have been set in the
previous sections. We limit the discussion here to a spherical shock
propagating into a homogeneous external matter. We consider two
extreme limits for the hydrodynamic evolution: fully radiative and
fully adiabatic.  If $\epsilon_e$ is somewhat less than unity during
the fast cooling phase $(t<t_{fs})$ then only a fraction of the
shock energy is lost to radiation. The scalings will be intermediate
between the two limits of fully radiative and fully adiabatic
discussed here.

\begin{figure}
\begin{center}
\includegraphics[width=9cm]{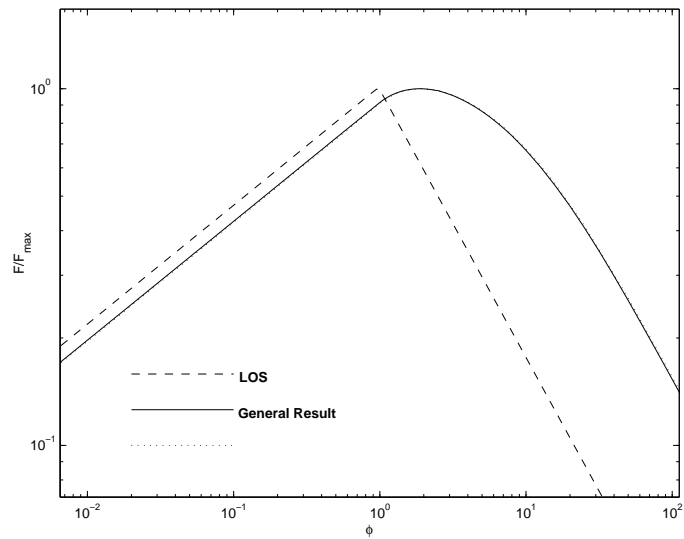}
\caption{\it  Calculated spectra or light curve from a Blandford-McKee
  solution. $F_\nu (t)$ is plotted as a function of $\phi \equiv
  Const. \times \nu t^{3/2}$. Thus if we consider a constant time this
  figure yields the spectrum while if we consider a fixed frequency it
  yields the light curve. The solid curve depictes emission from the
  full fireball. The  dashed line depictes the spectrum resultsing
   from emission  along the line of sight. From \cite{GPS98}.
}
\label{fig:spec_comp}
\end{center}
\end{figure}

For simplicity we assume that all the observed radiation reaches the
observer from the front of the shell and along the line of sight.
Actually to obtain the observed spectrum we should integrate over the
shell's profile and over different angles relative to the line of
sight. A full calculation \cite{GPS98} of the integrated spectrum
over a Blandford-McKee profile shows that  
that different radial points from which the radiation reaches the 
observers simultaneously  conspire to have practically the same  
 synchrotron frequency and therefore they emit the same spectrum.
Hence the radial integration over  a Blandford-McKee profile 
does not change the observed spectrum (note that this result
differs from the calculation for a homogeneous shell \cite{Waxman97c}). 
On the other hand  the contribution 
from angles away from the line of sight is important and it shapes
the observed spectrum, the light curve and the shape of the Afterglow (see Fig
\ref{fig:spec_comp}).

The instantaneous synchrotron spectra from a relativistic shock were
described in section \ref{sub_sec:syn_spectra}. They do  not depend on
the hydrodynamic evolution but rather on the instantaneous conditions
at the shock front, which determines the break energies $\nu_c$ and
$\nu_m$. The only assumption made is that the shock properties are
fairly constant over a time scale comparable to the observation time
$t$.

Using the adiabatic shell conditions (Eqs. \ref{R_t}-\ref{g_t}), Eqs.
\ref{nr3} for the shock conditions, Eq. \ref{syn_obs} for the
synchrotron energy and Eq. \ref{ga_c} for the ``cooling energy'' we
find:
\begin{eqnarray}
\label{abreaks}
\nu_c            & = & 2.7 \times 10^{12}~{\rm \ Hz} ~ \epsilon_B^{-3/2}
E_{52}^{-1/2} n_1^{-1} t_d^{-1/2}
                             , \cr
\nu_m           & = & 5.7 \times 10^{14} ~ {\rm \ Hz}~ \epsilon_B^{1/2}
\epsilon_e^2 E_{52}^{1/2} t_d^{-3/2}, \cr
F_{\nu,max}& = & 1.1 \times 10^5 ~ \mu{\rm J} ~ \epsilon_B^{1/2} E_{52}
n_1^{1/2} D_{28}^{-2} \ ,
\end{eqnarray}
where $t_d$ is the time in days, $D_{28}=D/10^{28}$ cm and we have
ignored cosmological redshift effects. Fig. \ref{fig:syn_spec} depicts the
instantaneous spectrum in this case.

For a fully radiative evolution we find:
\begin{eqnarray}
\label{rbreaks}
\nu_c   & = & 1.3 \times 10^{13} ~ {\rm \ Hz}~\epsilon_B^{-3/2}
  E_{52}^{-4/7} \gamma_{100}^{4/7}  n_1^{-13/14} t_d^{-2/7}, \cr
\nu_m   & = & 1.2 \times 10^{14}~ {\rm \ Hz}~ \epsilon_B^{1/2}
\epsilon_e^2 E_{52}^{4/7} \gamma_{100}^{-4/7} n_1^{-1/14}
t_d^{-12/7} , \cr
F_{\nu,max}& = & 4.5 \times 10^3  \mu{\rm J} \epsilon_B^{1/2} E_{52}^{8/7}
\gamma_{100}^{-8/7} n_1^{5/14} D_{28}^{-2}  t_d^{-3/7} \  ,
\end{eqnarray}
where we have scaled the initial Lorentz factor of the ejecta by a
factor of 100: $\gamma_{100}\equiv\gamma_0/100$.  These instantaneous
spectra are  also shown in Fig. \ref{fig:syn_spec}.

\subsubsection{Light Curves}
\label{sub_sec:lightcurves}

The light curves at a given frequency depend on the temporal
evolution of the break frequencies $\nu_m$ and $\nu_c$ and the peak
power $N_e P_{syn} (\ga_{e,min})$ (see Eq. \ref{Pslow-cool}).  These
depend, in turn, on how $\gamma$ and $N_e$ scale as a function of $t$.

\begin{figure}
\begin{center}
\includegraphics[width=9cm]{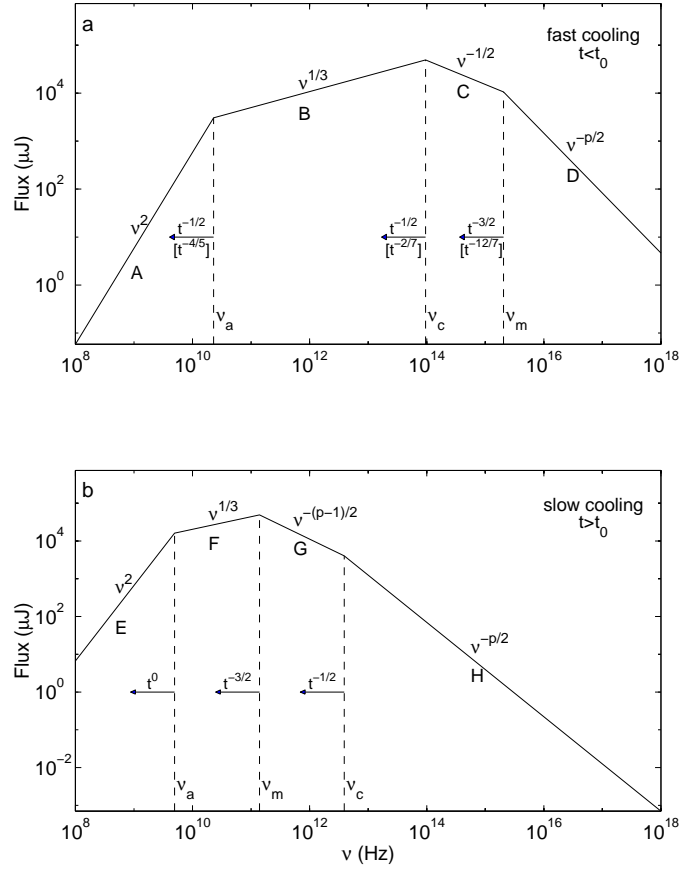}
\caption{\it   Synchrotron spectrum of a relativistic shock with a power-law
  electron distribution. (a) Fast cooling, which is expected at early
  times ($t<t_0$).  The spectrum consists of four segments, identified
  as A, B, C, D.  Self-absorption is important below $\nu_a$. The
  frequencies, $\nu_m$, $\nu_c$, $\nu_a$, decrease with time as
  indicated; the scalings above the arrows correspond to an adiabatic
  evolution, and the scalings below, in square brackets, to a fully
  radiative evolution. (b) Slow cooling, which is expected at late
  times ($t>t_0$). The evolution is always adiabatic. The four
  segments are identified as E, F, G, H.  From \cite{SPN98a}.}
\label{fig:syn_spec}
\end{center}
\end{figure}

The spectra presented in Fig. \ref{fig:syn_spec} show the positions of
$\nu_c$ and $\nu_m$ for typical parameters. In both the adiabatic and
radiative cases $\nu_c$ decreases more slowly with time than $\nu_m$.
Therefore, at sufficiently early times we have $\nu_c<\nu_m$,
i.e. fast cooling. At late times we have $\nu_c>\nu_m$, i.e., slow
cooling. The transition between the two occurs when $\nu_c=\nu_m$ at
$t_{fs}$ (see Eq. \ref{tfc}).  At $t=t_{fs}$, the spectrum changes
from fast cooling (Fig.
\ref{fig:syn_spec}a) to slow cooling (Fig. \ref{fig:syn_spec}b).  In addition,
if $\epsilon_e \approx 1$, the hydrodynamical evolution changes from
radiative to adiabatic. However, if $\epsilon_e \ll 1$, the evolution
remains adiabatic throughout.

Once we know how  the break frequencies, $\nu_c$, $\nu_m$, and the
peak flux $F_{\nu,max}$ vary with time, we can calculate the light
curve. Consider a fixed frequency (e.g. $\nu=10^{15}\nu_{15}$ Hz).
From the first two equations in (\ref{abreaks}) and (\ref{rbreaks}) we
see that there are two critical times, $t_c$ and $t_m$, when the break
frequencies, $\nu_c$ and $\nu_m$, cross the observed frequency $\nu$:
\begin{equation}
t_c=\cases{
7.3 \times 10^{-6} ~{\rm days}~\epsilon_B^{-3} E_{52}^{-1} n_1^{-2}
\nu_{15}^{-2} ,
         & ad, \cr
2.7 \times 10^{-7} ~{\rm days}~\epsilon_B^{-21/4} E_{52}^{-2} \gamma_{100}^{2}
n_1^{-13/4} \nu_{15}^{-7/2} ,
         & rad,}
\end{equation}
\begin{equation}
\label{tmin}
t_m=\cases{
0.69 ~{\rm days}~\epsilon_B^{1/3} \epsilon_e^{4/3} E_{52}^{1/3}
\nu_{15}^{-2/3} ,
         & ad, \cr
0.29 ~{\rm days}~\epsilon_B^{7/24} \epsilon_e^{7/6} E_{52}^{1/3}
\gamma_{100}^{-1/3}
\nu_{15}^{-7/12} n_1^{-1/24} ,
         & ra.}
\end{equation}

There are only two possible orderings of the three critical times,
$t_c$, $t_m$, $t_{fs}$, namely $t_{fs}>t_m>t_c$ and $t_{fs}<t_m<t_c$. We 
define the critical frequency, $\nu_0=\nu_c(t_{fs})=\nu_m(t_{fs})$:
\begin{equation}
\label{nu0}
\nu_0=\cases{ 1.8 \times 10^{11} \epsilon_B^{-5/2} \epsilon_e^{-1}
E_{52}^{-1} n_1^{-3/2} \ {\rm Hz}, & ad, \cr 8.5 \times 10^{12}
\epsilon_B^{-19/10} \epsilon_e^{-2/5} E_{52}^{-4/5}\gamma_{100}^{4/5}
n_1^{-11/10} \ {\rm Hz}, & rad.}
\end{equation}
When $\nu>\nu_0$, we have  $t_{fs}>t_m>t_c$  and we refer to
the corresponding light curve as the {\it high frequency light curve}.
Similarly, when $\nu<\nu_0$, we have $t_{fs}<t_m<t_c$, and we obtain the
{\it low frequency light curve}.

\begin{figure}
\begin{center}
\includegraphics[width=10cm]{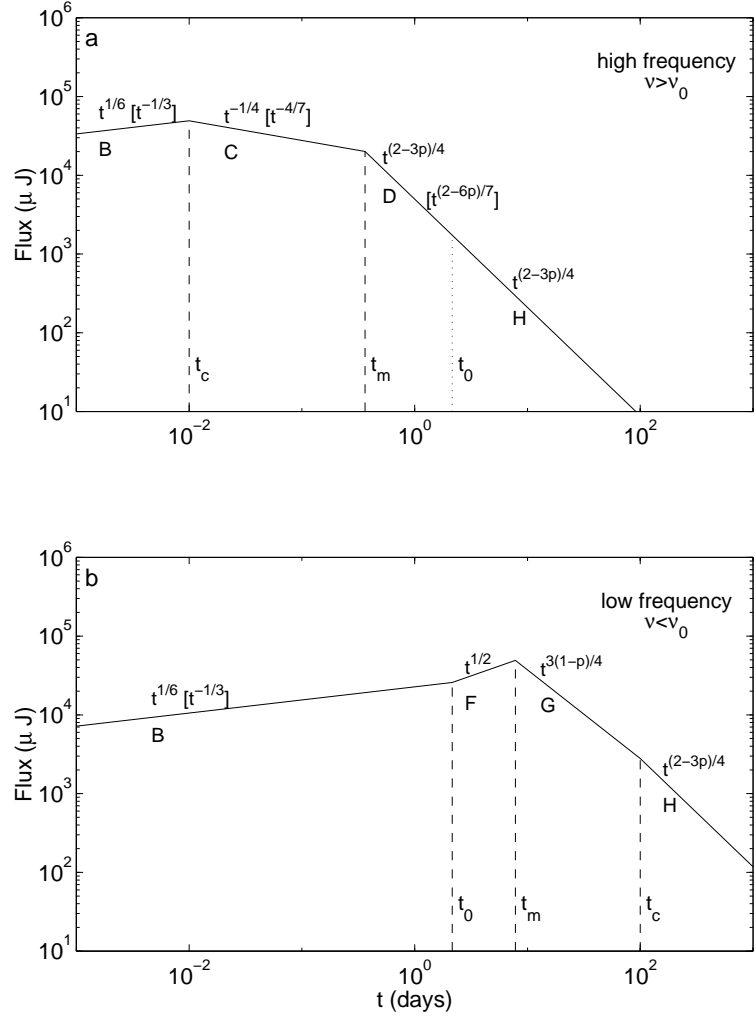}
\caption{\it Light curve due to synchrotron radiation from a spherical
  relativistic shock, ignoring the effect of self-absorption.  (a) The
  highfrequency case ($\nu>\nu_0$). The light curve has four segments,
  separated by the critical times, $t_c$, $t_m$, $t_0$.  The labels,
  B, C, D, H, indicate the correspondence with spectral segments in
  Fig.  \ref{fig:syn_spec}. The observed flux varies with time as
  indicated; the scalings within square brackets are for radiative
  evolution (which is restricted to $t<t_0$) and the other scalings
  are for adiabatic evolution. (b) The low frequency case
  ($\nu<\nu_0$). From \cite{SPN98a}. }
\label{fig:light_curve}
\end{center}
\end{figure}

Fig. \ref{fig:light_curve}a depicts a typical high frequency light curve.
At early times the electrons cool fast and $\nu < \nu_{m}$ and $\nu <
\nu_c$. Ignoring self absorption, the situation corresponds to segment
B in Fig. \ref{fig:syn_spec}, and the flux varies as $F_\nu\sim
F_{\nu,max}(\nu/\nu_c)^{1/3}$.  If the evolution is adiabatic,
$F_{\nu,max}$ is constant, and $F_\nu\sim t^{1/6}$. In the radiative
case, $F_{\nu,max}\sim t^{-3/7}$ and $F_\nu\sim t^{-1/3}$. The
scalings in the other segments, which correspond to C, D, H in Fig.
\ref{fig:syn_spec}, can be derived in a similar fashion and are shown in
Fig. \ref{fig:light_curve}a.

Fig. \ref{fig:light_curve}b shows the low frequency light curve,
corresponding to $\nu<\nu_0$. In this case, there are four phases in
the light curve, corresponding to segments B, F, G and H. The time
dependences of the flux are indicated on the plot for both the
adiabatic and the radiative cases. 

For a relativistic electron distribution with a power distribution
$\gamma^{-p}$ the uppermost spectral part behaves like
$\nu^{-p/2}$. The corresponding temporal index (for adiabatic
hydrodynamics) is $-3p/4$. In terms of the spectral index $\alpha$,
this yields the relation $F_\nu \propto t^{(1-3\alpha)/2}$.
Alternatively for slow cooling there is also another frequency range
(between $\nu_m$ and $\nu_c$) for which the spectrum is given by
$\nu^{-(p-1)/2}$ and the temporal decay is $-3(p-1)/4$.  Now we have
$F_\nu \propto t^{-3\alpha/2}$.  Note that in both cases there is a
specific relation between the spectral index and the temporal index
which could be tested by observations.

\subsubsection{Parameter Fitting for GRBs from Afterglow Observation and
  GRB970508}.
\label{sub_sec:GRB970508}

\begin{figure}
\begin{center}
\includegraphics[width=10cm]{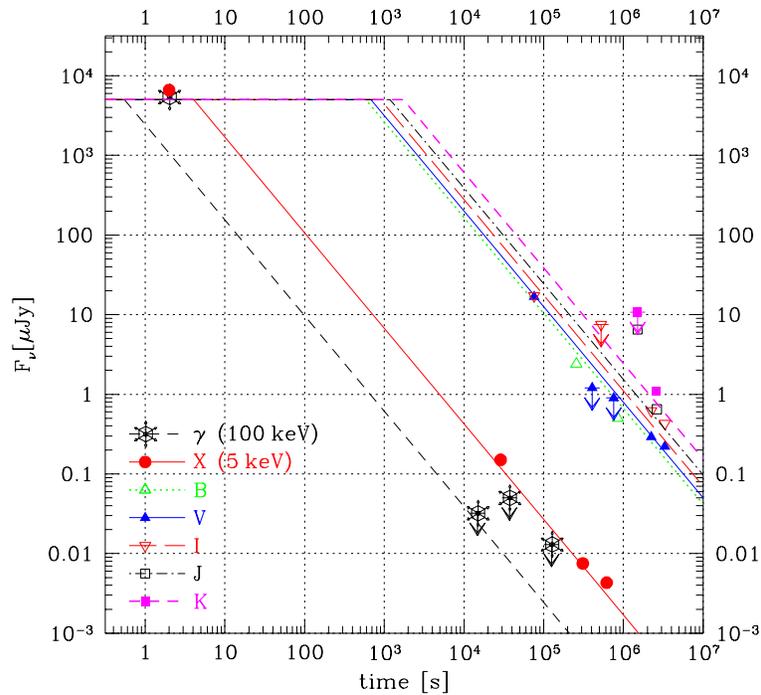}
\caption{\it Decline in the afterglow of GRB970228 in different wavelength and
theoretical model, from \cite{Wiejers_MR97}}
\label{afterglow_decline}
\end{center}
\end{figure}

Shortly after the observation of GRB970228 M\'esz\'aros \etall
\cite{Wiejers_MR97} showed that the decline in the intensity in
X-ray and several visual bands (from B to K) fit  the afterglow
model well (see fig. \ref{afterglow_decline}). The previous discussion
indicates that this agreement shows that the high energy tail (or late
time behavior) is produced by a synchrotron emission from a power law
distribution. 

\begin{figure}
\begin{center}
\includegraphics[width=10cm]{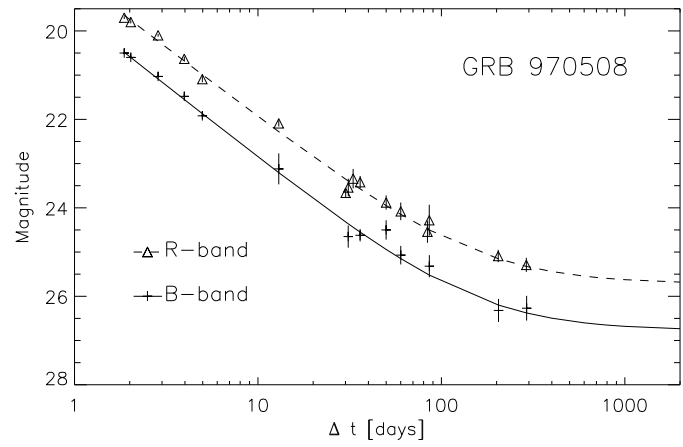}
\caption{\it Decay of the optical  afterglow in , GRB9700508.
A clear transition from a power law decay to a constant can be seen (from
\cite{Bloom98b}.}
\label{afterglow_optical_0508}
\end{center}
\end{figure}

\begin{figure}
\begin{center}
\includegraphics[width=10cm,angle=-90]{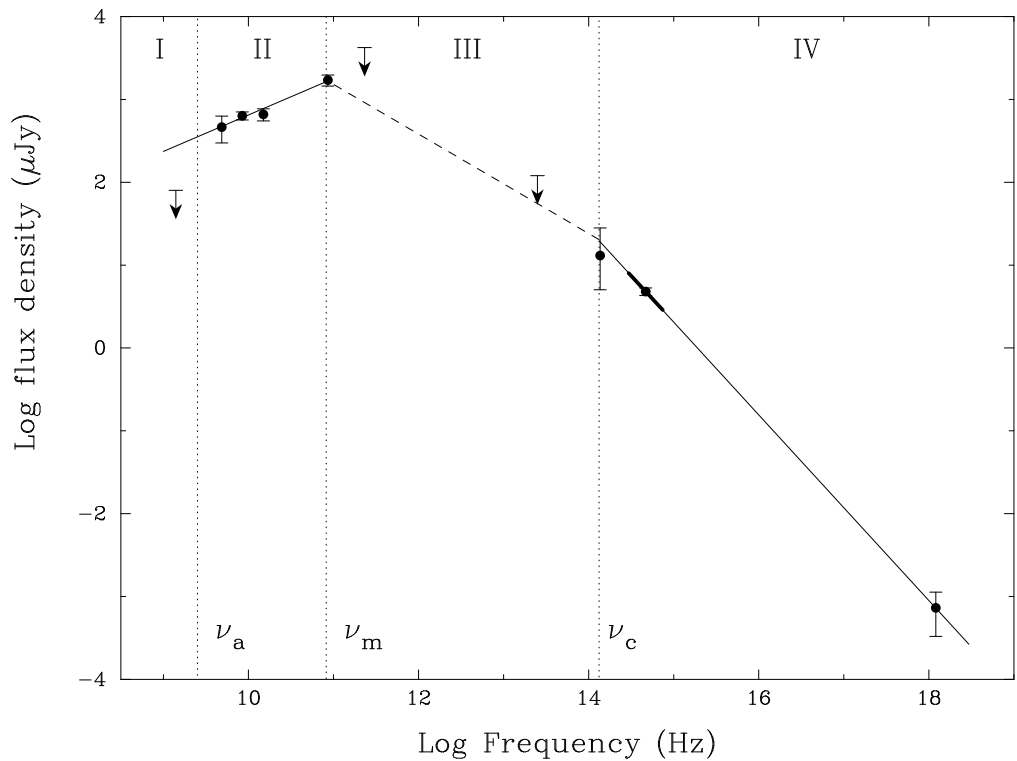}
\caption{\it The X-ray to radio spectrum of GRB 970508 on May 21.0 UT (12.1
days after the event). The fit to the low-frequency part,
$\alpha_{\rm 4.86-86 GHz}$ = 0.44
$\pm$ 0.07, is shown as well as the extrapolation from X-ray to
optical (solid lines). The local optical slope (2.1--5.0 days after
the event) is indicated by the thick solid line. Also indicated is
the extrapolation $F_{\nu} \propto
\nu^{-0.6}$ (lines). Indicated are the rough estimates of the break
frequencies $\nu_{\rm a}$, $\nu_{\rm m}$ and $\nu_{\rm c}$ for May
21.0 UT from \cite{Galama98c}}
\label{spec_0508}
\end{center}
\end{figure}

There are  much more data on the afterglow of GRB970508.  
The light curves in the different optical bands generally peak
around two days. There is a rather steep rise before the peak which is
followed by a long power law decay (see figs. 
\ref{afterglow_optical},\ref{afterglow_optical_0508}).
In the optical band 
the observed power law decay for GRB970508 is $-1.141 \pm 0.014$ 
\cite{Galama98b}. 
This implies for an adiabatic slow cooling model (which M\'esz\'aros
\etall \cite{Wiejers_MR97} use) a spectral index
$\alpha = -0.761 \pm 0.009$.  However, the observed spectral index is
$\alpha = - 1.12 \pm 0.04$ \cite{Galama98c}.  A fast cooling model,for
which the spectral index is $p/2$ and the temporal behavior is
$3p/4-1/2$, fits the data better as the temporal power law implies
that $p=2.188 \pm 0.019$ while the spectral index implies,
consistently, $p=2.24 \pm 0.08$ \cite{SPN98a,Galama98a} (see
Fig. \ref{spec_0508}.  Unfortunately, at present this fit does not
tell us much about the nature of the hydrodynamical processes and the
slowing down.

Using both the optical and the radio data on can try to fit the whole
spectrum and to obtain the unknown parameters that determine the
fireball evolution \cite{Wijers_Galama98,jgranot98b}. Wijers and
Galama \cite{Wijers_Galama98} have attempted to do so using the
spectrum of GRB970508.  They have obtained a reasonable set of
parameters.  However, more detailed analysis \cite{jgranot98b} reveals
that the solution is very sensitive to assumptions made on how to fit
the observational data to the theoretical curve. Moreover, the initial
phase of the light curve of GRB970508 does not fit any of the
theoretical curves. This suggests that at least initially an
additional process might be taking place.  Because of the inability to
obtain a good fit for this initial phase there is a large uncertainty
in the parameters obtained in this way.

No deviation in the observed decaying light curve from a single power
law was observed for GRB970508, until it faded below the level of
the surrounding nebula.  This suggests that there was no
significant beaming in this case. If the outflow is in the form of
a jet the temporal behavior will change drastically when the opening
angle of the jet equals $1/\gamma$ \cite{Rhodas97}.

\subsection{New Puzzles from Afterglow observations}

Afterglow observations fit well the fireball picture that was
developed for explaining the GRB phenomena. The available data is not
good enough to distinguish between different specific models. But in
the future we expect to be able to distinguish between those models
and even to be able to determine the parameters of the burst  $E$
and $\ga_0$ (if the data is taken early enough), the surrounding ISM
density and the intrinsic parameters of the relativistic shock
$\epsilon_e$, $\epsilon_B$ and $p$.  Still the current data is
sufficient to raise new puzzles and present us with new questions.

\begin{itemize}
\item{\bf Why  afterglow accompany some GRBs and not others?}

X-ray, Optical and radio afterglows have been observed in some bursts
but not in others. According to the current model afterglow is
produces when the ejecta that produced the GRB is shocked by the
surrounding matter.  Possible explanations to this puzzle invoke
environmental effects.  A detectable afterglow might be generated
efficiently  in some range of ISM densities and inefficiently in
another.  High ISM densities would slow down of the ejecta more
rapidly. This could make some afterglows detectable and others
undetectable.  ISM absorption is another alternative. While most
interstellar environments are optically thin to gamma-rays high
density ISM regions can absorb and attenuate efficiently x-rays and
optical radiation.

\item {\bf Jets and the Energy of GRB971214}

How can we explain the $ 10^{53}$ergs required for isotropic emission
in GRB971214?  As we discuss in the next section this amount is
marginal for most models that are based on the formation of a compact
source.  This problem can be resolved if we invoke beaming, with
$\theta \sim 0.1$.  However, such beaming would results in a break at
the light curve when the local Lorentz factor would reach a value of
$1/\theta$. Such a break was not seen in other afterglows for which
there are good data.  Note that recently Perna \& Loeb \cite{Perna_Loeb98}
inferred from the lack of radio transients that GRB beamns cannot be
very narrow. If typical GRBs are
beamed, the beam width $\theta$ should be larger than $6^o$.

\item{\bf GRB980425 and SN1998bw}

SN1998bw (and the associated GRB980425) is a factor of a hundred
nearer than a typical GRB (which are expected to be at $z \sim
1$). The corresponding (isotropic) gamma-ray energy, $\sim 5 \times
10^{47}$ergs, is four order of magnitude lower than a regular
burst. This can be in agreement with the peak flux distribution only
if the bursts with such a low luminosity compose a very small fraction
of GRBs. This leads naturally to the question is there an
observational coincidence between GRBs and SNs?  To which there are
conflicting answers \cite{Wang_Wheeler,Kippne98,Bloom98c}.

\end{itemize}

\section{Models of the Inner Engine}
\label{sec:models}

We turn now to the most difficult part: the nature of the beast that
produces the GRB-modeling of the Inner Engine. We examine a few
general considerations in section \ref{subsec:innerengine} and then we
turn to the Binary Neutron star merger model in \ref{sec:ns2m}.

\subsection{The ``Inner Engine''}
\label{subsec:innerengine}

The fireball model is based on an ``inner engine'' that supplies the
energy and accelerates the baryons.  This ``engine" is well hidden
from direct observations and it is impossible to determine what is it
from current observations. Unfortunately, the discovery of afterglow
does not shed  additional direct light on this issue. However it adds some
indirect evidence from the association of the 
location of the bursts in star forming regions.

Once the cosmological origin of GRBs was established we had two  
direct clues on the nature of the ``inner engine'': the rate
and the energy output. GRBs occur at a rate of about one per
$10^6$years per galaxy \cite{Pi92} and the total energy is $\sim
10^{52}$ergs. These estimates assume isotropic emission. Beaming with
an angle $\theta$ changes these estimates by a factor $4
\pi /\theta^2$ in the rate and $\theta^{2}/4\pi$ in the total 
energy involved.  These estimates are also
based on the assumption that the burst rate does not vary with cosmic
time.  The observations that GRB hosts are star forming galaxies
\cite{Djorgovski98a,Djorgovski98b,Bloom98b,Fruchter98,Hogg_Fruchter98} 
indicates that the rate of GRBs may follow the star formation rate
\cite{Totani,Sahuetal97a,Wijersetal98}. In this case the bursts are
further and they take place at a lower rate and have significantly
higher energy output.

The fireball model poses an additional constraint: the inner engine
should be capable of accelerating $\sim 10^{-5} M_\odot$ to
relativistic energies.  One can imagine various scenarios in which
$10^{52}$ergs are generated within a short time. The requirement that
this energy should be converted to a relativistic flow is much more
difficult as it requires a ``clean'' system with a very low but non
zero baryonic load. This requirement suggests a preference for  models
based on electromagnetic energy transfer or electromagnetic energy
generation as these could more naturally satisfy this condition (see
\cite{Usov92,Thom,Katz97,MR97a}). Paczy\'nski \cite{Pac97}
has recently suggested a unique hydrodynamical model in which
$10^{54}$ergs are dumped into an atmosphere with a decreasing density
profile. This is a cosmological variant of Colgate's \cite{Colgate74}
galactic model. 
This would lead to an acceleration of fewer and fewer baryons
and eventually to a relativistic velocities. Overall one could say
that the ``baryonic load'' problem is presently the most bothersome
open question in the ``fireball model''.

The recent realization that energy conversion is most likely via
internal shocks rather than via external shocks provides additional
information about the inner engine: The relativistic flow must be
irregular (to produce the internal shocks), it must be variable on a
short time scale (as this time scale is seen in the variability of the
bursts), and it must be active for up to a few hundred seconds and
possibly much longer \cite{KaPS97} - as this determines the observed
duration of the burst. These requirements rule out all explosive
models.  The engine must be compact ($\sim 10^7$cm) to produce the
observed variability and it must operate for a million light crossing
times to produce a few hundred-second signals.

There are more than a hundred GRB models
\cite{Nemirof}.  At a certain stage, before BATSE,  there were probably
more models than observed bursts.  Most of these models are, however,
galactic and those have been ruled out if we accept the cosmological origin of
GRBs. This leaves a rather modest list of viable GRB models: binary
neutron star mergers - NS$^2$Ms - \cite{Eichler89} (see also 
\cite{Belinikov,GDN,Pac86,Pac91,PNS}), failed
supernova \cite{Woosley93}, white dwarf collapse \cite{Usov92} and
hypernova \cite{Pac97}. All these are based on the formation of
a compact object of one type or another and the release of its binding
energy. With a binding energy of $\sim 5 \times 10^{53}$ ergs or higher, all
these models have, in principle, enough energy to power a GRB. However
they face  similar difficulties in channeling enough energy to a
relativistic flow. 
This would be particularly difficult if  indeed 
$10^{53}$ergs are needed, as some recent burst have indicated.
Paczy\'nski's hypernova is an exception as in this
model all the energy is channeled initially to a non-relativistic flow
and only later a small fraction of it is converted to relativistic
baryons.  All these models are consistent
with the possibility that GRBs are associated with star forming regions as
the life time of massive stars is quite short and even the typical life
time of a neutron star binary ($\sim 10^8$yr) is sufficiently short to 
allow for this coincidence. -

Other models are based on an association of GRBs with 
massive black holes associated with Quasars or AGNs
in galactic centers (e.g. \cite{Carter92}). These
are  ruled out as all  GRBs with optical afterglow
are not associated with such  objects.
Furthermore, such objects do not appear in other small GRB  error boxes
searched by Schaefer \etall \cite{Schaefer97}. From a theoretical
point of view it is difficult to explain the observed energy and time
scales with such objects.

\subsection{NS$^2$Ms: Binary Neutron Star Mergers}
\label{sec:ns2m}

Binary neutron star mergers (NS$^2$Ms) \cite{Eichler89} or, with a
small variant: neutron star-black hole mergers \cite{Pac91} are
probably the best candidates for GRB sources.  These mergers take
place because of the decay of the binary orbits due to gravitational
radiation emission.  A NS$^2$M results, most likely, in a rotating
black hole \cite{Dav}.  The process releases $\approx 5 \times
10^{53}$ ergs \cite{CE}.  Most of this energy escapes as neutrinos and
gravitational radiation, but a small fraction of this energy suffices
to power a GRB. The discovery of the famous binary pulsar PSR 1913+16
\cite{Hulse75} demonstrated that this decay is taking place \cite{Taywei}.
The discovery of other binary pulsars, and in particular of PSR
1534+12 \cite{Wol}, has shown that PSR 1913+16 is not unique and that
such systems are common. These observations suggest that NS$^2$Ms take
place at a rate of $\approx 10^{-6}$ events per year per galaxy
\cite{NPS,Phinney91,Heuvel_Lorimer}.  This rate is comparable to the 
simple estimate of the GRB event rate (assuming no beaming and no
cosmic evolution of the rate) \cite{Pi92,PNS,CP95}.

It has been suggested \cite{TY,Lipunov97} that most neutron star
binaries are born with very close orbits and hence with very short
lifetimes (see however, \cite{Yungelson98,Bethe_Brown}). If this idea
is correct, then the merger rate will be much higher. This will
destroy, of course, the nice agreement between the rates of GRBs and
NS$^2$Ms. Consistency can be restored if we invoke beaming, which
might even be advantageous as far as the energy budget is concerned.
Unfortunately, the short lifetime of those systems, which is the
essence of this idea means that at any given moment of time there are
only about a hundred such systems in the Galaxy (compared to about
$10^5$ wider neutron star binaries).This makes it very hard to confirm
or rule out this speculation.  We should be extremely lucky to detect
such a system.

It is not clear yet how NS$^2$Ms form. The question is how does the
system survive the second supernova event?  The binary system will be
disrupted it this explosion ejects more than half of its total mass.
There are two competing scenarios for the formation of NS$^2$Ms. In
one scenario the first neutron star that forms
sinks into the envelope of its giant companion and its motion within
this envelope lead to a strong wind that carries away most of the
secondary's mass.  When the secondary reaches core collapse it has only
a small envelope and the total mass ejected is rather small. In a
second scenario the second supernova explosion is asymmetric. The
asymmetric explosion gives a velocity of a few hundred km/sec to the
newborn neutron star. In a fraction of the cases this velocity is in
the right direction to keep the binary together. Such a binary system
will have a comparable center of mass velocity
\cite{NPP,Lyne_Lorimer94,White_vanP96,Fryer_Kalogera97}. This
second scenario has several advantages. First it explains both the
existence of binary neutron stars and the existence of high velocity
pulsars \cite{Lyne_Lorimer94,Cordes-Chernoff}. Second, and more
relevant to GRBs, with these kick velocities these binaries could
escape from their parent galaxy, provided that this galaxy is small
enough.  Such escaping binaries will travel a distance of $\sim
200~{\rm kpc}~(v/200~{\rm km/sec})~(T/10^9{\rm yr})$ before they
merge. The GRB will occur when the system is at a distance of the
order of hundred kpc from the parent galaxy.  Clearly there is no ``no
host'' problem in this case
\cite{NPP}.

While  a NS$^2$M has  enough energy available to power a
GRB it is not clear how  the GRB is produced.  A central question is,
of course, how does a NS$^2$M generate the relativistic wind required to
power a GRB. Most of the binding energy (which is around $5 \times
10^{53}$ergs escapes as neutrinos \cite{CE}.  Eichler \etall
\cite{Eichler89} suggested that about one thousandth of these
neutrinos annihilate and produce pairs that in turn produce gamma-rays
via $\nu \bar \nu \rightarrow e^+ e^- \rightarrow \gamma\gamma$.  This
idea was criticized on several grounds by different authors.
Jaroszynksi \cite{jaroszynksi96} pointed out that a large fraction of
the neutrinos will be swallowed by the black hole that forms.  Davies
\etall \cite{Dav} and Ruffert \& Janka
\cite{Ruffert_Janka96a,Ruffert_Janka96b,Ruffert_Janka97} who
simulated neutron star mergers suggested that the central object won't
be warm enough to produce a significant neutrino flux because the
merger is nearly adiabatic \cite{Katz97}. The neutrinos are also
emitted over a diffusion time of several seconds, too long to explain
the rapid variations observed in GRB \cite{Katz96}, but to short to
explain the observed GRB durations.  Wilson \& Mathews
\cite{Mathews_Wilson,Mathewsetal97} included approximate general
relativistic effects in a numerical simulation of a neutron star
merger. They found that the neutron stars collapse to a single black
hole before they collide with each other. This again will suppress the
neutrino emission from the merger.  However, the approximation that
they have used has been criticized by various authors and it is not
clear yet that the results are valid.  Others suggested that the
neutrino wind will carry too many baryons.  However, it seems that the
most severe problem with this model stems from the fact that the
prompt neutrino burst could produce only a single smooth pulse. This
explosive burst is incompatible with the internal shocks scenario.

An alternative source of energy within the NS$^2$M is the accretion
power of a disk that forms around the black hole
\cite{NPP,Katz97,MR97a}.  Various numerical simulations of neutron
star mergers
\cite{Dav,Ruffert_Janka96a,Ruffert_Janka96b,Ruffert_Janka97} find that
a $\sim 0.1 M_\odot$ forms around the central black hole.  Accretion
of this disk on the central black hole may take a few dozen seconds
\cite{Katz97}. It may produce the wind needed to produce internal
shocks that could produce, in turn a GRB.

How can one prove or disprove this, or any other, GRB model?
Theoretical studies concerning specific details of the model can, of
course, make it more or less appealing. But in view of the fact that
the observed radiation emerges from a distant region which is very far
from the inner ``engine" I doubt if this will ever be sufficient.  It
seems that the only way to confirm any GRB model will be via detecting
in time-coincidence another astronomical phenomenon, whose source
could be identified with certainty. Unfortunately while the recent afterglow
observations take us closer to this target
they do not tell us what are the sources of GRBs. We still
have to search for additional signals.

NS$^2$Ms have two accompanying signals, a neutrino signal and a
gravitational radiation signal. Both signals are extremely difficult
to detect. The  neutrino signal could be emitted by some of the 
other sources that are based on a core collapse. Furthermore, with 
present technology detection of neutrino signals from a cosmological
distance is impossible.
On the other hand the gravitational radiation signal has a
unique characteristic form. This provides a clear prediction of
coincidence that could be proved or falsified sometime in the not too
distant future when suitable gravitational radiation detectors will
become operational.

\subsection{Binary Neutron Stars vs. 
Black Hole - Neutron star Mergers}

We have grouped together binary neutron star mergers and black-hole
neutron star mergers. At present several neutron star binaries are
known while no black hole - neutron star binary was found. Still on
theoretical grounds one should expect a similar rate for both events
\cite{NPS}.  Some even suggest that there are more black hole - neutron 
star binaries than neutron star-neutron star binaries \cite{Bethe_Brown}.
There is a lot of similarity between the two processes which are both
driven by gravitational radiation emission and both result in a single
black hole.  First, unless the mass of the black hole is of rather
small the neutron star will not be tidally disrupted before it is
captured by the black hole. Even if such a tidal disruption will take
place then while in the binary neutron star merger we expect a
collision, in the black hole - neutron star merger we expect at most a
tidal disruption followed by infall of the debris on the black hole.

This could lead to a situation in which one of the two events
will produce a GRB and the other will not. Presently it is too
difficult to speculate which of the two is the right one. One should
recall, however, that there is a marked difference between the
gravitational signature of those events and thus hopefully when we
discover a coincidence between a GRB and a gravitational radiation
signal we would also be able to find which of the two mechanism is the
right one.

\section{Other Related Phenomena}
\label{sec:other}
It is quite likely that  other particles (in addition to $\gamma$-rays)
are emitted in these events. Let $f_{x\gamma}$ be the ratio of energy
emitted in other particles relative to $\gamma$-rays. These
particles will appear as a burst accompanying the GRB.  The total
fluence of a ``typical'' GRB observed by BATSE, $F_\gamma$ is
$10^{-7}$ergs/cm$^2$, and the fluence of a ``strong'' burst is about
hundred times larger.  Therefore we should expect accompanying bursts
with typical fluences of:
\begin{equation}
F_{x~|prompt} = 10^{-3} {{\rm particles}\over  {\rm cm}^2} f_{x\gamma}
\big({F_\gamma \over 10^{-7} {\rm ergs/cm}^2}\big)
\big({E_x \over {\rm GeV}}\big)^{-1} ,
\end{equation}
where $E_x$ is the energy of these particles.  This burst will be
spread in time and delayed relative to the GRB if the particles do not
move at the speed of light. Relativistic time delay will be
significant (larger than 10 seconds) if the particles are not massless
and their Lorentz factor is smaller than $10^{8}$!  similarly a
deflection angle of $10^{-8}$ will cause a significant time delay.

In addition to the prompt burst we should expect a continuous
background of these particles.  With one $10^{51}$ergs GRB per $10^6$
years per galaxy we expect $~\sim 10^4$ events per galaxy in a Hubble
time (provided of course that the event rate is constant in
time). This corresponds to a background flux of
\begin{eqnarray}
F_{x~|bg} = 3 \cdot 10^{-8} {
{\rm particles}\over {\rm cm}^2{\rm sec}} f_{x\gamma}
\big ({E_\gamma \over 10^{51} {\rm ergs}}\big)  \\ \nonumber 
\big({R \over 10^{-6} {\rm y/galaxy}}\big)
\big({E_x \over {\rm GeV}}\big)^{-1} .
\end{eqnarray}

For any specific particle that could be produced one should calculate
the ratio $f_{x\gamma}$ and then compare the expected fluxes with
fluxes from other sources and with the capabilities of current
detectors. One should distinguish between two types of predictions:
(i) Predictions of the generic fireball model which include low energy
cosmic rays \cite{SP}, UCHERs \cite{Waxman95a,Vietri95,UM} and high
energy neutrinos \cite{Waxman_Bahcall} and (ii) Predictions of
specific models and in particular the NS$^2$M model. These include low
energy neutrinos \cite{CE} and gravitational waves
\cite{Eichler89,Kochaneck_Piran}.

\subsection{Cosmic Rays}

Already in 1990, Shemi \& Piran \cite{SP} pointed out that fireball
model is closely related to Cosmic Rays.  A ``standard'' fireball
model involved the acceleration of $\sim 10^{-7} M_\odot$ of baryons
to a typical energy of 100GeV per baryon.  Protons that leak out of
the fireball will become low energy cosmic rays. However, a comparison
of the GRB rate (one per $10^6$ years per galaxy) with the observed
low energy cosmic rays flux, suggests that even if $f_{CR-\gamma}
\approx 1$ this will amount only to 1\% to 10\% of the observed cosmic
ray flux at these energies. Cosmic rays are believed to be produced by
SNRs. Since supernovae are ten thousand times more frequent than GRBs,
unless GRBs are much more efficient in producing Cosmic Rays in some
specific energy range their contribution will be swamped by the SNR
contribution.

\subsection{UCHERs - Ultra High Energy Cosmic Rays}

Waxman \cite{Waxman95a} and Vietri \cite{Vietri95} have shown that the
observed flux of UCHERs (above $10^{19}$eV) is consistent with the
idea that these are produced by the fireball shocks provided that
$f_{UCHERs-\gamma} \approx 1$.  Milgrom \& Usov \cite{UM} pointed out
that the error boxes of the two highest energy UCHERs contain strong
GRBs - suggesting an association between the two phenomena.  The
relativistic fireball shocks that appear in GRBs are among the few
astronomical objects that satisfy the conditions for shock
acceleration of UCHERs.  Waxman \cite{Waxman95b} has shown that the
spectrum of UCHERs is consistent with the one expected from Fermi
acceleration within those shocks.

\subsection{High Energy Neutrinos}
Waxman \&  Bahcall \cite{Waxman_Bahcall} suggested that collisions
between protons and photons within the relativistic fireball shocks
produce pions.  These pions produce high energy neutrinos with $E_\nu
\sim 10^{14}$eV and $f_{{\rm high~energy}~\nu-\gamma} > 0.1$.  The
flux of these neutrinos is comparable to the flux of atmospheric
neutrinos but those will be correlated with the position of strong
GRBs. This signal might be detected in future km$^2$ size neutrino
detectors.

\subsection{Gravitational Waves}

If GRBs are associated with NS$^2$Ms then they will be associated with
gravitational waves and low energy neutrinos.  The spiraling in phase
of a NS$^2$M produces a clean chirping gravitational radiation
signal. This signal is the prime target of LIGO 
\cite{Abramovichi} and VIRGO, the two \cite{Bardachia}
large interferometers that are build now in the USA and in Europe. The
observational scheme of these detectors is heavily dependent on
digging deeply into the noise.  Kochaneck \& Piran
\cite{Kochaneck_Piran} suggested that a coincidence between a
chirping gravitational radiation signal from a neutron star merger and
a GRB could enhance greatly the statistical significance of the
detection of the gravitational radiation signal. At the same time this
will also verify the NS$^2$M GRB model.

\subsection{Low Energy Neutrinos}

Most of the energy generated in any core collpase event and in
particular in NS$^2$M is released as low energy ($\sim 5-10$MeV)
neutrinos \cite{CE}. The total energy is quite large $\sim$ a few $
\times 10^{53}$ergs, leading to $f_{{\rm low~energy}~\nu-\gamma}
\approx 10$. However, this neutrino signal will be quite similar to a
supernova neutrino signal, and at present only galactic SN neutrinos
can be detected.  Supernovae are ten thousand times more frequent then
GRBs and therefore low energy neutrinos associated with GRB constitute
an insignificant contribution to the background at this energy range.

\subsection{Black Holes} 
An NS$^2$M results, inevitably, in a black hole \cite{Dav}.  Thus a
direct implication of the NS$^2$M model is that GRBs signal to us
(indirectly) that a black hole has just (with the appropriate time of
flight in mind) formed.

\section{Cosmological Implications}
\label{sec:cosm}
Cosmological GRBs seem to be a relatively homogeneous population of
sources with a narrow luminosity function (the peak luminosity of GRBs
varies by less than a factor of 10 \cite {CP95,HE}) that is located at
relatively high redshifts \cite{Pi92,MP,Dermer92,Wic,CP95}.  The universe and
our Galaxy are transparent to MeV $\gamma$-rays (see e.g.
\cite{ZS89}). Hence GRBs constitute a unique homogeneous population of
sources which does not suffer from any angular distortion due to
absorption by the Galaxy or by any other object.  Could GRBs be the
holy grail of Cosmology and provide us with the standard candles
needed to determine the cosmological parameters $H_0$, $\Omega$, and
$\Lambda$?  Lacking any spectral feature, there is no indication of
the redshift of individual bursts. The available number vs.  peak
luminosity distribution is is not suitable to distinguish between
different cosmological models even when the sources are perfect
standard candles with no source evolution \cite{CP95}.

The situation might be different if optical afterglow observations
would yield an independent redshift measurement of a large number of
bursts. If the GRB luminosity function is narrow enough this might
allow us, in the future, to determine the cosmological closure
parameter $\Omega$ using a peak-flux vs. red-shift diagram (or the
equivalent more common magnitude - red-shift diagram).  For example a
hundred bursts with a measured $z$ are needed to estimate $\Omega$
with an accuracy of $\sigma_{\Omega} = 0.2$, if
${\sigma_L / L}=1$ \cite{CP97a}.

Currently, the rate of detection of bursts with counter-parts is a few
per year and of those detected until now only two have a measured
red-shift. This rate is far too low for any cosmological
measurement. However, there is an enormous potential for
improvements. For example, systematic measurements of the red-shift of
all bursts observed by BATSE ($\approx 300$ per year) would yield an
independent estimate of $\Omega$, with $\sigma_\Omega=0.1$, even if
the luminosity function is wide, (${\sigma_L / L}=0.9$), within one
year.

Direct redshift measurements would also enable us to
determine  the cosmological evolution of the rate of GRBs
\cite{CP97a}. Most current cosmological GRB models suggest that
the GRB rate   follows (with a rather short time delay) the rate
of star formation \cite{Livio_hunts}. Consequently measurements of the
rate of GRBs as a function of the red-shift will provide an
independent tool to study star formation and galactic evolution.

It is also expected that the bursts' sources follow the matter
distribution.  Then GRBs can map the large scale structure of the
Universe on scales that cannot be spanned directly otherwise. Lamb \&
Quashnock \cite{LQ93} have pointed out that a population of several
thousand cosmological bursts should show angular deviations from
isotropy on a scale of a few degrees.  This would immediately lead to
new interesting cosmological limits. So far there is no detected
anisotropy in the 1112 bursts of the BATSE 3B catalog \cite{Teg}. But
the potential of this population is clear and quite promising. A more
ambitious project would be to measure the multipole moments of the GRB
distribution and from this to estimate cosmological parameters
\cite{Piran_Singh97}. However, it seems that too many bursts are
required to overcome the signal to noise ratio in such measurements.

GRBs can also serve to explore cosmology as a background population
which could be lensed by foreground objects \cite{Pac86}. While
standard gravitational lensed object appears as several images of the
same objects, the low angular resolution of GRB detectors is
insufficient to distinguish between the positions of different images
of a lensed GRB. However, the time delay along the different lines of
sight of a gravitationally lensed burst will cause such a burst to
appear as repeated bursts with the same time profile but different
intensities from practically the same position on the sky.    Mao
\cite{Mao} estimated that the probability for lensing of a GRB by a regular
foreground galaxy is 0.04\%-0.4\%.  Hence the lack of a confirmed
lensed event so far \cite{Nem94} is not problematic yet.  In the
future, the statistics of lensed bursts could probe the nature of the
lensing objects and the dark matter of the Universe
\cite{Blase92}. The fact that no lensed bursts have been  detected so far
is sufficient to rule out a critical density ($\Omega=1$) of $10^{6.5}
M_\odot$ to $ 10^{8.1} M_\odot$ black holes
\cite{Nemiroff93a}. Truly, this was not the leading candidate for
cosmological dark matter. Still this result is a demonstration of the
power of this technique and the potential of GRB lensing.  The
statistics of lensing depends on the distance to the lensed objects
which is quite uncertain at present.  The detection of a significant
number of counterparts whose red-shift could be measured would improve
significantly this technique as well.

\section{Summary and Conclusions}

Some thirty years after the discovery of GRBs a generic GRB model is
beginning to emerge.  The observations of isotropy, peak flux
distribution and time dilation indicated that GRBs are
cosmological. The measurement of a redshift provided a final
confirmation for this idea. All cosmological models are based on the
fireball mode. The discovery of the afterglow confirmed this general
scheme.

The Fireball-Internal-External Shocks model seems to have the
necessary ingredients to explain the observations.  Relativistic
motion, which is the key component of all Fireball Model provided the
solution for the compactness problem. The existence of such motion was
confirmed by the radio afterglow observations in GRB970508.  Energy
conversion via internal shocks can produce the observed highly
variable light curves while the external shock model agrees, at least
qualitatively, with afterglow observations.  There are some
indications that the two kind of shocks might combine and operate
within the GRB itself producing two different components of the
signal.  This fireball model has some fascinating immediate
implications on accompanying UCHER and high energy neutrino signals.
An observations of these phenomena in coincidence with a GRB could
provide a final confirmation of this model.

In spite of this progress we are still far from a complete
solution. There are many open questions that has to be
resolved. Within the internal-external shocks model there is a nagging
efficiency problem in conversion of the initial kinetic energy to the
observed radiation.  If the overall efficiency is too low the initial
energy required might be larger than $10^{53}$ ergs and it is
difficult to imagine a source that could provide so much energy.
Beaming might provide a solution to this energy crisis. However, so
far there is no indication for the corresponding break in the
afterglow light curve, which is essential in any relativistic beaming
model when $\gamma \sim \theta^{-1}$.  These last two facts might be
consistent if GRB970508 took place within a very low density ISM - an
issue that should be explored further. Afterglow observations agree
qualitatively, but not quantitatively with the model.  Better
observations and more detailed theoretical modeling are needed.
Another nagging open question is what determines the appearance of
afterglow. Why there was no X-ray afterglow in the very strong 970111?  
Why was optical afterglow observed in GRB970228 and in GRB970508
but it was not seen in others (in particular in GRB970828
\cite{Pac97})? Finally we turn to the GRB itself and wonder why is the
observed radiation always in the soft $\gamma$-ray band? Is there an
observational bias? Are there other bursts that are not observed by
current instruments? If there are none and we do observer all or most
of the bursts why is the emitted radiation always in the soft
gamma-ray range?  Why it is insensitive to a likely variability in the
Lorentz factors of the relativistic flow and to variability of other
parameters in the model.

While there are many open questions concerning the fireball and the
radiation emitting regions the first and foremost open question
concerning GRBs is what are the inner engines that power GRBs?  In
spite of all the recent progress we still don't know what produces
GRBs.  My personal impression is that binary neutron mergers are the
best candidates. But other models that are based on the formation of a
compact object and release a significant amount of its binding energy
on a short time scale are also viable.  A nagging question in all
these models is what produces the the ``observed'' ultra-relativistic
flow? How are $\sim 10^{-5} M_\odot$ of baryons accelerated to an
ultra-relativistic velocity with $\gamma \sim 100$ or larger? Why is
the baryonic load so low? Why isn't it lower?  There is no simple
model for that. An ingenious theoretical idea is clearly needed here.

However, I believe that theoretical reasoning won't be enough and only
observations can provide a final resolution of the questions what is
are the sources of GRBs? The binary neutron star merger model has one
specific observational prediction: A coincidence between a (near by and
therefore strong) GRB and a characteristic gravitational radiation signal. Luckily
these events have a unique gravitational radiation signature. The
detection of these gravitational radiation events is the prime target
of three gravitational radiation detector that are being built now.
Hopefully they will become operational within the next decade and their
observations might confirm or rule out this model. Such 
predictions, of an independently observed phenomena are clearly needed
for all other competing models.

GRBs seem to be the most relativistic phenomenon discovered so far.
They involve a macroscopic relativistic motion not found elsewhere 
before. As cosmological objects they display numerous  relativistic
cosmological phenomena. According the the NS$^2$M model they are
associated with the best sources for gravitational radiation emission
and more than that they signal, though in directly, the formation
of a new black hole.

I thank E. Cohen, J. Granot, J. I. Katz, S. Kobayashi, R. Narayan, and R. Sari
for many helpful discussions and D. Band and G. Blumenthal for helpful
remarks.  This work was supported by the US-Israel BSF grant 95-328
and by NASA grant NAG5-3516.


\clearpage

\clearpage

\end{document}